\documentclass[10pt]{article}
\usepackage{axodraw}
\usepackage{epsfig}
\usepackage{amsmath}
\newcommand{\eqnzero}{\setcounter{equation}{0}} 

\newcommand{\bq}{\begin{equation}}
\newcommand{\eq}{\end{equation}}
\newcommand{\ba}{\begin{eqnarray}}
\newcommand{\ea}{\end{eqnarray}}

\newcommand{\baa}[1]{\begin{array}{#1}}
\newcommand{\eaa}{\end{array}}

\newcommand{\nll}{\nonumber\\}

\newcommand{\Rw} {\mbox{$R_{\sss{W}}  $}}

\newcommand{\Rz} {\mbox{$R_{\sss{Z}}  $}}

\newcommand{\Litwo}{\mbox{${\rm{Li}}_{2}$}}

\def\gz{\Gamma_{\sss{Z}}}

\def\Pgg{\Pi_{\gamma\gamma}}
\def\gfd{\gamma_5}

\def\nl{\nonumber \\}
\def\nn{\nonumber}
\def\href#1#2{#2}

\newcommand{\bqa}{\begin{eqnarray}}
\newcommand{\eqa}{\end{eqnarray}}
\newcommand{\ds }{\displaystyle}
\newcommand{\sss}[1]{\scriptscriptstyle{#1}}
\newcommand{\QL}{\sss{QL}}
\newcommand{\LQ}{\sss{LQ}}
\newcommand{\zg}{{\sss{Z}}\ph}
\newcommand{\ph}{\gamma}
\newcommand{\ab}{A}
\newcommand{\zb}{Z}
\newcommand{\wb}{W}
\newcommand{\hb}{H}
\newcommand{\fe}{e}
\newcommand{\fbe}{{\bar{e}}}
\newcommand{\ff}{f}
\newcommand{\ffp}{f'}
\newcommand{\fep}{e^{+}}
\newcommand{\fem}{e^{-}}

\newcommand{\fu}{u}

\newcommand{\ft}{t}
\newcommand{\fb}{b}

\newcommand{\gap}{\gdp}
\newcommand{\gadi}[1]{\gamma_{#1}}
\newcommand{\cff}[6]{C_{0}\big( #1,#2,#3;#4,#5,#6 \big) }    
\newcommand{\sman}{s}
\newcommand{\mhl }{M_{\sss{H}}}

\newcommand{\mwl }{M_{\sss{W}}}
\newcommand{\mzl }{M_{\sss{Z}}}
\newcommand{\mfl }{m_f}
\newcommand{\mtl }{m_t}

\newcommand{\mbl }{m_b}
\newcommand{\mel }{m_e}
\newcommand{\mfpl}{m_{f'}}
\newcommand{\mfps}{m^2_{f'}}

\newcommand{\uml} {m_{t}}
\newcommand{\wml} {M_{\sss{W}}}
\newcommand{\zml}{ M  _{\sss{Z}}}
\newcommand{\dml}{ m   {_f}}

\newcommand{\mfs }{m^2_f}

\newcommand{\mul }{m  _t}
\newcommand{\mus }{m^2_t}
\newcommand{\mds }{m^2_b}
\newcommand{\mdf }{m^4_b}
\newcommand{\wmf} {M^4_{\sss{W}}}

\newcommand{\mws}{M^2_{\sss{W}}}
\newcommand{\mzs}{M^2_{\sss{Z}}}

\newcommand{\mzf}{M^4_{\sss{Z}}}
\newcommand{\mhs}{M^2_{\sss{H}}}
\newcommand{\mts}{m^2_{t}}

\newcommand{\mes}{m^2_e}

\newcommand{\mtq }{m^4_{t}}
\newcommand{\mfq }{m^4_{f}}

\newcommand{\asums}[1]{\sum_{#1}}
\newcommand{\cf}{c_f}
\newcommand{\Nf}{N_f}     
\newcommand{\fbf}{{\bar{f}}}

\newcommand{\tpfi}{\lpar 2\pi\rpar^4\ib}

\newcommand{\Vverti}[3]{V_{#1}^{#2}\lpar{#3}\rpar}

\newcommand{\lpar}{\left(}                            
\newcommand{\rpar}{\right)}
\newcommand{\lrbr}{\left[}
\newcommand{\rrbr}{\right]}

\newcommand{\ib  }{i}
\newcommand{\qf  }{Q_f  }
\newcommand{\qfm }{|Q_f|}

\newcommand{\qfs }{Q^2_f}

\newcommand{\qe  }{Q_e  }

\newcommand{\qes }{Q^2_e}

\newcommand{\qfp }{Q_{f'}}

\newcommand{\gbc }{g^3}
\newcommand{\gdp }{\gamma_{+  }}
\newcommand{\gdm }{\gamma_{-  }}
\newcommand{\gdpm}{\gamma_{\pm}}

\newcommand{\gdmu}{\gamma_{\mu}}
\newcommand{\Gverti}[3]{G_{#1}^{{#2}}\lpar{#3}\rpar}
\newcommand{\Zverti}[3]{Z_{#1}^{{#2}}\lpar{#3}\rpar}
\newcommand{\vvertil}[3]{F^{#1}_{#2}\lpar{#3}\rpar}
\newcommand{\vvertilt}[3]{\tilde{F}^{#1}_{#2}\lpar{#3}\rpar}
\newcommand{\cvetril}[3]{{\cal{F}}^{#1}_{#2}\lpar{#3}\rpar}
\newcommand{\cvertil}[3]{{\cal{F}}^{#1}_{#2}\lpar{#3}\rpar}

\newcommand{\fverti}[2]{F^{#1}_{#2}}

\newcommand{\gadu}[1]{\gamma_{#1}}
\newcommand{\siw }{s_{\sss{W}}}           
\newcommand{\cow }{c_{\sss{W}}}
\newcommand{\siws}{s^2_{\sss{W}}}
\newcommand{\cows}{c^2_{\sss{W}}}

\newcommand{\cowsc}{c^6_{\sss{W}}}
\newcommand{\siwf}{s^4_{\sss{W}}}
\newcommand{\cowf}{c^4_{\sss{W}}}
\newcommand{\vpa}[2]{\sigma_{#1}^{#2}}
\newcommand{\vpae }{\sigma_e}
\newcommand{\vma}[2]{\delta_{#1}^{#2}}

\newcommand{\tcie}{I^{(3)}_e}
\newcommand{\tcif}{I^{(3)}_f}
\newcommand{\tcit}{I^{(3)}_t}
\newcommand{\tcib}{I^{(3)}_b}
\newcommand{\saff}[1]{A_{#1}}             
                   
\newcommand{\sbff}[1]{B_{#1}}             
\newcommand{\sfbff}[1]{B^{F}_{#1}}
\newcommand{\bff}[4]{B_{#1}\big( #2;#3,#4\big)}             
\newcommand{\fbff}[4]{B^{F}_{#1}\big(#2;#3,#4\big)}        
\newcommand{\scff}[1]{C_{#1}}             
\newcommand{\sdff}[1]{D_{#1}}                 
\newcommand{\dffp}[6]{D_{0} \big( #1,#2,#3,#4,#5,#6;}       
\newcommand{\dffm}[4]{#1,#2,#3,#4 \big) }       
\newcommand{\delrho}[1]{{\Delta \rho}^{#1}}
\newcommand{\bdelrho}[1]{{\Delta\bar \rho}^{#1}}

\newcommand{\fbd}{{\overline{d}}}
\newcommand{\wbm}{W^{-}}
\newcommand{\wbp}{W^{+}}

\newcommand{\Lnrt}{\Lmmt}
\newcommand{\Lmme}{L_\mu(\mes)}
\newcommand{\Lmmt}{L_\mu(\mfs)}
\newcommand{\Lmmz}{L_\mu(\mzs)}
\newcommand{\Lmmh}{L_\mu(\mhs)}
\newcommand{\Lmmw}{L_\mu(\mws)}

\newcommand{\rhw}{r_{_{\hb\wb}}}
\newcommand{\rhws}{r^2_{_{\hb\wb}}}
\newcommand{\rhzs}{r^2_{_{\hb\zb}}}

\newcommand{\Dz}[2]{{\cal{D}}_{\sss Z}^{#1}\lpar{#2}\rpar}
\newcommand{\Pzg}{\Pi_{\zg}}

    \newcommand{\tHs}{\mu}
    \newcommand{\tHss}{\mu^2}
    
    \newcommand{\stwl}{s_{\sss{W}}  }
    \newcommand{\ctwl}{c_{\sss{W}}  }
    \newcommand{\stws}{s^2_{\sss{W}}}
    \newcommand{\stwf}{s^4_{\sss{W}}}
    \newcommand{\ctws}{c^2_{\sss{W}}}
    \newcommand{\ctwf}{c^4_{\sss{W}}}

  \newcommand{\vpau }{\sigma  _f}
  \newcommand{\vmau }{\delta  _f}
  \newcommand{\vmae }{\delta  _e}

\newcommand{\bos}{\rm{bos}}
\newcommand{\fer}{\rm{fer}}
\newcommand{\pole}{\dlt}

\newcommand{\Trqf}{ \sum_f c_f Q^2_f }

\newcommand{\ve }{v_e}  
\newcommand{\aee}{a_e}  
\newcommand{\asum}[3]{\sum_{#1=#2}^{#3}}
\newcommand{\minds}[1]{m^2_{#1}}
\newcommand{\Minds}[1]{M^2_{#1}}
\newcommand{\dlt}{\displaystyle{\frac{1}{\epsb}}}
\newcommand{\epsh}{\hat\varepsilon}
\newcommand{\epsb}{\bar\varepsilon}

\newcommand{\eqn}[1]{Eq.~(\ref{#1})}
\newcommand{\eqns}[2]{Eqs.~(\ref{#1})--(\ref{#2})}

\newcommand{\eqnsc}[2]{Eqs.~(\ref{#1}) and (\ref{#2})}

\newcommand{\tbn}[1]{Table~\ref{#1}}

\newcommand{\tbnsc}[2]{Tabs.~\ref{#1} and \ref{#2}}
\newcommand{\fig}[1]{Fig.~\ref{#1}}

\newcommand{\bPzga}[2]{{\bar\Pi}^{#1}_{_{\zb\gamma}}\lpar#2\rpar}

\newcommand{\qb}{Q_b}
\newcommand{\qt}{Q_t}

\newcommand{\qu}{Q_f}

\newcommand{\af}{a_f}
\newcommand{\vf}{v_f}
\newcommand{\ruz}{r_{f{\sss Z}}}
\newcommand{\rtz}{r_{f{\sss Z}}}
\newcommand{\rth}{r_{f{\sss H}}}
\newcommand{\rht}{r_{{\sss H}f}}

\newcommand{\Longab}[3]{L_{ab} \lpar #1,#2,#3\rpar}        
\newcommand{\Longna}[3]{L_{na} \lpar #1,#2,#3\rpar}        
\newcommand{\LongHi}[3]{L_{Hi} \lpar #1,#2,#3\rpar}

\newcommand{\rtw }{r_{f{\sss W}}}

\newcommand{\ruw }{r_{f{\sss W}}}
\newcommand{\ruh }{r_{f{\sss H}}}
\newcommand{\rhz }{r_{\sss HZ}}
\newcommand{\hkg}{\phi}
\newcommand{\hkn}{\phi^{0}}                 
\newcommand{\fbfp}{{\overline{f}}'}
\newcommand{\Rxi}{R_{\gpar}}
\newcommand{\gpar}{\xi}
\newcommand{\hkp}{\phi^{+}}
\newcommand{\fpxm}{X^-}
\newcommand{\fpyZA}{Y_{\ssZ,\gamma}}
\newcommand{\ssZ}{{\sss{Z}}}
\newcommand{\fpxp}{X^+}
\newcommand{\btp}{\beta_t}
\newcommand{\hkm}{\phi^{-}}
\newcommand{\sdfit} {\Delta_{4r}}
\newcommand{\sdtit} {\Delta_{3r}}

\newcommand{\FQLt }{{\tilde F}  _{\sss QL}\lpar s,t,u \rpar}
\newcommand{\FQLtc}{{\tilde F}^*_{\sss QL}\lpar s,t,u \rpar}
\newcommand{\FLQt }{{\tilde F}  _{\sss LQ}\lpar s,t,u \rpar}
\newcommand{\FLQtc}{{\tilde F}^*_{\sss LQ}\lpar s,t,u \rpar}
\newcommand{\FQQt }{{\tilde F}  _{\sss QQ}\lpar s,t,u \rpar}
\newcommand{\FQQtc}{{\tilde F}^*_{\sss QQ}\lpar s,t,u \rpar}
\newcommand{\FLLt }{{\tilde F}  _{\sss LL}\lpar s,t,u \rpar}
\newcommand{\FLLtc}{{\tilde F}^*_{\sss LL}\lpar s,t,u \rpar}
\newcommand{\FLDt }{{\tilde F}  _{\sss LD}\lpar s,t,u \rpar}

\newcommand{\FQDt }{{\tilde F}  _{\sss QD}\lpar s,t,u \rpar}
\newcommand{\FQDtc}{{\tilde F}^*_{\sss QD}\lpar s,t,u \rpar}
\newcommand{\ip }[1]{u\lpar{#1}        \rpar}    
\newcommand{\iap}[1]{{\bar{v}}\lpar{#1}\rpar}    
\newcommand{\op}[1]{{\bar{u}}\lpar{#1}\rpar}     
\newcommand{\oap}[1]{v\lpar{#1}\rpar}            
\newcommand{\qmomi}[1]{q_{#1}}
\newcommand{\rbw }{r_{\ffp{\sss W}}}
\newcommand{\rdw }{r_{\ffp{\sss W}}}
\newcommand{\rD  }{r^-}
\newcommand{\rP  }{r^+}
\newcommand{\mbs }{m^2_{b}}

\newcommand{\Lnwm}{L_\mu(\mws)}

\newcommand{\tHlas}{\lambda^2}

\newcommand{\tmi }{t_-}
\newcommand{\tmip }{t^{'}_-}
\newcommand{\tmis}{t^2_-}
\newcommand{\umi }{u_-}
\newcommand{\sqs}{\sqrt{s}}
\newcommand{\szmi}{s_-}
\newcommand{\szpl}{s_+}
\newcommand{\tpl}{t_+}
\newcommand{\tplp}{t^{'}_+}
\newcommand{\tpls}{t^2_+}

\newcommand{\jaat}{J_{\sss AA}(-s,-t;\mel,\mfl)}

\newcommand{\jazt}{J_{\sss AZ}(-s,-t;\mel,\mfl)}

\newcommand{\cesoeo}{\cff{-\mes}{-\mes}{-s}{0}{\mel}{0}}
\newcommand{\ceszeo}{\cff{-\mes}{-\mes}{-s}{\mzl}{\mel}{0}}
\newcommand{\ctszto}{\cff{-\mfs}{-\mfs}{-s}{\mzl}{\mfl}{0}}
\newcommand{\ctsoto}{\cff{-\mfs}{-\mfs}{-s}{0}{\mfl}{0}}
\newcommand{\cattze}{\cff{-\mfs}{-\mes}{-t}{\mfl}{\mzl}{\mel}}

\newcommand{\cateot}{\cff{-\mes}{-\mfs}{-t}{\mel}{0}{\mfl}}

\newcommand{\ctstot}{\cff{-\mfs}{-\mfs}{-s}{\mfl}{0}{\mfl}}
\newcommand{\boptot}{\bff{0p}{-\mfs}{0}{\mfl}}

\newcommand{\Faa}[1]{{\cal F}^{\sss AA}_{\sss #1}}
\newcommand{\Fza}[1]{{\cal F}^{\sss ZA}_{\sss #1}}

\newcommand{\bofsoo}{\fbff{0}{-s}{0}{0}}
\newcommand{\bofszo}{\fbff{0}{-s}{\mzl}{0}}
\newcommand{\boftzt}{\fbff{0}{-\mfs}{\mzl}{\mfl}}
\newcommand{\boftet}{\fbff{0}{-t}{\mel}{\mfl}}

\newcommand{\bofstt}{\fbff{0}{-s}{\mfl}{\mfl}}
\newcommand{\boftto}{\fbff{0}{-\mfs}{\mfl}{0}}
\newcommand{\bebeta}{\frac{(1+\beta^2)}{2 \beta}}      
\newcommand{\betatm}{\beta_-}
\newcommand{\betatp}{\beta_+}
\newcommand{\lelog}{{l}_e}
\newcommand{\tman}{t}
\newcommand{\uman}{u}
\newcommand{\pmomi}[1]{p_{#1}}

\newcommand{\pmoms}{p^2}
\newcommand{\pmom}{p}
\newcommand{\mf }{m^2_f}
\newcommand{\Lmmb}{L_\mu(\mfps)}
\newcommand{\sz}{s_z}
\newcommand{\szs}{s^2_z}
\newcommand{\ums}{\mfs}
\newcommand{\tciu}{I^{(3)}_f}
\newcommand{\cusdwd}{\cff{-\ums}{-\ums}{-s}{\dml}{\wml}{\dml}}

\newcommand{\cosozo}{\cff{0}{0}{-s}{0}{\zml}{0}}
 
\newcommand{\coswdw}{\cff{0}{0}{-s}{\wml}{\dml}{\wml}}
\newcommand{\bofsww}{\fbff{0}{-s}{\mwl}{\mwl}}
\newcommand{\bofsdd}{\fbff{0}{-s}{\dml}{\dml}}
\newcommand{\Lnzm}{\Lmmz}
\newcommand{\lrbw}{d_{\sss W}}
\newcommand{\wms}{\mws}
\newcommand{\Lndm}{L_\mu(\mfps)}

\begin{document}
\def\theequation{\arabic{section}.\arabic{equation}}
\newcommand{\chic}{\chi^*}
\def\href#1#2{#2}

\setcounter{page}{0}
\thispagestyle{empty}

\begin{flushright}
{\bf
{\tt hep-ph/0207156} \\
}
\end{flushright}
\vspace*{\fill}
\begin{center}

{\LARGE\bf
Update of one-loop corrections for $e^{+}e^{-}\to\ff\bar{f}$, \\[2.5mm]
first run of {\tt SANC} system}
\vspace*{1.5cm}

{\bf 
A.~Andonov, D.~Bardin, S.~Bondarenko$^*$,  \\[1mm]
P.~Christova, L.~Kalinovskaya, and G.~Nanava
}

\vspace*{3mm}
{\normalsize
{\it Laboratory for Nuclear Problems, JINR,\\[1mm]
$^*$ Bogoluobov Laboratory of Theoretical Physics, JINR,\\[1mm] 
     ul. Joliot-Curie 6,
     RU-141980 Dubna, Russia}}
\vspace*{1.5cm}

\end{center}

\begin{abstract}

We present a description of calculations of the amplitude for $e^+e^-\to\ff\bar{f}$ process
with account of electroweak and QED one-loop corrections.  
This study is performed within the framework of the project {\tt SANC}.
The calculations are done within the OMS (on-mass-shell) renormalization scheme
in two gauges: in $\Rxi$, which allows an explicit control of gauge invariance
by examining cancellation of gauge parameters
and search for gauge-invariant subsets of diagrams,
and in the unitary gauge as a cross-check.
The formulae we derived are realized in two independent {\tt FORTRAN} codes,
{\tt eeffLib}, which was written in an old fashioned way, i.e. manually,
and another one, created automatically with an aid of {\tt s2n.f} (symbols to numbers) 
software --- a part of  {\tt SANC} system.
We present a comprehensive comparison with the results of the well-known program 
{\tt ZFITTER} for all the light fermion production channels, as well as with the results 
existing in the world literature for the process $e^+e^-\to\ft\bar{t}$.

\end{abstract}
\vspace*{1mm}
\vfill
\centerline{Submitted to {\it Particles and Nuclei}}
\vspace*{1mm}
\vfill
\bigskip
\footnoterule
\noindent
{\footnotesize \noindent
Work supported by INTAS $N^{o}$ 00-00313.
\\
E-mails: andonov@nusun.jinr.ru, bardin@nusun.jinr.ru, bondarenko@jinr.ru \\ 
\phantom{XXXXX}
penchris@nusun.jinr.ru, kalinov@nusun.jinr.ru, nanava@nusun.jinr.ru
}
\clearpage
\tableofcontents
\clearpage
\listoffigures
\listoftables    
\clearpage
\addcontentsline{toc}{section}{Introduction}
\section*{Introduction\label{eett_introduction}}
\eqnzero
The process $e^+e^-\to\ff\bar{f}$ with taking into account of the higher-order 
Standard Model (SM) corrections has been studied already several decades, 
yet in pre-LEP times.
We mention papers from late seventies,~\cite{kn:pv},~\cite{kn:borig},
~\cite{sord} and \cite{kn:woh}, which eventually lead to 
producing of precision theoretical predictions realized in the well-known 
computer codes of LEP-era: 
{\tt ZFITTER}~\cite{Bardin:1999yd}, {\tt BHM}~\cite{kn:bhm}
{\tt ALIBABA}~\cite{kn:ali} and {\tt TOPAZ0}~\cite{Montagna:1999kp}.
These codes played a very important role throughout the full life-time of
LEP providing experimental community with precision tools for fitting
of experimental data to the SM predictions, see, 
for instance~\cite{martin_hab}--\cite{LEPEWWG99}.
A comprehensive review of underlying theory and methods which has been used 
for creation of these codes may be found in the monograph~\cite{Bardin:1999ak}.

At LEP, the higher-order corrections for process $e^+e^-\to\ff\bar{f}$ 
might be studied ignoring external fermion masses, since the cms energy was
far below $\ft\bar{t}$ threshold. However, studies of finite mass effects
has been started more than ten years ago \cite{2oftwo},
\cite{Beenakker:1991ca} and \cite{Beenakker:1991kh}
in connection with experiments at future linear 
colliders (see, for instance, the reviews \cite{TDR} and
\cite{Beneke:2000hk}).

New wave of papers on finite mass effect was triggered by studies of 
the electroweak radiative corrections (EWRC) in the MSSM
which may involve additional heavy fermions and sfermions,
\cite{Beenakker:1993js}--\cite{Driesen:1996tn}.

\begin{figure}[!h]
\[
\begin{picture}(132,132)(0,0)
\ArrowLine(-40,22)(-2,66)
 \Text(-30,22)[lb]{$e^-$}
\ArrowLine(-2,66)(-40,110)
 \Text(-30,110)[lb]{$e^+$}
 \Photon(-2,66)(44,66){2}{7}
 \Text(12, 70)[lb]{$\gamma, Z$}
\ArrowLine(44,66)(66,88)
 \Text(45, 80)[lb]{$t$}
\ArrowLine(66,44)(44,66)
 \Text(45,47)[lb]{$\bar{t}$}
\Photon(66,44)(88,22){2}{5}
 \Text( 60, 99)[lb]{$W$}
\ArrowLine(88,22)(110,44)
\ArrowLine(110,0)(88,22)
 \Text(114,44)[lb]{$l$}
 \Text(114,0)[lb]{$\bar{\nu}$}
\Photon(66,88)(88,110){2}{5}

 \Text( 60,25)[lb]{$W$}
\ArrowLine(110,88)(88,110)
\ArrowLine(88,110)(110,132)
 \Text(114,132)[lb]{$f_1$}
 \Text(114, 88)[lb]{$\bar{f_2}$}
\ArrowLine(66,88)(90,68)
\ArrowLine(90,64)(66,44)
 \Text(85 , 80)[lb]{$b$}
 \Text(85 ,47)[lb]{$\bar{b}$}
\end{picture}
\]
\caption[The six-fermion $e^+e^-\to\ft\bar{t}$ process.]
{The six-fermion $e^+e^-\to\ft\bar{t}$ process.\label{six-fermion}}
\end{figure}
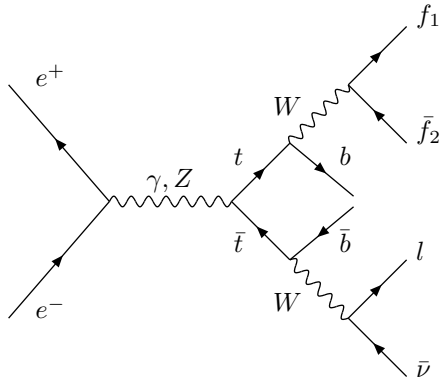

The process $e^+e^-\to\ft\bar{t}$ in reality
is a six-fermion process (see \cite{Piccinini:2000ib}); 
one of the channels is shown in~\fig{six-fermion}.

However, the cross-section of a hard subprocess, 
$\sigma(\fep\fem\to\ft\bar{t})$,
with tops on the mass shell is an ingredient within various approaches, such as 
DPA~\cite{Denner:2000bj} or the so-called Modified Perturbation Theory (MPT),
see~\cite{Nekrasov:1999cj}.

One should emphasise that the treatment of even one-loop corrections with
taking into account of finite-state fermion masses is extremely cumbersome 
and practically undouble `manually'. Nowadays, all the calculations of
such a kind are being performed with the aid of automatic computer systems,
among which one should mention first of all {\tt FeynArts} package, which 
exists already more than ten years~\cite{Hahn:2000jm} and which is able 
to compute the cross-section of the process $\sigma(\fep\fem\to\ft\bar{t})$ 
at the one-loop level.

In this article, we present a review of a new calculation 
(based on preprints \cite{eett_subm}--\cite{CalcPHEP:2000})
of the $e^+e^-\to \ff \bar\ff $ process at the one-loop level
made with an aid of the computer system {\tt SANC}
where all the calculations from the Lagrangians up to numbers
are going to be eventually automatized. 
This system is being created at the site {\tt brg.jinr.ru}.
It roots back to dozens of supporting {\tt form}~\cite{Vermaseren:2000f}
codes written by authors of the book~\cite{Bardin:1999ak} while working on it. 
Later on, the idea came up to collect, 
order, unify and upgrade these codes up to the level of a `computer system'.
Its first phase is described in~\cite{CalcPHEP:2000}.
 
One of main goals of this new calculation was to 
create a platform for the treatment of any $2\ff\to2\ff$ process within
the {\tt SANC} and to cross-check its
results against the results obtained with the other 
existing codes for a rather complicated $2\ff\to2\ff$ process 
at one-loop level taking into account final state masses.

At present, {\tt SANC}, like {\tt FeynArts},
uses the OMS renormalization sche\-me, a complete presentation 
of which was made in \cite{Bardin:1999ak}. However, for the first
time the calculations are performed in two gauges simultaneously: 
$\Rxi$ and the unitary gauge. 

  Note, that there was wide experience of calculations in the $R_\xi$ gauge
for processes such as  
$H\to\ff\bar{f},WW,ZZ,\gamma Z,\gamma\gamma$, or $e^+e^-\to ZH,WW$.
So, in \cite{Fleischer:1981ub} and \cite{Jegerlehner:1983bf}
a complete set of one-loop counterterms for the SM is given.
 Electromagnetic form factors for arbitrary $\xi$ are discussed in
\cite{Jegerlehner:1985ch} and \cite{Jegerlehner:1986ia}.
 Explicit expressions can be found in the CERN library program EEWW 
\cite{Fleischer:1987xa}.

 However, we are not aware of the existence of calculations in the $\Rxi$ gauge
for the $e^+e^-\to\ft\bar{t}$ process, although there are many studies in 
the $\xi=1$ gauge \cite{2oftwo}--\cite{Beenakker:1991kh} and
\cite{Beenakker:1993js}--\cite{Driesen:1996tn}.

\vspace*{3mm}

 Additional purposes of this review are: 

\begin{itemize}

\item[$-$] to explicitly demonstrate gauge invariance in $\Rxi$ by examining
cancellation of gauge parameters for gauge-invariant subsets of diagrams;

\item[$-$] to offer a possibility to compare the results with those in the 
unitary gauge, as a cross-check;

\item[$-$] to present a self-contained list of results for one-loop amplitude
in terms of Passarino--Veltman functions $\saff{0},\;\sbff{0},\;\scff{0}$ 
and $\sdff{0}$ (as well as {\em special} functions $a_0,\,b_0,\,c_0$ and $d_0$
which originate because of a particular form of the photonic propagator in the $\Rxi$
gauge) and their combinations in the spirit of the 
book~\cite{Bardin:1999ak}, where the process $e^+e^-\to\ft\bar{t}$ was
not covered; this review may thus be considered as an Annex to the book,
which completes its pedagogical aims for $2\ff\to2\ff$ processes at one loop; 

\item[$-$] to provide a {\tt FORTRAN} code, {\tt eeffLib}~\cite{eeffLib:2001}
for the calculation of the cross-section of this process for a complementary 
use within the MPT framework;

\item[$-$] to compare the results derived with the aid of {\tt eeffLib} 
with the results of another code, which was created automatically 
using the {\tt s2n.f} software --- a part of {\tt SANC} system,
thereby benchmarking {\tt s2n.f} software.

\end{itemize}

This review consists of seven sections. 

\vspace*{2mm}

In Section 1, we present the Born amplitude of the process, basically to 
introduce our notation and then define {\em the basis} in which the one-loop 
amplitude was calculated. 
We explain {\em the splitting} between QED and EW corrections and between 
`dressed' $\ph$ and $\zb$ exchanges.

\vspace*{2mm}

Section 2 contains explicit expressions for all {\em the building blocks}: 
self-energies, vertices and boxes, both QED and EW.
Note that no diagram was computed by hand.
They all were supplied by the {\tt SANC} system.

\vspace*{2mm}

In Section 3, we describe the procedure of construction of 
{\em the scalar form factors} of the one-loop amplitudes out of the building 
blocks. One of the aims of this section was to create a frame for a subsequent
realization of the renormalization procedure within the {\tt SANC} project.

\vspace*{2mm}

Section 4 contains explicit expressions for the improved Born approximation
(IBA) cross-section and the explicit expressions for helicity amplitudes 
made of the scalar form factors at the one-loop level.

\vspace*{2mm}

Section 5 contains some additional expressions for different QED contributions
that might be derived analytically. They are not in the main stream of our 
approach:
Lagrangian $\rightarrow$ scalar form factors $\rightarrow$ helicity amplitudes 
$\rightarrow$ one-loop differential cross-section. However, they are  
useful for the pedagogical reasons, and their coding in complementary 
{\tt FORTRAN} branches
of {\tt eeffLib} provided us with the powerful internal cross-checks of our 
codes for numerical calculations.
In reality, the {\tt eeffLib} version of February 2002 has three QED branches.

\vspace*{2mm}

Finally, in Section 6 we present results of a comprehensive numerical 
comparison between {\tt eeffLib} and {\tt ZFITTER}.
Here we also present a comparison with  another of our codes, which was
created automatically using the {\tt s2n.f} software.
We also present a comprehensive comparison between the results derived with 
our two codes and the results existing in the world literature. 
In particular, we found a high-precision agreement with {\tt FeynArts} 
results up to 11 
digits for the differential cross-sections with virtual corrections,
and with recent results of~\cite{Zeuthen:2001} within 8 digits, 
even when the soft photon radiation is included, 
see~\cite{K-BZcomparison:2002}--\cite{BZ_Japan}.

\section{Amplitudes\label{amplitudes}}
\subsection{Born amplitudes\label{B-amplitudes}}
We begin with the Born amplitudes for the process 
$\fep(\pmomi{+})\fem(\pmomi{-})\to$ \linebreak
$\ff(\qmomi{-}){\bar f}(\qmomi{+})$, 
which is described by the two Feynman diagrams with $\ph$ and $\zb$ exchange.
The Born amplitudes are:
\bqa
A^{\sss{B}}_{\ph} &=&e\qe\,e\qf
                  \gadu{\mu} \otimes \gadu{\mu} \frac{-\ib}{Q^2}
               = -\ib\,4\pi\alpha(0)\frac{\qe\qf}{Q^2}
                  \gadu{\mu} \otimes \gadu{\mu}\,,
\\ 
\label{amplborn}
A^{\sss{B}}_{\sss{\zb}} &=&
       \lpar\frac{e}{2\stwl\ctwl}\rpar^2
                         \gadu{\mu}
       \lrbr \tcie \gdp -2\qe \stws \rrbr 
                 \otimes \gadu{\mu}  
       \lrbr \tcif \gdp -2\qf \stws \rrbr 
       \frac{-\ib}{Q^2+\mzs}                                      
\nll &=& -                   
\ib e^2\frac{1}{4\stws\ctws(Q^2 +\mzs)}
\biggl[
  \tcie\tcif \gadu{\mu}\gdp\otimes\gadu{\mu}\gdp 
+\vmae \tcif \gadu{\mu}\otimes\gadu{\mu}\gdp 
\nll &&
+\tcie \vmau \gadu{\mu}\gdp\otimes\gadu{\mu}
+\vmae \vmau \gadu{\mu}\otimes\gadu{\mu} 
\biggr]\,,
\eqa
where $\;\gdpm=1\pm\gfd\;$ and the symbol $\otimes$ is used in the following 
short-hand notation:
\bqa
&&\gadu{\mu}\lpar L_{1}\gdp+Q_{1} \rpar\otimes
\gadu{\nu}\lpar L_{2}\gdp+Q_{2} \rpar
\nl
&&=
\iap{\pmomi{+}}\gadu{\mu}\lpar L_{1}\gdp+Q_{1} \rpar\ip{\pmomi{-}} 
\op{\qmomi{-}}\gadu{\nu}\lpar L_{2}\gdp+Q_{2} \rpar\oap{\qmomi{+}};
\label{dlia_Lidy}
\eqa
furthermore
\bqa
\delta_f &=& \vf - \af = - 2 \qf \stws\,.
\eqa

Introducing the $LL$, $QL$, $LQ$, and $QQ$ structures, correspondingly
(see last \eqn{amplborn}), we have four structures to which the 
complete Born amplitude may be reduced: one for the $\ph$ exchange amplitude 
and four for the $\zb$ exchange amplitude ($\gadu{\mu}\otimes\gadu{\mu}$
structure presents in both amplitudes).
\subsection{One-loop amplitude for $e^+e^- \to\ff{\bar f}$}
 For the $e^+e^- \to\ff{\bar f}$ process at one loop,
it is possible to consider a gauge-invariant subset
of {\em electromagnetic corrections} separately: QED vertices, $\ab\ab$ and 
$\zb\ab$ boxes. Together with QED bremsstrahlung diagrams, it is free of 
infrared divergences. 
The total electroweak amplitude is a sum of `dressed' $\gamma$ and $Z$
exchange amplitudes, plus the contribution from the weak box diagrams
($WW$ and $ZZ$ boxes).

 Contrary  to the Born amplitude, the one-loop amplitude may be paramet\-rized
by 6 form factors, a number equal to the number of independent helicity 
amplitudes for this process if the electron mass is ignored and the
unpolarized case is studied\footnote{
If the initial-state masses were not ignored too, we would have ten 
independent helicity
amplitudes, ten structures and ten scalar form factors.}.

 We work in the so-called $LQD$ basis, in which 
the amplitude may be schematically represented as:
\bqa
  \lrbr i \gadu{\mu}\gdp \vvertil{e}{\sss{L}}{\sman} 
       +i \gadu{\mu}                   \vvertil{e}{\sss{Q}}{\sman} \rrbr
  \otimes
  \lrbr 
        i \gadu{\mu}\gdp \vvertil{f}{\sss{L}}{\sman} 
       +i \gadu{\mu}                   \vvertil{f}{\sss{Q}}{\sman} 
        + \mfl I D_\mu                  \vvertil{f}{\sss{D}}{\sman} \rrbr,
\eqa
with 
\bqa
D_\mu=(\qmomi{+}-\qmomi{-})_{\mu}\,.
\label{difference}
\eqa
Every form factor in the $\Rxi$ gauge could be represented as a sum of two 
terms:
\bqa
\vvertil{\xi}{\sss{L,Q,D}}{\sman}=
\vvertil{(1)}{\sss{L,Q,D}}{\sman} + \vvertil{\rm add}{\sss{L,Q,D}}{\sman}.
\eqa
The first term corresponds to the $\xi=1$ gauge and the second contains
all $\xi$ dependences and vanishes for $\xi=1$ by construction. 

 The $LQD$ basis was found to be particularly convenient to 
explicitly demonstrate the cancellation of all $\xi$-dependent terms.
We checked the cancellation of these terms in several groups of diagrams
separately:
the so-called $\ab$, $\zb$, and $\hb$ clusters, defined below; 
the $\wb$ cluster together with the self-energies and the $\wb\wb$ box; 
and the $\ab\ab$, $\zb\ab$ and $\zb\zb$ boxes. 
Therefore, for our process we found seven separately 
gauge-invariant subgroups of diagrams: three in the QED sector, and four in 
the EW sector.

 The `dressed' $\ph$ exchange amplitude is
\bqa
A^{\sss{\rm{IBA}}}_{\ph} = \ib\frac{4\pi\qe\qf}{\sman}
\alpha(\sman) \gadu{\mu} \otimes \gadu{\mu}\,,
\label{Born_modulo-old}
\eqa
which is identical to the Born amplitude of \eqn{amplborn} 
modulo the replacement of $\alpha(0)$ by the
running electromagnetic coupling $\alpha(\sman)$:
\bqa
\alpha(\sman)=\frac{\alpha}
{\ds{1-\frac{\alpha}{4\pi}\Bigl[\Pgg^{\fer}(\sman)-\Pgg^{\fer}(0)\Bigr]}}\,.
\label{alpha_fer-old}
\eqa

 In the $LQD$ basis the $\zb$ exchange amplitude has the following Born-like 
structure in terms of six ($LL$, $QL$, $LQ$, $QQ$, $LD$ and $QD$) form factors:
\begin{align}
\label{structures-old}
{\cal A}^{\sss{\rm{IBA}}}_{\sss{\zb}}=&
\ib\,e^2
\frac{\chi_{\sss{Z}}(\sman)}{\sman}
   \biggl\{\gadu{\mu} {\gdp } \otimes
        \gadu{\mu} {\gdp } \vvertilt{}{\sss{LL}}{\sman,\tman} 
+\gadu{\mu}     
      \otimes \gadu{\mu} {\gdp} \vvertilt{}{\sss{QL}}{\sman,\tman} 
\nll[2mm] &
+\gadu{\mu}{\gdp}\otimes\gadu{\mu} \vvertilt{}{\sss{LQ}}{\sman,\tman}
+\gadu{\mu}\otimes\gadu{\mu}
 \vvertilt{}{\sss{QQ}}{\sman,\tman}
\\[2mm] &
+\gadu{\mu}{\gdp}\otimes\lpar 
     -\ib\mfl D_{\mu} \rpar  \vvertilt{}{\sss{LD}}{\sman,\tman}
+\gadu{\mu} \otimes \lpar  
     -\ib\mfl D_{\mu} \rpar  \vvertilt{}{\sss{QD}}{\sman,\tman}
\biggr\},\qquad
\nonumber
\end{align}
where we introduce the notation for $\vvertilt{}{ij}{\sman,\tman}$ :
\bqa
\vvertilt{}{\sss{LL}}{\sman,\tman} &=& \tcie\tcif\vvertil{}{\sss{LL}}{\sman,\tman},
\nll     
\vvertilt{}{\sss{QL}}{\sman,\tman} &=& \vmae\tcif\vvertil{}{\sss{QL}}{\sman,\tman},
\nll
\vvertilt{}{\sss{LQ}}{\sman,\tman} &=& \tcie\vmau\vvertil{}{\sss{LQ}}{\sman,\tman},
\nll
\vvertilt{}{\sss{QQ}}{\sman,\tman} &=& \vmae\vmau\vvertil{}{\sss{QQ}}{\sman,\tman},
\nll
\vvertilt{}{\sss{LD}}{\sman,\tman} &=& \tcie\tcif\vvertil{}{\sss{LD}}{\sman,\tman},
\nll
\vvertilt{}{\sss{QD}}{\sman,\tman} &=& \vmae\tcif\vvertil{}{\sss{QD}}{\sman,\tman}.
\eqa
Note that {\it tilded} form factors absorb couplings, which leads to a compactification 
of formulae for the amplitude and IBA cross-section, while explicit expressions will be 
given for {\it untilded} quantities.
The representation of \eqn{structures-old} is very convenient for the subsequent 
discussion of one-loop amplitudes.
 
Furthermore, in \eqn{structures-old} we 
use the $\zb/\ph$ propagator ratio with an $\sman$-dependent (or constant) $\zb$ width:   
\bqa
\chi_{\sss{Z}}(\sman)&=&\frac{1}{4\siws\cows}\frac{\sman}
{\ds{\sman - \mzs + \ib\frac{\gz}{\mzl}\sman}}\,.
\label{propagators}
\eqa
\section{ Building Blocks in the OMS Approach\label{building-blocks}}
\eqnzero
We start our discussion by presenting various {\it building blocks}, used 
to construct the one-loop form factors of the processes $\fep\fem\to\ff\fbf$ 
in terms of the $\saff{0},\;\sbff{0},\;\scff{0}$ and $\sdff{0}$ functions. 
They are shown in order of increasing complexity: self-energies, 
vertices, and boxes.
 
\subsection{Bosonic self-energies \label{bosonicse}}
\subsubsection{$\zb,\ph$ bosonic self-energies and $\zb$--$\ph$ transition
\label{bsetran}}
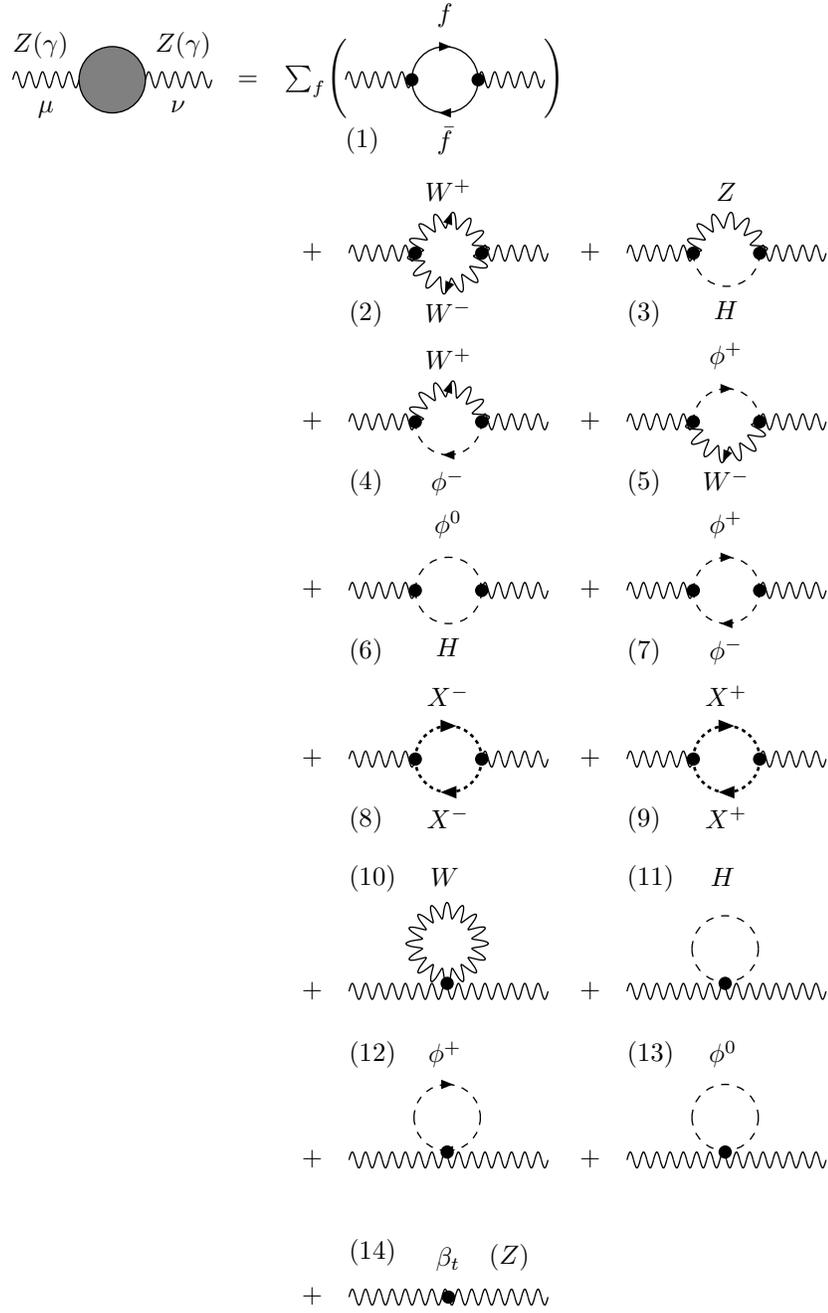
\begin{figure}[h]
\vspace*{-5mm}
\[
\baa{ccccccc}
\begin{picture}(75,20)(0,8)
  \Photon(0,10)(25,10){3}{5}
    \GCirc(37.5,10){12.5}{0.5}
  \Photon(50,10)(75,10){3}{5}
\Text(0,18)[bl]{$\zb(\gamma)$}
\Text(12.5, 2)[tc]{$\mu$}
\Text(75,18)[br]{$\zb(\gamma)$}
\Text(62.5, 2)[tc]{$\nu$}
\end{picture}
&=&{\LARGE{\sum_f\Biggl(\phantom{\Biggr)} }}
\hspace*{-5mm}
\begin{picture}(75,20)(0,8)
  \Photon(0,10)(25,10){3}{5}
  \ArrowArcn(37.5,10)(12.5,0,180)
  \ArrowArcn(37.5,10)(12.5,180,0)
    \Vertex(25,10){2.5}
    \Vertex(50,10){2.5}
  \Photon(50,10)(75,10){3}{5}
\Text(37.5,-8)[tc]{${\bar f}$}
\Text(37.5,30)[bc]{$\ff$}
\Text(0,-8)[lt]{$(1)$}
\end{picture}
\Biggr)
\nl \nl \nl
&&+\quad
\begin{picture}(75,20)(0,8)
  \Photon(0,10)(25,10){3}{5}
  \PhotonArc(37.5,10)(12.5,0,180){3}{7}
  \PhotonArc(37.5,10)(12.5,180,0){3}{7}
    \Vertex(25,10){2.5}
    \Vertex(50,10){2.5}
  \Photon(50,10)(75,10){3}{5}
    \ArrowLine(37.6,22.2)(38.3,24.2)
    \ArrowLine(37.5,-2.2)(36.7,-4)
\Text(37.5,-8)[tc]{$\wbm$}
\Text(37.5,30)[bc]{$\wbp$}
\Text(0,-8)[lt]{$(2)$}
\end{picture}
&&\hspace*{-5mm}+\quad
\begin{picture}(75,20)(0,8)
  \Photon(0,10)(25,10){3}{5}
  \PhotonArc(37.5,10)(12.5,0,180){3}{7}
  \DashCArc(37.5,10)(12.5,180,0){3.}
    \Vertex(25,10){2.5}
    \Vertex(50,10){2.5}
  \Photon(50,10)(75,10){3}{5}
\Text(37.5,-8)[tc]{$\hb$}
\Text(37.5,30)[bc]{$\zb$}
\Text(0,-8)[lt]{$(3)$}
\end{picture}
\nl \nl \nl \nl[1.5mm]
&&+\quad
\begin{picture}(75,20)(0,8)
  \Photon(0,10)(25,10){3}{5}
  \PhotonArc(37.5,10)(12.5,0,180){3}{7}
  \DashArrowArcn(37.5,10)(12.5,0,180){3.}
    \Vertex(25,10){2.5}
    \Vertex(50,10){2.5}
  \Photon(50,10)(75,10){3}{5}
    \ArrowLine(37.6,22.2)(38.3,24.2)
\Text(37.5,-8)[tc]{$\hkm$}
\Text(37.5,30)[bc]{$\wbp$}
\Text(0,-8)[lt]{$(4)$}
\end{picture}
&&\hspace*{-5mm}+\quad
\begin{picture}(75,20)(0,8)
  \Photon(0,10)(25,10){3}{5}
  \DashArrowArcn(37.5,10)(12.5,180,0){3.}
  \PhotonArc(37.5,10)(12.5,180,0){3}{7}
    \Vertex(25,10){2.5}
    \Vertex(50,10){2.5}
  \Photon(50,10)(75,10){3}{5}
    \ArrowLine(37.5,-2.2)(36.7,-4)
\Text(37.5,-8)[tc]{$\wbm$}
\Text(37.5,30)[bc]{$\hkp$}
\Text(0,-8)[lt]{$(5)$}
\end{picture}
&&
\nl \nl \nl \nl[1.5mm] 
&&+\quad
\begin{picture}(75,20)(0,8)
  \Photon(0,10)(25,10){3}{5}
  \DashCArc(37.5,10)(12.5,0,180){3.}
  \DashCArc(37.5,10)(12.5,180,0){3.}
    \Vertex(25,10){2.5}
    \Vertex(50,10){2.5}
  \Photon(50,10)(75,10){3}{5}
\Text(37.5,-8)[tc]{$\hb$}
\Text(37.5,30)[bc]{$\hkn$}
\Text(0,-8)[lt]{$(6)$}
\end{picture}
&&\hspace*{-5mm}+\quad
\begin{picture}(75,20)(0,8)
  \Photon(0,10)(25,10){3}{5}
  \DashArrowArcn(37.5,10)(12.5,180,0){3.}
  \DashArrowArcn(37.5,10)(12.5,0,180){3.}
    \Vertex(25,10){2.5}
    \Vertex(50,10){2.5}
  \Photon(50,10)(75,10){3}{5}
\Text(37.5,-8)[tc]{$\hkm$}
\Text(37.5,30)[bc]{$\hkp$}
\Text(0,-8)[lt]{$(7)$}
\end{picture}
& &
\nl \nl \nl \nl[1.5mm] 
&&+\quad
\begin{picture}(75,20)(0,8)
  \Photon(0,10)(25,10){3}{5}
\SetWidth{1.}
  \DashArrowArcn(37.5,10)(12.5,0,180){1.5}
  \DashArrowArcn(37.5,10)(12.5,180,0){1.5}
\SetWidth{.5}
    \Vertex(25,10){2.5}
    \Vertex(50,10){2.5}
  \Photon(50,10)(75,10){3}{5}
\Text(37.5,-8)[tc]{$\fpxm$}
\Text(37.5,30)[bc]{$\fpxm$}
\Text(0,-8)[lt]{$(8)$}
\end{picture}
&&\hspace*{-5mm}+\quad
\begin{picture}(75,20)(0,8)
  \Photon(0,10)(25,10){3}{5}
\SetWidth{1.}
  \DashArrowArcn(37.5,10)(12.5,0,180){1.5}
  \DashArrowArcn(37.5,10)(12.5,180,0){1.5}
\SetWidth{.5}
    \Vertex(25,10){2.5}
    \Vertex(50,10){2.5}
  \Photon(50,10)(75,10){3}{5}
\Text(37.5,-8)[tc]{$\fpxp$}
\Text(37.5,30)[bc]{$\fpxp$}
\Text(0,-8)[lt]{$(9)$}
\end{picture}
& &
\nl \nl \nl \nl \nl \nl[1.5mm] 
&&+\quad
\begin{picture}(75,20)(0,8)
  \Photon(0,10)(75,10){3}{15}
  \PhotonArc(37,28)(12.5,17,197){3}{8}
  \PhotonArc(37,28)(12.5,197,17){3}{8}
     \Vertex(37,13.){2.5}
\Text(36.26,50)[bc]{$\wb$}
\Text(0,48)[lb]{$(10)$}
\end{picture}
&&\hspace*{-5mm}+\quad
\begin{picture}(75,20)(0,8)
  \Photon(0,10)(75,10){3}{15}
  \DashCArc(37,26)(12.5,0,180){3}
  \DashCArc(37,26)(12.5,180,0){3}
    \Vertex(37,13){2.5}
\Text(36.26,50)[bc]{$\hb$}
\Text(0,48)[lb]{$(11)$}
\end{picture}
&&
\nl \nl \nl \nl[1.5mm] 
&&+\quad
\begin{picture}(75,20)(0,8)
  \Photon(0,10)(75,10){3}{15}
  \DashArrowArcn(37,26)(12.5,0,180){3}
  \DashArrowArcn(37,26)(12.5,180,0){3}
    \Vertex(37,13.){2.5}
\Text(36.26,45)[bc]{$\hkp$}
\Text(0,45)[lb]{$(12)$}
\end{picture}
&&\hspace*{-5mm}+\quad
\begin{picture}(75,20)(0,8)
  \Photon(0,10)(75,10){3}{15}
  \DashCArc(37,26)(12.5,0,180){3}
  \DashCArc(37,26)(12.5,180,0){3}
    \Vertex(37,13.){2.5}
\Text(36.26,45)[bc]{$\hkn$}
\Text(0,45)[lb]{$(13)$}
\end{picture}
\nl \nl \nl[1.5mm]
&&+\quad
\begin{picture}(75,20)(0,8)
  \Photon(0,10)(75,10){3}{15}
    \Vertex(37.5,10.){2.5}
\Text(37.5,20)[bc]{$\btp$}
\Text(0,22)[lb]{$(14)$}
\Text(53,20)[lb]{$(\zb)$}
\end{picture}
& &
\eaa
\]
\caption{$\zb(\gamma)$ boson self-energy;~~$\zb$--$\gamma$~~transition.
\label{se_zgamma}}
\vspace*{-40mm}
\end{figure}

In the $\Rxi$ gauge there are 14 diagrams that contribute to the
{\em total} $\zb$ and $\ph$ bosonic self-energies and to the $\zb$--$\ph$ 
transition. They are shown in \fig{se_zgamma}.

With $S_{\sss{\zb\zb}}$, $S_{\zg}$ and $S_{\ph\ph}$ standing for the sum of all
diagrams, depicted by a grey circle in \fig{se_zgamma}, we define the three 
corresponding self-energy functions $\Sigma_{\sss{AB}}$:
\ba
S_{\sss{\zb\zb}}&=&(2\pi)^4\ib\frac{g^2}{16\pi^2\cows}\Sigma_{\sss{\zb\zb}}\,,
\\
S_{\zg}   &=&(2\pi)^4\ib\frac{g^2\siw}{16\pi^2\cow}\Sigma_{\zg}\,,
\\
S_{\ph\ph}   &=&(2\pi)^4\ib\frac{g^2\siws}{16\pi^2}\Sigma_{\ph\ph}\,.
\label{sefunct}
\ea
\newpage

\clearpage

All {\bf bosonic} self-energies and transitions may be naturally split into
{\em boso\-nic} and {\em fermio\-nic} components.
\begin{itemize}
\item {Bosonic components of $\zb,\ph$ self-energies and
        $\zb$--$\ph$ transitions} (see diagrams \fig{se_zgamma})
\end{itemize}
\bqa
\Sigma^{\bos}_{\sss{\zb\zb}}(\sman)  &=& 
 \mzs \biggl\{
 \frac{ 1}{3}\frac{1}{\Rz}  \lpar  \frac{1}{2} - \ctws - 9 \ctwf \rpar
\\ &&
   - \frac{3}{2} \lrbr \lpar 1 + 2 \ctwf \rpar\frac{1}{\rhz}
   - \frac{1}{2} - \ctws + \frac{8}{3}\ctwf + \frac{1}{2} \rhz \rrbr 
      \biggr\}
\pole + \Sigma^{{\bos},F}_{\sss{\zb\zb}}(\sman),
\nonumber
\label{self_zz_finite}
\eqa
\bqa
\Sigma^{{\bos},F}_{\sss{\zb\zb}}(\sman)  &=& 
\frac{\mzs}{12}
\biggl\{
 \lrbr 4 \ctws \lpar 5 - 8 \ctws - 12\ctwf \rpar
              +\lpar 1 - 4 \ctws - 36\ctwf \rpar \frac{1}{\Rz} \rrbr
\nll  &&\times
 \fbff{0}{-\sman}{\mwl}{\mwl}
\nll   &&
 +\lrbr \frac{1}{\Rz} + 10 - 2 \rhz + \lpar \rhz-1 \rpar^2 \Rz \rrbr   
 \fbff{0}{-\sman}{\mhl}{\mzl}
\nll   &&
  +  \lrbr \frac{18}{\rhz} + 1 +  \lpar 1-\rhz \rpar \Rz \rrbr   \Lmmz
\nll   &&
  +  \rhz \Big[ 7 - \lpar 1 - \rhz \rpar \Rz \Big] \Lmmh
\nll   &&
  +  2 \ctws \lpar \frac{18}{\rhw} + 1 +  8 \ctws -24 \ctwf \rpar \Lmmw
\nll &&
  + \frac{4}{3}
 \lpar 1 - 2\ctws \rpar \frac{1}{\Rz} 
        - 6 \lpar 1+2\ctwf \rpar \frac{1}{\rhz}
\nll   &&
        - 3 (1+2 \ctws) - 9\rhz - \lpar 1-\rhz \rpar^2 \Rz 
  \biggr\}.
\eqa
 Here $L_\mu(M^2)$ denotes the log containing the 't Hooft scale $\mu$:
\bqa
L_\mu(M^2)=\ln\frac{M^2}{\mu^2}\;,
\eqa
and it should be understood that, contrary to the one used in 
\cite{Bardin:1999ak}, we define here
\bqa
 \bff{0}{-\sman}{M_1}{M_2}=\pole+\fbff{0}{-\sman}{M_1}{M_2},
\label{b0_finite}
\eqa
meaning that $B_0^F$ also depends on the scale $\mu$.
We will not explicitly maintain $\mu$ in the arguments list 
of $L_\mu$ and $B_0^F$. Leaving $\tHs$ unfixed, we retain an opportunity 
to control $\tHs$ ~independence (and therefore UV finiteness) in numerical
realization of one-loop form factors, providing thereby an additional 
cross-check.

Next, it is convenient to introduce the dimensionless quantities
$\Pzg^{\bos}(\sman)$ and $\Pgg^{\bos}(\sman)$ (vacuum polarizations):
\bqa
\Sigma^{\bos}_{{\sss \zb}\ph}(\sman) &=& - \sman \Pzg^{\bos}(\sman),
\label{sig_Zg} 
\\[2mm]
\Sigma^{\bos}_{\ph\ph}(\sman) &=& - \sman \Pgg^{\bos}(\sman).
\label{sig_gg} 
\eqa

In \eqnsc{self_zz_finite}{b0_finite}
and below, the following abbreviations are used:
\bqa
\cows=\frac{\mws}{\mzs}\,,
\qquad
r_{ij}=\frac{m^2_i}{m^2_j}\,,
\qquad
\Rw=\frac{\mws}{\sman}\,,
\qquad
\Rz=\frac{\mzs}{\sman}\,.
\qquad
\label{abbrev-old}
\eqa
 Since only finite parts will contribute to resulting expressions for
physical amplitudes, which should be free from ultraviolet poles, 
it is convenient to split every divergent function into singular and 
finite parts:
\bqa
\Pgg^{\bos}(\sman) & = &  3 \pole + \Pgg^{{\bos},F}(\sman),
\label{pigg_pf}
\\[2mm]   
\label{pigg_pf1}
\Pgg^{{\bos},F}(\sman)&=& 
 \lpar 3 + 4 \Rw \rpar \fbff{0}{-s}{\mwl}{\mwl} + 4 \Rw \Lmmw,
\eqa
and
\bqa
\Pzg^{\bos}(\sman) & = &  \lpar \frac{1}{6} + 3 \ctws + 2\Rw\rpar \pole
                           + \Pzg^{{\bos},F}(\sman),
\label{pizg_pf}
\\   
\Pzg^{{\bos},F}(\sman)  &=&
   \lrbr
  \frac{1}{6} + 3 \cows 
 + 4 \lpar \frac{1}{3}+\cows \rpar \Rw
\rrbr 
\fbff{0}{-s}{\mwl}{\mwl}
\nll &&
 +\frac{1}{9}
-\lpar \frac{2}{3} - 4 \ctws \rpar \Rw \Lmmw.
\label{pizg_f}
\eqa

With the $\zb$ boson self-energy, $\Sigma_{\sss{\zb\zb}}$, we construct
a useful ratio:
\bqa
\Dz{}{\sman}&=&\frac{1}{\cows}
            \frac{\Sigma^{}_{\sss{\zb\zb}}\lpar s\rpar
                 -\Sigma^{}_{\sss{\zb\zb}}\lpar \mzs\rpar}{\mzs-\sman}\,,
\label{usefulratio}
\eqa
which also has bosonic and fermionic parts.
The bosonic component is:
\bqa
\Dz{\bos}{\sman}&=&
 \frac{1}{\ctws} \lpar - \frac{1}{6} + \frac{1}{3}\ctws + 3\ctwf \rpar 
                  \pole
+\Dz{{\bos},F}{\sman},
\eqa
with the finite part:
\bqa
\label{Dz_bF}
\Dz{{\bos},F}{\sman} &=&
 \frac{1}{\cows}
\biggl\{
     \lpar \frac{1}{12} + \frac{4}{3} \cows -
                        \frac{17}{3} \cowf - 4 \cowsc \rpar 
\\ &&
\times \frac{\mzs}{\mzs-s}
\Bigl[
\fbff{0}{-s}{\mwl}{\mwl} - \fbff{0}{-\mzs}{\mwl}{\mwl} \Bigr]
\nn
\eqa
\bqa
&&  -\lpar \frac{1}{12} - \frac{1}{3}\cows - 3 \cowf \rpar 
      \fbff{0}{-s}{\mwl}{\mwl}
\nll  &&
  + \lpar  1 - \frac{1}{3} \rhz + \frac{1}{12} \rhzs \rpar  
\nll &&
\times\frac{\mzs}{\mzs-s} 
    \Bigl[\fbff{0}{-s}{\mhl}{\mzl} 
        - \fbff{0}{-\mzs}{\mhl}{\mzl} \Bigr]
\nll &&
  -   \frac{1}{12}
  \lrbr 1 - \lpar 1 - \rhz \rpar^2 \Rz  \rrbr 
\fbff{0}{-s}{\mhl}{\mzl}
\nll &&
 -\frac{1}{12} \Rz
   \lpar  1 - \rhz \rpar 
   \Bigl[  \rhz \lpar \Lmmh -1 \rpar - \Lmmz + 1 \Bigr] 
  - \frac{1}{9}\lpar 1 - 2 \cows  \rpar 
\biggr\}.
\nn
\eqa
\begin{itemize}
 \item{Fermionic components of the $\zb$ and $\ph$ bosonic self-energies and
of the $\zb$--$\ph$ transition}
\end{itemize}
These are represented as sums over all fermions of
the theory, $\sum_f$. They, of course, depend on vector and axial couplings
of fermions to the $\zb$ boson, $\vf$ and $\af$, and to the photon, 
electric charge $e Q_f$,
as well as on the colour factor $c_f$ and fermion mass $\mfl$.
The couplings are defined as usual:
\ba
\vf=I^{(3)}_f-2Q_f\siws\,,
\qquad
\af=I^{(3)}_f,
\ea
with weak isospin $I^{(3)}_f$, and
\bqa
Q_f&=&-1\quad\mbox{for leptons},\quad+\frac{2}{3}\quad\mbox{for up quarks},
\quad-\frac{1}{3}\quad\mbox{for down quarks},
\nll
c_f&=&1\quad\mbox{for leptons},\quad 3\quad\mbox{for quarks}.
\eqa

The three main self-energy functions are: 
\bqa
\Sigma^{\fer}_{\sss{\zb\zb}}(\sman) \hspace*{-2mm}&=&\hspace*{-2mm} 
-\hspace*{-1mm}\sum_f c_f \biggl[
\lpar v^2_f+a^2_f \rpar\sman  
\bff{f}{-\sman}{\mfl}{\mfl}
\hspace*{-1mm}+\hspace*{-1mm}2a^2_f\mf\bff{0}{-\sman}{\mfl}{\mfl}
           \biggr],
\label{Sigma_zz_fer}
\nll
\\
\Sigma^{\fer}_{\ph\ph}(\sman) &=& -\sman 
\Pgg^{\fer}(\sman),
\label{Sgg_Pgg}
\\[2mm]
\Sigma^{\fer}_{\zg}(\sman)& = & -\sman 
\Pzg^{\fer}\lpar\sman\rpar. 
\label{Szg_Pzg}
\eqa

The quantities $\Pgg^{\fer}$ and $\Pzg^{\fer}$ are different according to
different couplings, but proportional to one function $\sbff{f}$
(see Eq. (5.252) of \cite{Bardin:1999ak} for its definition):
\bqa
\Pgg^{\fer}(\sman) &=& 4 \sum_f c_f\,Q^2_f\,   \bff{f}{-\sman}{\mfl}{\mfl}\,,
\label{Pi_gg_fer}
\\
\Pzg^{\fer}(\sman) &=& 2 \sum_f c_f\,Q_f\,\vf\,\bff{f}{-\sman}{\mfl}{\mfl}\,.
\label{Pi_zg_fer}
\eqa
As usual, we subdivided them into singular and finite parts:
\begin{align}
\label{fermionic-split}
\Pzg^{\fer}(\sman)  =&
  - \hspace{-1.5mm} \frac{1}{3} 
     \biggl(\frac{1}{2}\Nf-4\siws\asums{\ff}\cf\qfs\biggr)
     \pole
  +  \Pzg^{{\fer},F}(\sman),
\\
\Sigma^{\fer}_{\sss{\zb\zb}}(\sman) = &
\biggl\{-\frac{1}{2}\asums{\ff}\cf\mfs
\hspace{-1mm} + \hspace{-1mm}
 \frac{\sman}{3}
\biggl[\lpar\frac{1}{2}-\siws\rpar\Nf
\hspace{-1mm}+\hspace{-1mm}
4\siwf\asums{\ff}\cf\qfs\biggr]
\biggr\} \pole
 + \Sigma^{{\fer},F}_{\sss{\zb\zb}}(\sman).
\nonumber
\end{align}
In \eqn{fermionic-split}, $\Nf=24$ is the total number of fermions in the SM.
We do not show explicit expressions for finite parts, marked with superscript 
$F$, because these might be trivially derived from 
\eqn{Sigma_zz_fer} and 
Eqs.~(\ref{Pi_gg_fer})--(\ref{Pi_zg_fer}) by replacing complete 
expressions for $\sbff{f}$ and $\sbff{0}$ with their finite parts
$\sfbff{f}$ and $\sfbff{0}$, correspondingly.

\subsubsection{$\wb$ boson self-energy}

\begin{figure}[t]
\vspace*{-35mm}
\[
\baa{ccccccc}
\begin{picture}(75,20)(0,8)
  \Photon(0,10)(25,10){3}{5}
\ArrowLine(-25,10)(-10,10)
\Text(-17.5,2)[tc]{$\pmom$}
    \GCirc(37.5,10){12.5}{0.5}
  \Photon(50,10)(75,10){3}{5}
    \ArrowLine(10,9.6)(10.55,11.6)
    \ArrowLine(65,9.6)(65.55,11.6)
\Text(12.5,18)[bc]{$\wbp$}
\Text(12.5, 2)[tc]{$\mu$}
\Text(62.5,18)[bc]{$\wbm$}
\Text(62.5, 2)[tc]{$\nu$}
\end{picture}
&=&{\LARGE{\sum_d\Biggl(\phantom{\Biggr)} }}
\hspace*{-5mm}
\begin{picture}(75,20)(0,8)
  \Photon(0,10)(25,10){3}{5}
  \ArrowArcn(37.5,10)(12.5,0,180)
  \ArrowArcn(37.5,10)(12.5,180,0)
    \Vertex(25,10){2.5}
    \Vertex(50,10){2.5}
  \Photon(50,10)(75,10){3}{5}
    \ArrowLine(10,9.6)(10.55,11.6)
    \ArrowLine(65,9.6)(65.55,11.6)
\Text(37.5,-8)[tc]{$\fbd$}
\Text(37.5,30)[bc]{$\fu$}
\Text(0,-8)[lt]{$(1)$}
\end{picture}
\Biggr)
 &&\hspace*{-5mm}+\quad
\begin{picture}(75,20)(0,8)
  \Photon(0,10)(25,10){3}{5}
  \PhotonArc(37.5,10)(12.5,0,180){3}{7}
  \PhotonArc(37.5,10)(12.5,180,0){3}{7}
    \Vertex(25,10){2.5}
    \Vertex(50,10){2.5}
  \Photon(50,10)(75,10){3}{5}
    \ArrowLine(10,9.6)(10.55,11.6)
    \ArrowLine(65,9.6)(65.55,11.6)
    \ArrowLine(37.6,22.2)(38.3,24.2)
\Text(37.5,-8)[tc]{$\zb$}
\Text(37.5,30)[bc]{$\wbp$}
\Text(0,-8)[lt]{$(2)$}
\end{picture}
\nl \nl[2mm]
&&+\quad
\begin{picture}(75,20)(0,8)
  \Photon(0,10)(25,10){3}{5}
  \PhotonArc(37.5,10)(12.5,0,180){3}{7}
  \PhotonArc(37.5,10)(12.5,180,0){3}{13}
    \Vertex(25,10){2.5}
    \Vertex(50,10){2.5}
  \Photon(50,10)(75,10){3}{5}
    \ArrowLine(10,9.6)(10.55,11.6)
    \ArrowLine(65,9.6)(65.55,11.6)
    \ArrowLine(37.6,22.2)(38.3,24.2)
\Text(37.5,-8)[tc]{$\gamma$}
\Text(37.5,30)[bc]{$\wbp$}
\Text(0,-8)[lt]{$(3)$}
\end{picture}
&&\hspace*{-5mm}+\quad
\begin{picture}(75,20)(0,8)
  \Photon(0,10)(25,10){3}{5}
  \PhotonArc(37.5,10)(12.5,0,180){3}{7}
  \DashCArc(37.5,10)(12.5,180,0){3.}
    \Vertex(25,10){2.5}
    \Vertex(50,10){2.5}
  \Photon(50,10)(75,10){3}{5}
    \ArrowLine(10,9.6)(10.55,11.6)
    \ArrowLine(65,9.6)(65.55,11.6)
    \ArrowLine(37.6,22.2)(38.3,24.2)
\Text(37.5,-8)[tc]{$\hb$}
\Text(37.5,30)[bc]{$\wbp$}
\Text(0,-8)[lt]{$(4)$}
\end{picture}
\nl \nl \nl \nl
&&+\quad
\begin{picture}(75,20)(0,8)
  \Photon(0,10)(25,10){3}{5}
  \DashArrowArcn(37.5,10)(12.5,180,0){3.}
  \PhotonArc(37.5,10)(12.5,180,0){3}{7}
    \Vertex(25,10){2.5}
    \Vertex(50,10){2.5}
  \Photon(50,10)(75,10){3}{5}
    \ArrowLine(10,9.6)(10.55,11.6)
    \ArrowLine(65,9.6)(65.55,11.6)
\Text(37.5,-8)[tc]{$\zb$}
\Text(37.5,30)[bc]{$\hkp$}
\Text(0,-8)[lt]{$(5)$}
\end{picture}
&&\hspace*{-5mm}+\quad
\begin{picture}(75,20)(0,8)
  \Photon(0,10)(25,10){3}{5}
  \DashArrowArcn(37.5,10)(12.5,180,0){3.}
  \PhotonArc(37.5,10)(12.5,180,0){3}{13}
    \Vertex(25,10){2.5}
    \Vertex(50,10){2.5}
  \Photon(50,10)(75,10){3}{5}
    \ArrowLine(10,9.6)(10.55,11.6)
    \ArrowLine(65,9.6)(65.55,11.6)
\Text(37.5,-8)[tc]{$\gamma$}
\Text(37.5,30)[bc]{$\hkp$}
\Text(0,-8)[lt]{$(6)$}
\end{picture}
\nl \nl \nl \nl
&&+\quad
\begin{picture}(75,20)(0,8)
  \Photon(0,10)(25,10){3}{5}
  \DashArrowArcn(37.5,10)(12.5,180,0){3.}
  \DashCArc(37.5,10)(12.5,180,0){3.}
    \Vertex(25,10){2.5}
    \Vertex(50,10){2.5}
  \Photon(50,10)(75,10){3}{5}
    \ArrowLine(10,9.6)(10.55,11.6)
    \ArrowLine(65,9.6)(65.55,11.6)
\Text(37.5,-8)[tc]{$\hb$}
\Text(37.5,30)[bc]{$\hkp$}
\Text(0,-8)[lt]{$(7)$}
\end{picture}
&&\hspace*{-5mm}+\quad
\begin{picture}(75,20)(0,8)
  \Photon(0,10)(25,10){3}{5}
  \DashArrowArcn(37.5,10)(12.5,180,0){3.}
  \DashCArc(37.5,10)(12.5,180,0){3.}
    \Vertex(25,10){2.5}
    \Vertex(50,10){2.5}
  \Photon(50,10)(75,10){3}{5}
    \ArrowLine(10,9.6)(10.55,11.6)
    \ArrowLine(65,9.6)(65.55,11.6)
\Text(37.5,-8)[tc]{$\hkn$}
\Text(37.5,30)[bc]{$\hkp$}
\Text(0,-8)[lt]{$(8)$}
\end{picture}
& &
\nl \nl \nl \nl
&&+\quad
\begin{picture}(75,20)(0,8)
  \Photon(0,10)(25,10){3}{5}
\SetWidth{1.}
  \DashArrowArcn(37.5,10)(12.5,0,180){1.5}
  \DashArrowArcn(37.5,10)(12.5,180,0){1.5}
\SetWidth{.5}
    \Vertex(25,10){2.5}
    \Vertex(50,10){2.5}
  \Photon(50,10)(75,10){3}{5}
    \ArrowLine(10,9.6)(10.55,11.6)
    \ArrowLine(65,9.6)(65.55,11.6)
\Text(37.5,-8)[tc]{$\fpxm$}
\Text(37.5,30)[bc]{$\fpyZA$}
\Text(0,-8)[lt]{$(9)$}
\end{picture}
&&\hspace*{-5mm}+\quad
\begin{picture}(75,20)(0,8)
  \Photon(0,10)(25,10){3}{5}
\SetWidth{1.}
  \DashArrowArcn(37.5,10)(12.5,0,180){1.5}
  \DashArrowArcn(37.5,10)(12.5,180,0){1.5}
\SetWidth{.5}
    \Vertex(25,10){2.5}
    \Vertex(50,10){2.5}
  \Photon(50,10)(75,10){3}{5}
    \ArrowLine(10,9.6)(10.55,11.6)
    \ArrowLine(65,9.6)(65.55,11.6)
\Text(37.5,-8)[tc]{$\fpyZA$}
\Text(37.5,30)[bc]{$\fpxp$}
\Text(0,-8)[lt]{$(10)$}
\end{picture}
& &
\nl \nl \nl \nl \nl \nl
&&+\quad
\begin{picture}(75,20)(0,8)
  \Photon(0,10)(75,10){3}{15}
  \PhotonArc(37,28)(12.5,17,197){3}{8}
  \PhotonArc(37,28)(12.5,197,17){3}{8}
     \Vertex(37,13.){2.5}
    \ArrowLine(10,9.6)(10.55,11.6)
    \ArrowLine(65,9.6)(65.55,11.6)
\Text(36.26,50)[bc]{$\wb$}
\Text(0,48)[lb]{$(11)$}
\end{picture}
&&\hspace*{-5mm}+\quad
\begin{picture}(75,20)(0,8)
  \Photon(0,10)(75,10){3}{15}
  \PhotonArc(37,28)(12.5,17,197){3}{8}
  \PhotonArc(37,28)(12.5,197,17){3}{8}
     \Vertex(37,13.){2.5}
    \ArrowLine(10,9.6)(10.55,11.6)
    \ArrowLine(65,9.6)(65.55,11.6)
\Text(36.26,50)[bc]{$\zb$}
\Text(0,48)[lb]{$(12)$}
\end{picture}
\nl \nl \nl \nl
&&+\quad
\begin{picture}(75,20)(0,8)
  \Photon(0,10)(75,10){3}{15}
  \PhotonArc(37,28)(12.5,3,183){3}{13}
  \PhotonArc(37,28)(12.5,183,3){3}{13}
     \Vertex(37,13.){2.5}
    \ArrowLine(10,9.6)(10.55,11.6)
    \ArrowLine(65,9.6)(65.55,11.6)
\Text(36.26,50)[bc]{$\gamma$}
\Text(0,48)[lb]{$(13)$}
\end{picture}
&&\hspace*{-5mm}+\quad
\begin{picture}(75,20)(0,8)
  \Photon(0,10)(75,10){3}{15}
  \DashCArc(37,26)(12.5,0,180){3}
  \DashCArc(37,26)(12.5,180,0){3}
    \Vertex(37,13){2.5}
    \ArrowLine(10,9.6)(10.55,11.6)
    \ArrowLine(65,9.6)(65.55,11.6)
\Text(36.26,45)[bc]{$\hb$}
\Text(0,45)[lb]{$(14)$}
\end{picture}
\nl \nl \nl \nl
&&+\quad
\begin{picture}(75,20)(0,8)
  \Photon(0,10)(75,10){3}{15}
  \DashArrowArcn(37,26)(12.5,0,180){3}
  \DashArrowArcn(37,26)(12.5,180,0){3}
    \Vertex(37,13.){2.5}
    \ArrowLine(10,9.6)(10.55,11.6)
    \ArrowLine(65,9.6)(65.55,11.6)
\Text(36.26,45)[bc]{$\hkp$}
\Text(0,45)[lb]{$(15)$}
\end{picture}
&&\hspace*{-5mm}+\quad
\begin{picture}(75,20)(0,8)
  \Photon(0,10)(75,10){3}{15}
  \DashCArc(37,26)(12.5,0,180){3}
  \DashCArc(37,26)(12.5,180,0){3}
    \Vertex(37,13.){2.5}
    \ArrowLine(10,9.6)(10.55,11.6)
    \ArrowLine(65,9.6)(65.55,11.6)
\Text(36.26,45)[bc]{$\hkn$}
\Text(0,45)[lb]{$(16)$}
\end{picture}
\nl \nl[1mm]
&&+\quad
\begin{picture}(75,20)(0,8)
  \Photon(0,10)(75,10){3}{15}
    \Vertex(37.5,10.){2.5}
\Text(37.5,20)[bc]{$\btp$}
\Text(0,22)[lb]{$(17)$}
\end{picture}
\eaa
\]
\vspace*{-4mm}
\caption{$\wb$ boson self-energy.\label{se_wb}}
\vspace*{-35mm}
\end{figure}
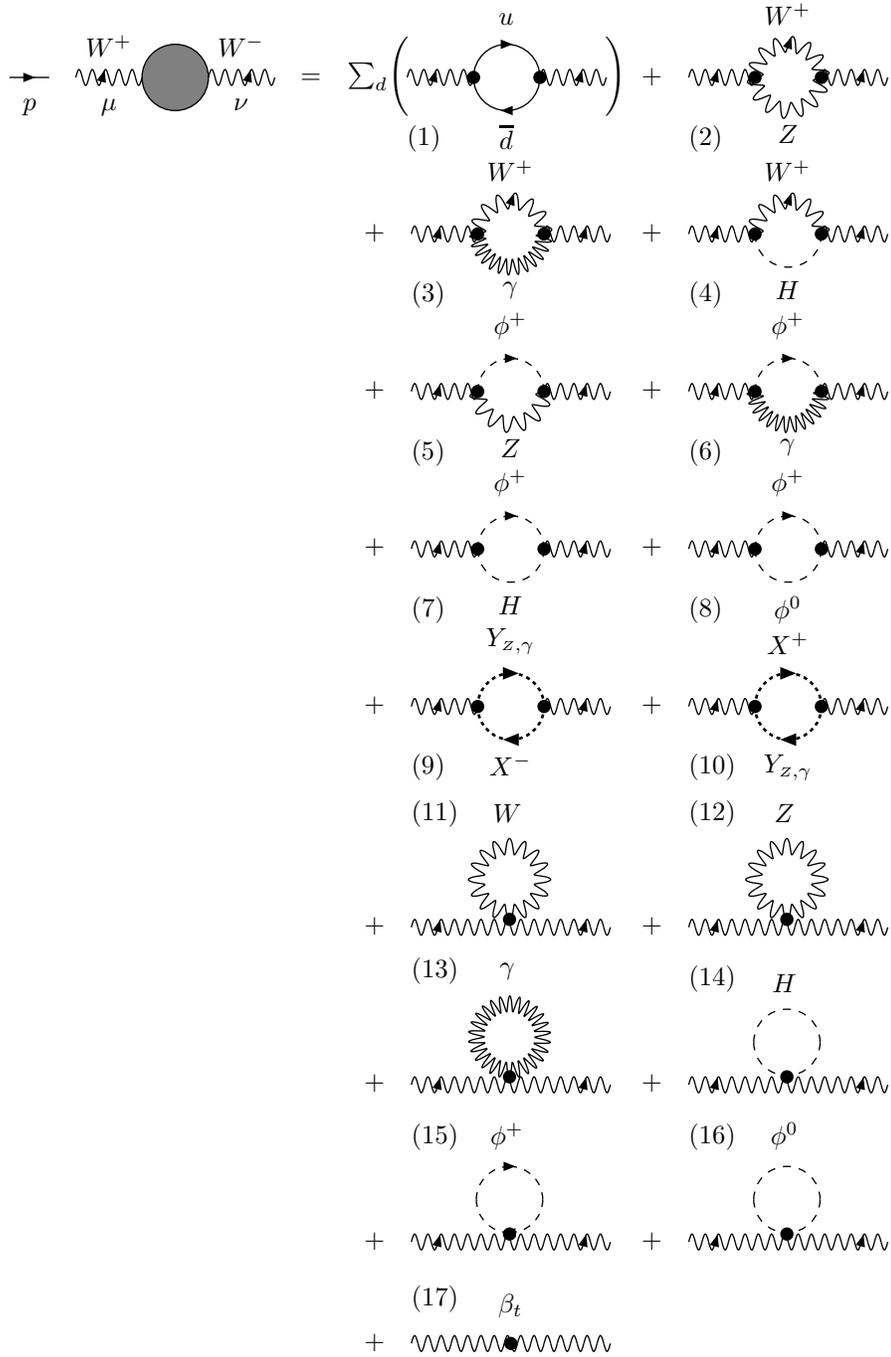

Next we consider the $\wb$ boson self-energy, which
is described, in the $\Rxi$  gauge, by 17 diagrams, shown in \fig{se_wb}.

First, we present an explicit expression for its bosonic component:
\bqa
\Sigma^{\bos}_{_{\wb\wb}}(\sman)&=&
\label{lost-label}
\mws\biggl\{ 
  - \frac{19}{6} \frac{1}{\Rw} 
  - \frac{1}{4} \biggl[ 
 \frac{6}{\rhw} \lpar \frac{1}{\ctwf} + 2 \rpar 
  - \frac{3}{\ctws}
 + 10 + 3 \rhw \biggr] \biggr\}
          \pole
\nll &&
      + \Sigma^{{\bos},F}_{_{\wb\wb}}(\sman),
\eqa
where
\bqa
\Sigma^{{\bos},F}_{_{\wb\wb}}(\sman)&=&
  \frac{\mws}{12} 
        \biggl\{
        \biggl[ \lpar 1 - 40\ctws \rpar \frac{1}{\Rw} 
         + 2 \lpar \frac{5}{\ctws} - 27 - 8\ctws \rpar 
\nll &&
    +\frac{\stwf}{\ctws} 
    \lpar\frac{1}{\ctws} + 8 \rpar \Rw \biggr] 
\fbff{0}{-\sman}{\mwl}{\mzl}
\nll && 
  + \lrbr \frac{1}{\Rw}+2 \lpar 5-\rhw \rpar + \lpar 1-\rhw \rpar^2 \Rw \rrbr 
\fbff{0}{-\sman}{\mwl}{\mhl}
\nll &&
  - 8\stws \lpar \frac{5}{\Rw} + 2 - \Rw \rpar
\fbff{0}{-\sman}{\mwl}{0}
\nll &&
  + \rhw \Big[ 7 - \lpar 1 - \rhw \rpar \Rw \Big] \Lmmh 
\nll &&
  + \frac{1}{\ctws}\lrbr  \frac{18}{\rhw} \frac{1}{\ctws} 
  + 1 - 16 \ctws 
  + \stws \lpar \frac{1}{\ctws}+8 \rpar \Rw \rrbr \Lmmz 
\nll &&
  + \lrbr 2 \lpar\frac{18}{\rhw}-7\rpar-\lpar\frac{1}{\ctws}-2+\rhw\rpar\Rw\rrbr\Lmmw
\nll &&
  - \frac{4}{3}\frac{1}{\Rw}
  - 12 \lpar   \frac{1}{2}\frac{1}{\ctwf} + 1 \rpar \frac{1}{\rhw}
   - 3 \lpar   \frac{1}{\ctws}            + 2 \rpar - 9\rhw   
\nll &&
  - \lrbr\lpar \frac{1}{\ctws} + 6\stws \rpar \frac{1}{\ctws}
   - \rhw \lpar 2 - \rhw \rpar \rrbr \Rw 
      \biggr\}. 
\label{sig_ww}
\eqa
\clearpage

\clearpage

Secondly, we give its fermionic component:
\begin{equation}
\label{sig_wwf}
\hspace*{-4mm}
\Sigma^{\fer}_{_{\wb\wb}}(\sman)=
    -\sman\asum{f}{d}{} c_f\bff{f}{-\sman}{\mfpl}{\mfl} 
         +\asums{\ff} c_f\minds{\ff}\bff{1}{-\sman}{\mfpl}{\mfl},
\end{equation}
where summation in the first term extends to all {\em doublets} of the SM.

\subsubsection{Bosonic self-energies and counterterms\label{bosonic_ct}}
Bosonic self-energies and transitions enter one-loop amplitudes either
directly through the functions $\Dz{}{\sman}$, $\Pgg^{}(\sman)$ and $\Pzg^{}(\sman)$,
or by means of bosonic counterterms, which are made of self-energy functions at
zero argument, owing to {\em electric charge renormalization}, or  
at $\pmoms=-\Minds{}$, that is on a mass shell, owing to {\em 
on-mass-shell renormalization} (OMS scheme).
\begin{itemize}
\item Electric charge renormalization
\end{itemize}

The electric charge renormalization introduces the quantity $z_{\ph}-1$:
\bqa
\lpar z_{\ph}-1 \rpar = 
    \stws   \bigg[ \Pgg(0)-\frac{2}{\mws}{\overline{\Sigma}}_{3Q}(0)\bigg],
\eqa
with bosonic (see Eq. (6.161) of \cite{Bardin:1999ak}):
\bqa
\lpar z_{\ph}-1 \rpar^{\bos}=
\stws   \bigg[ 3 \bigg(\pole-\Lmmw \bigg) + \frac{2}{3}\bigg]\,,  
\eqa
and fer\-mio\-nic
\bqa
\lpar z_{\ph}-1 \rpar^{\fer}&=&
\stws   \bigg[ \bigg( -\frac{4}{3}\Trqf \bigg) \pole + \Pgg^{{\fer},F}(0) \bigg]
\label{elcharen}
\eqa
components.

\begin{itemize}
\item $\rho$-parameter
\end{itemize}

Finally, two self-energy functions enter Veltman's parameter $\Delta\rho$,
a gauge-invariant combination of self-energies, which naturally appears 
in the one-loop calculations:
\bqa
\Delta \rho &=&
\frac{1}{\mws}
\Bigl[ \Sigma_{_{WW}}(\mws) - \Sigma_{_{ZZ}}(\mzs) \Bigr],
\label{rhodef-old}
\eqa
with individual components where we explicitly show the pole parts: 
\bqa
\Delta \rho^{\bos}
&=& \lpar-\frac{1}{ 6\ctws} - \frac{41}{6} + 7 \ctws \rpar
      \pole + \Delta \rho^{{\bos},F},
\\
\Delta \rho^{\fer} &=& \frac{1}{3} \frac{\stws}{\ctws} 
\Biggl( \frac{1}{2}  N_f -4 \stws  \Trqf \Biggr)
         \pole + \Delta \rho^{{\fer},F}.\qquad
\eqa
The finite part of $\Delta \rho^{\bos}$ is given explicitly by
\bqa
\Delta \rho^{{\bos},F}
     &=&
   \lpar \frac{1}{ 12 \cowf}+\frac{4}{3\cows}
              -\frac{17}{3}-4\cows    \rpar
\nll &&\times
\Bigl[  \fbff{0}{-\mws}{\mwl}{\mzl}
                  -\ctws       \fbff{0}{-\mzs}{\mwl}{\mwl} \Bigr]
\nll &&
+\lpar 1-\frac{1}{3}\rhw
 +\frac{1}{12} \rhws  
 \rpar                                       \fbff{0}{-\mws}{\mwl}{\mhl}
\nll &&
-\lpar 1 - \frac{1}{3} \rhz
         + \frac{1}{12}\rhzs \rpar \frac{1}{\cows}
                                            \fbff{0}{-\mzs}{\mzl}{\mhl} 
\nll &&
 -4 \stws                                   \fbff{0}{-\mws}{\mwl}{0}
\nll && 
 + \frac{1}{12} \biggl[ 
          \lpar 
 \frac{1}{\cowf} + \frac{6}{\cows} - 24 + \rhw \rpar  \Lmmz 
\nll &&
    -\lpar \frac{1}{\ctws} + 14 + 16 \ctws - 48 \ctwf + \rhw \rpar \Lmmw 
\nll &&
       + \stws \rhws \Bigl[ \Lmmh - 1 \Bigr]
    -\frac{1}{\cowf} - \frac{19}{3 \cows} + \frac{22}{3}
         \biggr],
\eqa
while the finite part of $\Delta \rho^{\fer}$ is not shown, since  
it is trivially derived from the defining equation 
(\ref{rhodef-old}) by replacing the total self-energies with their finite parts.
\vspace*{-2mm}

\subsection{Fermionic self-energies}
\subsubsection{Fermionic self-energy diagrams}
\begin{figure}[b]
\vspace*{-30mm}
\[
\baa{cccccccc}
&
\begin{picture}(75,20)(0,7.5)
\Text(12.5,13)[bc]{$\ff$}
\Text(62.5,13)[bc]{$\ff$}
  \ArrowLine(0,10)(25,10)
  \GCirc(37.5,10){12.5}{0.5}
  \ArrowLine(50,10)(75,10)
\end{picture}
&=&&&&&
\nl \nl[4mm]
&
\begin{picture}(75,20)(0,7.5)
\Text(12.5,13)[bc]{$\ff$}
\Text(62.5,13)[bc]{$\ff$}
\Text(37.5,26)[bc]{$\ff$}
\Text(37.5,-8)[tc]{$\gamma$}
\Text(0,-7)[lt]{$(1)$}
  \ArrowLine(0,10)(25,10)
  \ArrowArcn(37.5,10)(12.5,180,0)
  \PhotonArc(37.5,10)(12.5,180,0){3}{15}
  \Vertex(25,10){2.5}
  \Vertex(50,10){2.5}
  \ArrowLine(50,10)(75,10)
\end{picture}
&+&
\begin{picture}(75,20)(0,7.5)
\Text(12.5,13)[bc]{$\ff$}
\Text(62.5,13)[bc]{$\ff$}
\Text(37.5,26)[bc]{$\ff$}
\Text(37.5,-8)[tc]{$\zb$}
\Text(0,-7)[lt]{$(2)$}
  \ArrowLine(0,10)(25,10)
  \ArrowArcn(37.5,10)(12.5,180,0)
  \PhotonArc(37.5,10)(12.5,180,0){3}{9}
  \Vertex(25,10){2.5}
  \Vertex(50,10){2.5}
  \ArrowLine(50,10)(75,10)
\end{picture}
&+&
\begin{picture}(75,20)(0,7.5)
\Text(12.5,13)[bc]{$\ff$}
\Text(62.5,13)[bc]{$\ff$}
\Text(37.5,26)[bc]{$\ffp$}
\Text(37.5,-8)[tc]{$\wb$}
\Text(0,-7)[lt]{$(3)$}
  \ArrowLine(0,10)(25,10)
  \ArrowArcn(37.5,10)(12.5,180,0)
  \PhotonArc(37.5,10)(12.5,180,0){3}{7}
  \Vertex(25,10){2.5}
  \Vertex(50,10){2.5}
  \ArrowLine(50,10)(75,10)
\end{picture}
&+&
\nl \nl \nl[4mm]
&\begin{picture}(75,20)(0,7.5)
\Text(12.5,13)[bc]{$\ff$}
\Text(62.5,13)[bc]{$\ff$}
\Text(37.5,26)[bc]{$\ff$}
\Text(37.5,-8)[tc]{$\hb$}
\Text(0,-7)[lt]{$(4)$}
  \ArrowLine(0,10)(25,10)
  \ArrowArcn(37.5,10)(12.5,180,0)
  \DashCArc(37.5,10)(12.5,180,0){3}
  \Vertex(25,10){2.5}
  \Vertex(50,10){2.5}
  \ArrowLine(50,10)(75,10)
\end{picture}
&+&
\begin{picture}(75,20)(0,7.5)
\Text(12.5,13)[bc]{$\ff$}
\Text(62.5,13)[bc]{$\ff$}
\Text(37.5,26)[bc]{$\ff$}
\Text(37.5,-8)[tc]{$\hkn$}
\Text(0,-7)[lt]{$(5)$}
  \ArrowLine(0,10)(25,10)
  \ArrowArcn(37.5,10)(12.5,180,0)
  \DashCArc(37.5,10)(12.5,180,0){3}
  \Vertex(25,10){2.5}
  \Vertex(50,10){2.5}
  \ArrowLine(50,10)(75,10)
\end{picture}
&+&
\begin{picture}(75,20)(0,7.5)
\Text(12.5,13)[bc]{$\ff$}
\Text(62.5,13)[bc]{$\ff$}
\Text(37.5,26)[bc]{$\ffp$}
\Text(37.5,-8)[tc]{$\hkg$}
\Text(0,-7)[lt]{$(6)$}
  \ArrowLine(0,10)(25,10)
  \ArrowArcn(37.5,10)(12.5,180,0)
  \DashCArc(37.5,10)(12.5,180,0){3}
  \Vertex(25,10){2.5}
  \Vertex(50,10){2.5}
  \ArrowLine(50,10)(75,10)
\end{picture}
&&
\label{fsesm}
\eaa
\]
\vspace*{2mm}
\caption{Fermionic self-energy diagrams. \label{f_se}}
\end{figure}
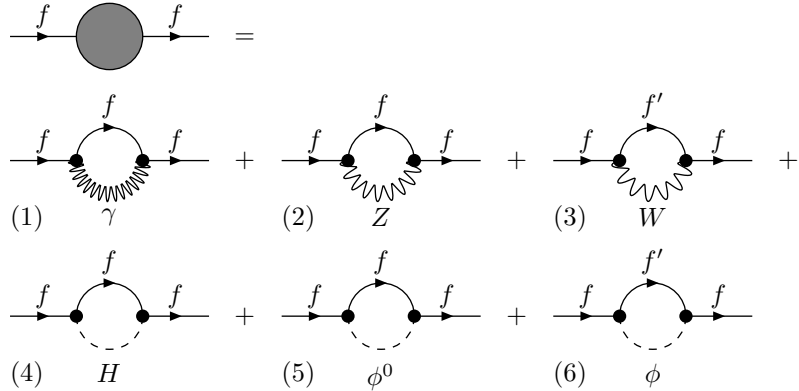

 The total self-energy function of a fermion 
in the $\Rxi$ gauge is described by the six diagrams of \fig{f_se}.
\vspace*{-5mm}
\clearpage

A detailed calculation of the wave-function renormalization factors
$\sqrt{z_{_{L,R}}}$ associated with these diagrams may be found, 
for instance, in \cite{Bardin:1999yd}.

It is convenient to distinguish the electromagnetic components,
\begin{equation}
\lpar \sqrt{z_{_{L}}} - I \rpar^{em}_f  =  
          \lpar \sqrt{z_{_{R}}} - I \rpar^{em}_f
        = \stws Q^2_f \lpar
        - \frac{1}{2\epsb} + \frac{1}{\epsh}
        + \frac{3}{2} \ln \frac{\mf}{\tHss} - 2 \rpar,
\label{wfren_em}
\end{equation}
and the weak components,
\bqa
 \Big| \sqrt{z_{\sss L,R}} \Big| - I  =  \lpar w_v \pm w_a \rpar,                   
\eqa
where five non-zero contributions are: 
\begin{align}
w_v^{\sss Z} =&
- \frac{1}{8}\frac{1}{\ctws}
\biggl\{ 
  \lpar \vf^2 + \af^2 + 2 \af^2 \rtz \rpar \pole
+ \lpar \vf^2 + \af^2 \rpar
            \biggl( 
     \frac{1}{\rtz} \Bigl[\fbff{0}{- \mfs}{\mzl}{\mfl}
\nll &
+\Lmmz-1 \Bigr]
    + 2 \lpar 1 + 2 \rtz \rpar \mzs \bff{0p}{-\mfs}{\mfl}{\mzl} 
    - \Lmmt \biggr)
\nll &
  +2 \af^2 \rtz \biggl(
    \frac{1}{\rtz} \Bigl[ \fbff{0}{-\mfs}{\mzl}{\mfl} + \Lmmz - 1 \Bigr]
\nll &
    - 6 \mzs \bff{0p}{ -\mfs}{\mfl}{\mzl} - \Lmmt + 1  \biggr)   
\biggr\}, 
\\[.1mm]
w_v^{\sss W} =& -\frac{1}{16} 
\biggl\{  \pole\lpar 2+\rP \rpar - \rP
  +\frac{1}{\ruw}\biggl[\lpar 2+\rP \rpar\Bigl[\Lmmw-\rdw\Lmmb\Bigr]
\nll &
      +\lpar 2+2\ruw+\rD+\rP\rD\rpar\Bigl[\fbff{0}{-\mfs}{\mwl}{\mfps}-1\Bigr]
                 \biggr]
\nll &
      +2\mws\Bigl[ 2-\rP-\lpar\rD\rpar^2\Bigr]\bff{0p}{ -\mfs}{\mwl}{\mfps} 
\biggr\},
\\[.1mm]
w_v^{\sss H} =&
  - \frac{1}{16}\rtw 
\biggl\{ \pole           
 + \rht \lrbr \fbff{0}{-\mfs}{\mhl}{\mfl} + \Lmmh - 1 \rrbr 
\nll &
 -  2 \lpar 4 \rth 
       - 1 \rpar \mhs \bff{0p}{ -\mfs}{\mfl}{\mhl} 
 -  \Lmmt + 1 
\biggr\},
\\[.1mm]
w_a^{\sss Z} =&
- \frac{1}{4\ctws} \vf \af 
            \biggl\{
         \pole
  - \frac{1}{\rtz}  \lrbr \fbff{0}{-\mfs}{\mzl}{\mfl}
 + \Lmmz - 1 \rrbr
\nll &
  + 2 \fbff{0}{-\mfs}{\mzl}{\mfl} + \Lnrt - 2 \biggr\},
\\[.1mm]
w_a^{\sss W} =&\frac{1}{16}
    \biggl\{ -\pole\lpar 2-\rD \rpar + \rD
  +\frac{1}{\ruw}\biggl[\lpar 2-\rD \rpar\Bigl[\Lmmw-\rdw\Lmmb\Bigr]
\nll &
      +\lpar 2-2\ruw-\rP+\rP\rD\rpar\Bigl[\fbff{0}{-\mfs}{\mwl}{\mfps}-1\Bigr]
                 \biggr]
\biggr\},
\end{align}
with $r^\pm=\ruw\pm\rdw$ and $\ffp$ being the weak isospin partner of fermion $\ff$.
\subsection{The $\zb\ff\fbf$ and $\ph\ff\fbf$ vertices}
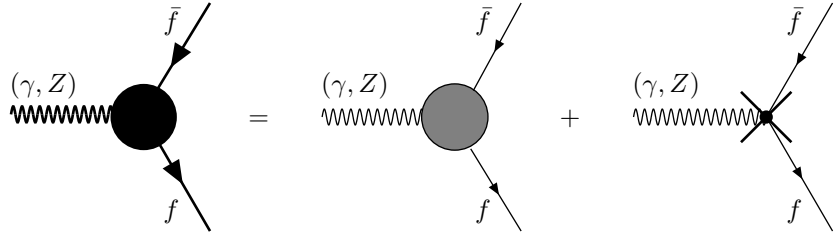
\begin{figure}[!h]
\vspace*{-17.5mm}
\[
\begin{array}{ccccc}
  \begin{picture}(90,86)(40,40)
  \Vertex(100,43){12.5}
\SetScale{2.}
  \Photon(25,22)(50,22){1.5}{15}
  \ArrowLine(50,21.5)(62.5,0)
  \ArrowLine(62.5,43)(50,22.5)
  \Text(108,74)[lb]{$\fbf$}
  \Text(62.5,50)[bc]{$(\ph,\zb)$}
  \Text(108,12)[lt]{$\ff$}
\end{picture}
&=&
\begin{picture}(90,86)(40,40)
  \Photon(50,43)(100,43){3}{15}
  \GCirc(100,43){12.5}{0.5}
  \ArrowLine(125,86)(107,53)
  \ArrowLine(106,31)(125,0)
  \Text(108,74)[lb]{$\fbf$}
  \Text(62.5,50)[bc]{$(\ph,\zb)$}
  \Text(108,12)[lt]{$\ff$}
\end{picture}
&+&
\begin{picture}(90,86)(40,40)
  \Photon(50,43)(100,43){3}{15}
\SetScale{2.0}
  \Line(45.25,16.75)(54.75,26.25)
  \Line(45.25,26.25)(54.75,16.75)
\SetScale{1.0}
  \ArrowLine(125,86)(100,43)
  \Vertex(100,43){2.5}
  \ArrowLine(100,43)(125,0)
  \Text(108,74)[lb]{$\fbf$}
  \Text(62.5,50)[bc]{$(\ph,\zb)$}
  \Text(108,12)[lt]{$\ff$}
\end{picture}
\end{array}
\]
\vspace{10mm}
\caption[$\zb\ff\fbf$ and $\ph\ff\fbf$ vertices with counterterms.]
{\it
$\zb\ff\fbf$ and $\ph\ff\fbf$ vertices with fermionic counterterms.
\label{zavert1}}
\end{figure}

 Consider now the sum of all vertices and corresponding counterterms 
whose contribution originates from the fermionic self-energy diagrams of 
Fig.~\ref{f_se}. This sum is shown in Fig.~\ref{zavert1}.

%

The formulae which determine the counterterms are:
\bqa
 F^{\gamma,ct}_{\sss Q} & = &  2 \lpar\sqrt{z_{\sss R}}-I \rpar, \\[1mm]
 F^{\gamma,ct}_{\sss L} & = &  \lpar\sqrt{z_{\sss L}}-I \rpar
                       - \lpar\sqrt{z_{\sss R}}-I \rpar,         \\[1mm]
 F^{z,ct}_{\sss Q}  & = &  \delta^2_f
                         \lpar\sqrt{z_{\sss R}}-I \rpar,         \\[1mm]
 F^{z,ct}_{\sss L}  & = &  \sigma^2_f
                          \lpar\sqrt{z_{\sss R}}-I \rpar
                         -\delta^2_f
                          \lpar\sqrt{z_{\sss R}}-I \rpar,    
\eqa
where
\bqa
\delta_f=\vf-\af,\qquad
\sigma_f=\vf+\af.
\eqa
 For the sum of all $\ph\to\ff\fbf$ and $\zb\to\ff\fbf$ vertices 
(the {\em total} $\ph(\zb)\ff\fbf$ vertex depicted by a grey circle
in \fig{zavert1}) 
we use the standard
normalization
\begin{equation}
\ib\pi^2=\tpfi\frac{1}{16\pi^2}\,,
\end{equation}
and define
\bqa
\Vverti{\mu}{\ph}{\sman} &=&
\tpfi\frac{1}{16\pi^2}\Gverti{\mu}{ }{\sman},
\\
\Vverti{\mu}{\zb}{\sman} &=&
\tpfi\frac{1}{16\pi^2}\Zverti{\mu}{ }{\sman},
\eqa
while we denote the individual vertices as follows:
\bqa
\Gverti{\mu}{}{\sman}&=&
 \Gverti{\mu}{\sss\ab}{\sman}+\Gverti{\mu}{\sss\zb}{\sman}
+\Gverti{\mu}{\sss\wb}{\sman}+\Gverti{\mu}{\sss\hb}{\sman},
\\
\Zverti{\mu}{}{\sman}&=&
 \Zverti{\mu}{\sss\ab}{\sman}+\Zverti{\mu}{\sss\zb}{\sman}
+\Zverti{\mu}{\sss\wb}{\sman}+\Zverti{\mu}{\sss\hb}{\sman}.
\eqa
All vertices  have three components in our $LQD$  basis.

\subsubsection{Scalar form factors}
Now we construct the $24=(4:B=\ab,\zb,\hb,\wb\,\mbox{-virtual})
\otimes(3:L,Q,D)\otimes(2:\ph,\zb\,\mbox{-incoming})$
scalar form factors, originating from the diagrams of \fig{zavert1}. 
They are derived from the following six equations --- three projections 
for $\ph\ff\fbf$ vertices:
\bqa
\vvertil{\gamma {\sss B}}{\sss L}{\sman}
&=&\frac{2}{\siw \tcif}
\biggl\{ \Gverti{\mu}{\sss B}{\sman}[\ib\gbc \gadu{\mu}\gap]
+\siw\qf\fverti{\gamma,ct}{\sss L} \biggr\},
\\
\vvertil{\gamma {\sss B}}{\sss Q}{\sman}
&=&\frac{1}{\siw\qf}
\biggl\{\Gverti{\mu}{\sss B}{\sman}[\ib\gbc \gadu{\mu}]    
+\siw\qf
\fverti{\gamma,ct}{\sss Q} \biggr\},
\\ 
\vvertil{\gamma {\sss B}}{\sss D}{\sman}
&=&
 \frac{2}{\stwl\tcif} 
  \Gverti{\mu}{\sss B}{\sman}[\gbc \mtl I D_\mu]\,,
\eqa
and three projections for $\zb\ff\fbf$ vertices:
\bqa
\vvertil{z{\sss B}}{\sss L}{\sman}&=&\frac{2\cow}{\tcif}
\biggl\{\Zverti{\mu}{\sss B}{\sman}[\ib \gbc \gadu{\mu} \gap]
+\frac{1}{\cow}\fverti{z,ct}{\sss L}
\biggr\},
\\
\vvertil{z{\sss B}}{\sss Q}{\sman}&=&
\frac{2\cow}{\delta_f}
\biggl\{
\Zverti{\mu}{\sss B}{\sman}[\ib \gbc \gadu{\mu}]
+\frac{1}{\cow}\fverti{z,ct}{\sss Q}
\biggr\},
\\
\vvertil{z{\sss B}}{\sss D}{\sman}
& = & \frac{2\ctwl}{\tcif} 
 \Zverti{\mu}{\sss B}{\sman}[\gbc \mtl I D_\mu]\,.
\eqa
Here we introduce the symbol $[\dots]$ for the definition
of the procedure of the projection of $\Gverti{\mu}{}{\sman}$
and $\Zverti{\mu}{}{\sman}$ to our basis.
It has the same meaning as in {\tt form} language \cite{Vermaseren:2000f},
namely, e.g. $\Gverti{\mu}{\sss B}{\sman}[\ib\gbc \gadu{\mu}\gap]$
means that only the coefficient of $[\ib\gbc \gadu{\mu}\gap]$ of
the whole expression $\Gverti{\mu}{\sss B}{\sman}$ is taken
({\em projected}).

 The factors $ 1/\bigl(\qf \siw \bigr)$, $ 2/\bigl(\siw\tcif\bigr)$ 
and $ 2 \ctws /\bigl(\stwl\tcif \bigr)$
for $\ph\ff\fbf$ vertices, and the 
factors $2\cow/\tcif$, $2\cow/\delta_f$
and  $ 2\ctwl/\tcif $ 
for $\zb\ff\fbf$ are due to the form factor definitions of
\eqn{structures-old}.

The total $\gamma\ff\bar{f}$ and $\zb\ff\bar{f}$ form factors are sums over 
four bosonic contributions $B =\ab,\zb,\wb,\hb$.
 All 24 components of the total scalar form factors in the $LQD$ basis look like:
\bqa
  \vvertil{\gamma(z) ff}{\sss{L,Q,D}}{\sman}  =
  \vvertil{\gamma(z){\sss A}}{\sss{L,Q,D}}{\sman} 
+ \vvertil{\gamma(z){\sss Z}}{\sss{L,Q,D}}{\sman} 
+ \vvertil{\gamma(z){\sss W}}{\sss{L,Q,D}}{\sman}
+ \vvertil{\gamma(z){\sss H}}{\sss{L,Q,D}}{\sman}.
\eqa
The quantities $\vvertil{\ph(z){\sss B}}{\sss L,Q,D}{\sman}$
originate from groups of diagrams, which we will call {\it clusters}. 
\subsection{Library of form factors for $Bff$ clusters}
Here we present a complete collection of scalar form factors
$\vvertil{\ph(z){\sss B}}{\sss L,Q,D}{\sman}$ originating 
from a vertex diagram with a virtual vector boson, contribution of 
a scalar partner of this vector boson, and relevant counterterms.

Actually three gauge-invariant subsets of diagrams of this kind,
$\ab$, $\zb$ and $\hb$, appear in our calculation.
They may be termed {\it clusters}, since they are 
natural building blocks of the complete scalar form factors, which are
the aim of our calculation. 
Again, in the spirit of our presentation, we write down their pole and finite 
parts.

The remaining vertices with virtual $\wb$ and $\hkp,\hkm$ with relevant 
counterterms we also define as the $\wb$ cluster. However, the latter diagrams
do not form a gauge-invariant subset.

We note, that $\vvertil{\gamma{\sss{A}}}{\sss{L}}{\sman}$ and
$\vvertil{\gamma{\sss{H}}}{\sss{L}}{\sman}$ are equal to zero.
\subsubsection{Library of QED form factors for $Att$ clusters}
 
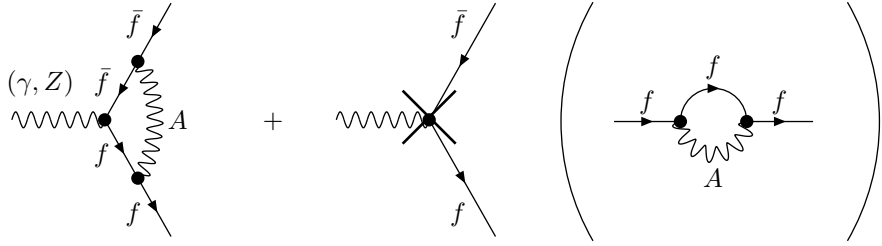
\begin{figure}[!h]
\vspace{-19mm}
\[
\baa{ccccc}
\begin{picture}(55,88)(0,41)
  \Photon(-10,44)(25,44){3}{6}
  \Vertex(25,44){2.5}
  \PhotonArc(0,44)(44,-28,28){3}{9}
  \Vertex(37.5,22){2.5}
  \Vertex(37.5,66){2.5}
  \ArrowLine(50,88)(37.5,66)
  \ArrowLine(37.5,66)(25,44)
  \ArrowLine(25,44)(37.5,22)
  \ArrowLine(37.5,22)(50,0)
  \Text(-12,53)[lb]{$(\gamma,Z)$}
  \Text(33,75)[lb]{$\fbf$}
  \Text(21,53)[lb]{$\fbf$}
  \Text(49,44)[lc]{$A$}
  \Text(21,35)[lt]{$\ff$}
  \Text(33,13)[lt]{$\ff$}
  \Text(1,1)[lb]{}
\end{picture}
\qquad  &+& \qquad
\begin{picture}(55,88)(0,41)
  \Photon(-10,44)(25,44){3}{6}
  \Vertex(25,44){2.5}
  \ArrowLine(50,88)(25,44)
  \ArrowLine(25,44)(50,0)
  \Text(33,75)[lb]{$\fbf$}
  \Text(33,13)[lt]{$\ff$}
  \Text(1,1)[lb]{}
\SetScale{2.0}
  \Line(17.5,17.5)(7.5,27.5)
  \Line(7.5,17.5)(17.5,27.5)
\end{picture}
&& \qquad 
\begin{picture}(75,20)(0,7.5)
\Text(12.5,13)[bc]{$\ff$}
\Text(62.5,13)[bc]{$\ff$}
\Text(37.5,26)[bc]{$\ff$}
\Text(37.5,-8)[tc]{$A$}
\Text(0,-7)[lt]{$ $}
  \ArrowLine(0,10)(25,10)
  \CArc(10,10)(90, -30, 30)
  \CArc(70,10)(90, -210 ,-150)
  \ArrowArcn(37.5,10)(12.5,180,0)
  \PhotonArc(37.5,10)(12.5,180,0){3}{7}
  \Vertex(25,10){2.5}
  \Vertex(50,10){2.5}
  \ArrowLine(50,10)(75,10)
\end{picture}
\eaa
\]
\vspace{8mm}
\caption
[$\ab$ cluster.] 
{$\ab$ cluster. 
The fermionic self-energy diagram in brackets
gives rise to the counter term contribution depicted by the solid cross.
\label{QEDcluster} }
\end{figure}

  Up to the one-loop level, there are two diagrams that contribute 
to the $\ab$ cluster, see ~\fig{QEDcluster}.

 Separating out pole contributions $1/{\bar\varepsilon}$, we define finite
(calligraphic) quantities. We note that, if a form factor $F_{\sss A}^{ij}(s)$
has a pole, then the corresponding finite part ${\cal F}_{\sss A}^{ij}(s)$
is $\mu$-dependent.

Since the scalar form factors of $\ab$ cluster
become UV-finite after wave function renormalization, 
for all form factors, which are also separately
gauge-invariant, we have:
\bqa
F^{\gamma(z){\sss A}}_{\sss{L,Q,D}} = {\cal F}^{\gamma(z){\sss A}}_{\sss{L,Q,D}}.
\eqa
Individual components are:
\begin{align}
{\cal F}^{\gamma{\sss A}}_{\sss L}  =& 0,
\nll[1mm]
{\cal F}^{\gamma{\sss A}}_{\sss Q}  =& \qf^2 \stws \biggl\{
  2 \lpar s-2 \mfs\rpar  \ctstot
\nll &
 -3 \bofstt + 3 \boftto  - 4 \mfs \boptot \biggr\},
\nll[1mm]
{\cal F}^{\gamma{\sss A}}_{\sss D}  =&
-\frac{\qf^3\stws}{\tcif} \frac{4}{\sdtit}\bigg[ \bofstt-\boftto \bigg],
\nn
\end{align}
\begin{align}
{\cal F}^{z{\sss A}}_{\sss L}  =& {\cal F}^{\gamma \sss A}_{Q} + \qf^2 \stws 
    \frac{8\mfs}{\sdtit}\bigg[ \bofstt-\boftto\bigg],
\nll[1mm]
{\cal F}^{z{\sss A}}_{\sss Q}  =& 
{\cal F}^{\gamma \sss A}_{Q} - \qf^2 \stws 
    \frac{8\mfs}{\sdtit} \frac{\tcif}{\delta_f} \bigg[ \bofstt-\boftto \bigg],
\nll[1mm]
{\cal F}^{z{\sss A}}_{\sss D}  =& 
-\frac{\qf^2\stws}{\tcif}\frac{2\vf}{\sdtit} \bigg[ \bofstt-\boftto \bigg],
\end{align}
with
\bqa
\sdtit&=& 4 \mfs - \sman\,.
\eqa

\subsubsection{Form factors of the $Z$  cluster}
\begin{figure}[!h]
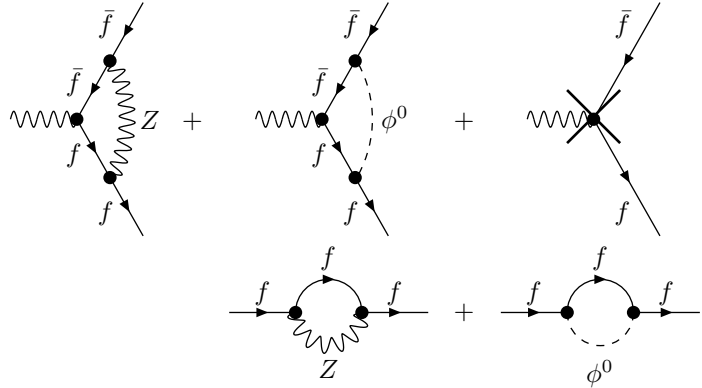

\caption
[$Z$ cluster.]
{$Z$ cluster. The two fermionic self-energy diagrams in the second row
give rise to the counterterm contribution depicted by the solid cross
in the last diagram of the first row.
\label{zavert}}
\end{figure}

The diagrams shown in \fig{zavert} contribute to the $\zb$ cluster.

\vspace{-17mm}
\[
\baa{ccccccccc}
&&
\begin{picture}(55,88)(0,41)
  \Photon(0,44)(25,44){3}{5}
  \Vertex(25,44){2.5}
  \PhotonArc(0,44)(44,-28,28){3}{9}
  \Vertex(37.5,22){2.5}
  \Vertex(37.5,66){2.5}
  \ArrowLine(50,88)(37.5,66)
  \ArrowLine(37.5,66)(25,44)
  \ArrowLine(25,44)(37.5,22)
  \ArrowLine(37.5,22)(50,0)
  \Text(33,75)[lb]{$\fbf$}
  \Text(21,53)[lb]{$\fbf$}
  \Text(49,44)[lc]{$\zb$}
  \Text(21,35)[lt]{$\ff$}
  \Text(33,13)[lt]{$\ff$}
  \Text(1,1)[lb]{}
\end{picture}
&+&
\begin{picture}(55,88)(0,41)
  \Photon(0,44)(25,44){3}{5}
  \Vertex(25,44){2.5}
  \DashCArc(0,44)(44,-28,28){3}
  \Vertex(37.5,22){2.5}
  \Vertex(37.5,66){2.5}
  \ArrowLine(50,88)(37.5,66)
  \ArrowLine(37.5,66)(25,44)
  \ArrowLine(25,44)(37.5,22)
  \ArrowLine(37.5,22)(50,0)
  \Text(33,75)[lb]{$\fbf$}
  \Text(21,53)[lb]{$\fbf$}
  \Text(47.5,44)[lc]{$\hkn$}
  \Text(21,35)[lt]{$\ff$}
  \Text(33,13)[lt]{$\ff$}
  \Text(1,1)[lb]{}
\end{picture}
&+&
\begin{picture}(55,88)(0,41)
  \Photon(0,44)(25,44){3}{5}
  \Vertex(25,44){2.5}
  \ArrowLine(50,88)(25,44)
  \ArrowLine(25,44)(50,0)
  \Text(33,75)[lb]{$\fbf$}
  \Text(33,13)[lt]{$\ff$}
  \Text(1,1)[lb]{}
\SetScale{2.0}
  \Line(17.5,17.5)(7.5,27.5)
  \Line(7.5,17.5)(17.5,27.5)
\end{picture}
\nl \nl  [13mm] 
& &
& &
\begin{picture}(75,20)(0,7.5)
\Text(12.5,13)[bc]{$\ff$}
\Text(62.5,13)[bc]{$\ff$}
\Text(37.5,26)[bc]{$\ff$}
\Text(37.5,-8)[tc]{$\zb$}
\Text(0,-7)[lt]{$ $}
  \ArrowLine(0,10)(25,10)
  \ArrowArcn(37.5,10)(12.5,180,0)
  \PhotonArc(37.5,10)(12.5,180,0){3}{7}
  \Vertex(25,10){2.5}
  \Vertex(50,10){2.5}
  \ArrowLine(50,10)(75,10)
\end{picture}
&+&
\begin{picture}(75,20)(0,7.5)
\Text(12.5,13)[bc]{$\ff$}
\Text(62.5,13)[bc]{$\ff$}
\Text(37.5,26)[bc]{$\ff$}
\Text(37.5,-8)[tc]{$\hkn$}
\Text(0,-7)[lt]{$ $}
  \ArrowLine(0,10)(25,10)
  \ArrowArcn(37.5,10)(12.5,180,0)
  \DashCArc(37.5,10)(12.5,180,0){3}
  \Vertex(25,10){2.5}
  \Vertex(50,10){2.5}
  \ArrowLine(50,10)(75,10)
\end{picture}
\eaa
\vspace*{-1mm}
\]


 Separating out pole contributions $1/{\bar\varepsilon}$, we define finite
quantities:
\begin{align}
 \vvertil{\gamma{\sss Z}}{\sss{L}}{\sman}  = & 
 \cvertil{\gamma{\sss Z}}{\sss{L}}{\sman}, 
\nll[2mm]
\vvertil{\gamma{\sss Z}}{\sss{Q}}{\sman}   = &
  \cvertil{\gamma{\sss Z}}{\sss{Q}}{\sman},
\nll[2mm]
\vvertil{\gamma{\sss Z}}{\sss{D}}{\sman}   = &
  \cvertil{\gamma{\sss Z}}{\sss{D}}{\sman},
\nll[1mm]
 \vvertil{z{\sss Z}}{\sss L}{\sman}   = & 
 -\frac{1}{4} \rtw \pole 
 +\cvertil{z{\sss Z}}{\sss L}{\sman}, 
\nll
  \vvertil{z{\sss Z}}{\sss{Q}}{\sman} = &
 -\frac{1}{16} \frac{1}{\qfm \stws} \rtw  \pole
 +\cvertil{z{\sss Z}}{\sss{Q}}{\sman},
\nll[1mm]
\vvertil{z{\sss Z}}{\sss{D}}{\sman}   = &
  \cvertil{z{\sss Z}}{\sss{D}}{\sman}.
\end{align}
Here the finite parts are:
\begin{align}
{\cal F }^{\gamma {\sss Z}}_{\sss L} \lpar \sman \rpar   =&
 \frac{1}{\ctws} \qf \vf \biggl\{
            2 \lpar 2+\frac{1}{\Rz} \rpar
                        \mzs \cff{-\mfs}{-\mfs}{-s}{\mfl}{\mzl}{\mfl}
\nll &
           -3 \fbff{0}{-s}{\mfl}{\mfl} + 2 \fbff{0}{-\mfs}{\mfl}{\mzl} - \Lmmt
\\   &
   + \fbff{d1}{-\mfs}{\mfl}{\zml} -2 \lpar 1+4 \ruz \rpar
     \frac{\mzs}{\sdtit} \Longab{\mfl}{\mfl}{\mzl}  \biggr\},
\nonumber
\label{b0d_containing}
\end{align}
\vspace*{-1mm}
\begin{align}
 {\cal F }^{\gamma {\sss Z}}_{\sss Q}\lpar \sman \rpar
 =&
 \frac{1}{4 \cows} \bigg\{
\delta_f^2 \biggl[ \biggl(4 \lpar 1-\ruz \rpar+\frac{2}{\Rz} \biggr) 
  \mzs  \cff{-\mfs}{-\mfs}{-s}{\mfl}{\mzl}{\mfl}
\nll &
                -3 \fbff{0}{-s}{\mfl}{\mfl}+4 \fbff{0}{-\mfs}{\mfl}{\mzl}+\Lmmz-\frac{5}{2}
\nll &
                -\ruz \fbff{d2}{-\mfs}{\mfl}{\zml}
                -2 \lpar 1+2 \ruz \rpar 
                 \mzs \bff{0p}{-\mfs}{\mfl}{\mzl} \biggr]
\nll & 
 + 2 \vf \af \ruz \biggl[-4 \mzs \cff{-\mfs}{-\mfs}{-s}{\mfl}{\mzl}{\mfl}
\nll[1mm] & 
       +2\lpar \Lmmt - \Lmmz \rpar
               - 2 \lpar 1 - \ruz \rpar \fbff{d2}{-\mfs}{\mfl}{\mzl}+1
\nll & 
               - \frac{2}{\ruz} \biggl(  \lpar 1 + 2 \ruz\rpar\mzs 
                 \bff{0p}{-\mfs}{\mfl}{\mzl} + \frac{1}{2} \biggr) \biggr]
\nll & 
 +2 \af^2 \ruz   \biggl[ \fbff{0}{-s}{\mfl}{\mfl}  + \Lmmz
               - \ruz \fbff{d2}{-\mfs}{\uml}{\zml}
\nll & 
                +6\mzs \bff{0p}{-\mfs}{\uml}{\zml} - \frac{5}{2} \biggr]
\nll &  
 -4 \biggl[\frac{ \delta_f^2}{2}-\lpar 4\vf \af-\af^2 \rpar \ruz \biggr]
  \frac{\mzs}{\sdtit} \Longab{\mfl}{\mfl}{\mzl}
                  \biggr\}, 
\end{align}
\vspace*{-1mm}
\begin{align}
{\cal F}^{\gamma{\sss Z}}_{\sss D} \lpar \sman \rpar    = &
  -\frac{2\qf}{\tcif\ctws}
\frac{1}{\sdtit} \biggl\{
      \frac{\vf^2+\af^2}{2}
 \biggl[ -4 \mzs 
 \cff{-\mfs}{-\mfs}{-s}{\mfl}{\mzl}{\mfl}
\nll[1mm] &
 + \fbff{0}{-\sman}{\mfl}{\mfl}-2 \fbff{0}{-\mfs}{\mfl}{\mzl}-\Lmmt
\nll &
 + \fbff{d1}{-\mfs}{\mfl}{\zml} 
           +2  +6 \frac{\mzs}{\sdtit} \Longab{\mfl}{\mfl}{\mzl}
                 \biggr]
\nll &
         +\af^2 \biggl[ 2 \lpar 3 \rtz-\frac{1}{\Rz} \rpar \mzs 
 \cff{-\mfs}{-\mfs}{-s}{\mfl}{\mzl}{\mfl}
\nll &
                +\fbff{0}{-\mfs}{\mfl}{\mzl} + \Lmmz - 1
\nll &
                -\rtz \Bigl[\fbff{0}{-\sman}{\mfl}{\mfl} + \Lmmt - 2 \Bigr]
\nll &
          -2\biggl( 2 - 3 \frac{\mfs}{\sdtit} \biggr) \Longab{\mfl}{\mfl}{\mzl}
               \biggr] \biggr\},
\end{align}

\clearpage

\begin{align}
{\cal F}^{z{\sss Z}}_{\sss L}\lpar \sman \rpar =&
\frac{1}{4 \cows} \biggl\{
\frac{3\vf^2+\af^2}{3} \biggl[ 2 \bigg( 3 \lpar 2+\frac{1}{\Rz} \rpar-2 \ruz 
\bigg) 
\nll[-1mm]&\hspace*{4.1cm}\times
\mzs 
     \cff{-\mfs}{-\mfs}{-s}{\mfl}{\mzl}{\mfl}
\nll &
     -9 \fbff{0}{-s}{\mfl}{\mfl}+8 \fbff{0}{-\mfs}{\mfl}{\mzl}-\Lmmt-2
\nll &
 + \fbff{d1}{-\mfs}{\mfl}{\zml} 
    -2 \lpar 1+2 \ruz \rpar \mzs \bff{0p}{-\mfs}{\mfl}{\mzl} \bigg]
\nll &
    -\frac{2}{3} \af^2 \biggl[ 4 \mfs \cff{-\mfs}{-\mfs}{-s}{\mfl}{\mzl}{\mfl}
\\[-.5mm]   &
     +3 \ruz \Big[ \fbff{0}{-s}{\mfl}{\mfl} + \Lmmt \Big] 
     +\fbff{0}{-\mfs}{\mfl}{\mzl} 
\nll &
+ 3 \Lmmz + 2 \Lmmt-1
     + 2\fbff{d1}{-\mfs}{\mfl}{\zml}
\nll &
     + 2 \lpar 1-7 \ruz \rpar \mzs \bff{0p}{-\mfs}{\mfl}{\mzl}  \biggr] 
\nll[-2mm] &
     -2 \biggl[ \lpar 3\vf^2+\af^2\rpar \lpar 1 + 4 \ruz \rpar-2 \af^2 \ruz \biggr] 
 \frac{\mzs} {\sdtit} \Longab{\mfl}{\mfl}{\mzl}  
                \biggr\}, 
\nll[1mm]
{\cal F}^{z{\sss Z}}_{\sss Q}\lpar \sman \rpar =&
 \frac{1}{\cows} \biggl\{ 
 \frac{1}{4} \delta_f^2 
  \Bigl[ 2 \lpar 2-2 \ruz+\frac{1}{\Rz}\rpar
            \mzs \cff{-\mfs}{-\mfs}{-s}{\mfl}{\mzl}{\mfl}
\nll &
  - 3 \fbff{0}{-s}{\mfl}{\mfl}+4 \fbff{0}{-\mfs}{\mfl}{\mzl}+\Lmmt-2
\nll[1mm] &
  - \fbff{d1}{-\mfs}{\mfl}{\zml}
                -2 \lpar 1+2 \ruz \rpar \mzs \bff{0p}{-\mfs}{\mfl}{\mzl} 
\nll &
                -2\frac{\mzs}{\sdtit} \Longab{\mfl}{\mfl}{\mzl} \biggr]
\nll &
 +\af \ruz \biggl( \vf \biggl[-2 \mzs \cff{-\mfs}{-\mfs}{-s}{\mfl}{\mzl}{\mfl}
\nll &
 +\frac{1}{\ruz} \Bigl[ \fbff{0}{-\mfs}{\mfl}{\mzl} + \Lmmt
 - 1 -\fbff{d1}{-\mfs}{\mfl}{\zml}
\nll &
               -\lpar 1 + 2 \ruz \rpar \mzs \bff{0p}{-\mfs}{\mfl}{\mzl} \Bigr]
       +6 \frac{\mzs}{\sdtit} \Longab{\mfl}{\mfl}{\mzl} \Bigr]
\nll &
       -\frac{1}{2}\af \biggl[ 8 \mzs \cff{-\mfs}{-\mfs}{-s}{\mfl}{\mzl}{\mfl}
\nll[1mm] &
       - \fbff{0}{-s}{\mfl}{\mfl} - \Lmmt + 2 
       +\fbff{d1}{-\mfs}{\mfl}{\zml}
\nll[-.5mm] &
       - 6 \mzs \bff{0p}{-\mfs}{\mfl}{\mzl} 
       - 10 \frac{\mzs}{\sdtit} \Longab{\mfl}{\mfl}{\mzl} 
                      \biggr]
\nll[-1mm] & 
 -\frac{\af^2}{\delta_f} \biggl[
                 4 \mzs \cff{-\mfs}{-\mfs}{-s}{\mfl}{\mzl}{\mfl}
\nll[-1.5mm] &
                         - \fbff{0}{-s}{\mfl}{\mfl}+1 
                -6 \frac{\mzs}{\sdtit} \Longab{\mfl}{\mfl}{\mzl} \biggr]
           \biggr)             \biggr\},
\end{align}
\begin{align}
{\cal F}^{z{\sss Z}}_{\sss D}\lpar \sman \rpar  = &
-\frac{1}{2\tcif\ctws} 
\frac{1}{\sdtit}
   \biggl\{\lpar 3\af^2+\vf^2\rpar \vf
\nll &
   \times  \biggl[-4 \mzs \cff{-\mfs}{-\mfs}{-s}{\mfl}{\mzl}{\mfl}
\nll[1mm] &
      + \fbff{0}{-\sman}{\mfl}{\mfl}-2 \fbff{0}{-\mfs}{\mfl}{\mzl}-\Lmmt+2
\nll[1mm] &
      + \fbff{d1}{-\mfs}{\mfl}{\zml}
      + 6 \frac{\mzs}{\sdtit} \Longab{\mfl}{\mfl}{\mzl}
               \biggr]
\nll &
    +2 \vf \af^2 \biggl[ 2 \lpar 7 \rtz-2 \frac{1}{\Rz} \rpar \mzs 
 \cff{-\mfs}{-\mfs}{-s}{\mfl}{\mzl}{\mfl}
\nll &
                +\fbff{0}{-\mfs}{\mfl}{\mzl}+\Lmmz-1
\nll &
                -\rtz \Bigl[ \fbff{0}{-\sman}{\mfl}{\mfl}+\Lmmt-2 \Bigr] 
\nll &
                -2 \bigg( 4-3 \frac{\mfs}{\sdtit} \bigg) 
         \Longab{\mfl}{\mfl}{\mzl}
               \biggr]                        
    \biggr\}.
\end{align}
\noindent
In \eqn{b0d_containing} the `once and twice subtracted' functions $\sfbff{d1}$ 
and $\sfbff{d2}$ are met:
\begin{align}
\fbff{d1}{-\mfs}{\mfl}{\zml} =&
 \frac{1}{\rtz} \Bigl[
 \fbff{0}{-\mfs}{\mfl}{\zml}+\Lmmz-1\Bigr],
\\
\fbff{d2}{-\mfs}{\mfl}{\zml} =&
 \frac{1}{\rtz^2} \biggl[
 \fbff{0}{-\mfs}{\mfl}{\zml}+\Lmmz-1
\nll &\hspace{1cm}
 -\rtz \lpar\Lmmt-\Lmmz+\frac{1}{2} \rpar\biggr].
\nn
\end{align}
They remain finite in the limit $\mfl\to 0$.

We note that, for the $\zb$ cluster, all the six scalar form factors
$\vvertil{\ph(z){\sss Z}}{\sss L,Q,D}{\sman}$ are {\em separately}
gauge-invariant.
\subsubsection{Form factors of the $H$ cluster}
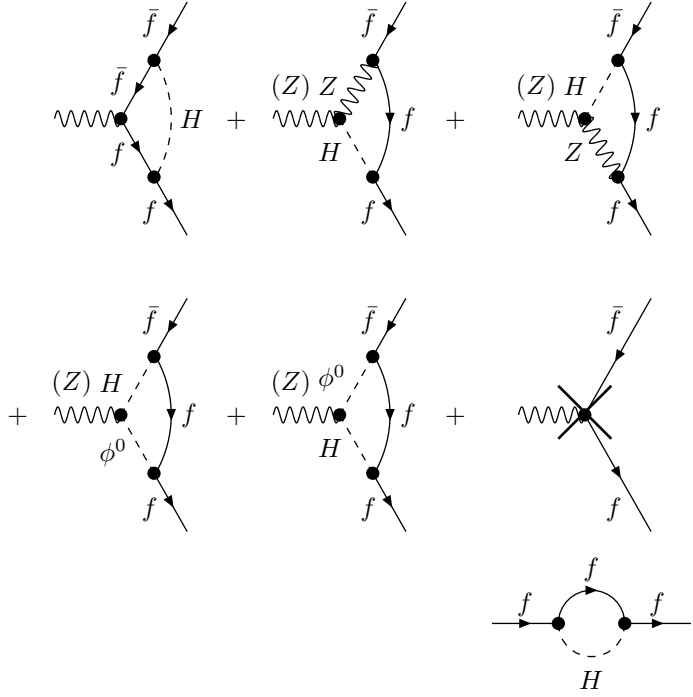
\begin{figure}[h]
\vspace{-15mm}
\[
\hspace{-1cm}
\baa{cccccccc} 
& &
\begin{picture}(55,88)(0,41)
  \Photon(0,44)(25,44){3}{5}
  \Vertex(25,44){2.5}
  \DashCArc(0,44)(44,-28,28){3}
  \Vertex(37.5,22){2.5}
  \Vertex(37.5,66){2.5}
  \ArrowLine(50,88)(37.5,66)
  \ArrowLine(37.5,66)(25,44)
  \ArrowLine(25,44)(37.5,22)
  \ArrowLine(37.5,22)(50,0)
  \Text(33,75)[lb]{$\fbf$}
  \Text(21,53)[lb]{$\fbf$}
  \Text(47.5,44)[lc]{$\hb$}
  \Text(21,35)[lt]{$\ff$}
  \Text(33,13)[lt]{$\ff$}
  \Text(1,1)[lb]{}
\end{picture}
& + &
\begin{picture}(55,88)(0,41)
  \Photon(0,44)(25,44){3}{5}
  \Vertex(25,44){2.5}
  \ArrowArcn(0,44)(44,28,-28)
  \Vertex(37.5,22){2.5}
  \Vertex(37.5,66){2.5}
  \ArrowLine(50,88)(37.5,66)
  \ArrowLine(37.5,22)(50,0)
  \Photon(25,44)(37.5,66){3}{5}
  \DashLine(37.5,22)(25,44){3}
  \Text(33,75)[lb]{$\fbf$}
  \Text(-1,51)[lb]{$(\zb)$}
  \Text(17,53)[lb]{$\zb$}
  \Text(48,44)[lc]{$\ff$}
  \Text(17,35)[lt]{$\hb$}
  \Text(33,13)[lt]{$\ff$}
  \Text(1,1)[lb]{}
\end{picture}
& + &
\begin{picture}(55,88)(0,41)
  \Photon(0,44)(25,44){3}{5}
  \Vertex(25,44){2.5}
  \ArrowArcn(0,44)(44,28,-28)
  \Vertex(37.5,22){2.5}
  \Vertex(37.5,66){2.5}
  \ArrowLine(50,88)(37.5,66)
  \ArrowLine(37.5,22)(50,0)
  \DashLine(25,44)(37.5,66){3}
  \Photon(37.5,22)(25,44){3}{5}
  \Text(33,75)[lb]{$\fbf$}
  \Text(-1,51)[lb]{$(\zb)$}
  \Text(17,53)[lb]{$\hb$}
  \Text(48,44)[lc]{$\ff$}
  \Text(17,35)[lt]{$\zb$}
  \Text(33,13)[lt]{$\ff$}
  \Text(1,1)[lb]{}
\end{picture}
 \nl \nl [3mm]
& + & 
\begin{picture}(55,88)(0,41)
  \Photon(0,44)(25,44){3}{5}
  \Vertex(25,44){2.5}
  \ArrowArcn(0,44)(44,28,-28)
  \Vertex(37.5,22){2.5}
  \Vertex(37.5,66){2.5}
  \ArrowLine(50,88)(37.5,66)
  \ArrowLine(37.5,22)(50,0)
  \DashLine(25,44)(37.5,66){3}
  \DashLine(37.5,22)(25,44){3}
  \Text(33,75)[lb]{$\fbf$}
  \Text(-1,51)[lb]{$(\zb)$}
  \Text(17,53)[lb]{$\hb$}
  \Text(48,44)[lc]{$\ff$}
  \Text(17,35)[lt]{$\hkn$}
  \Text(33,13)[lt]{$\ff$}
  \Text(1,1)[lb]{}
\end{picture}
& + &
\begin{picture}(55,88)(0,41)
  \Photon(0,44)(25,44){3}{5}
  \Vertex(25,44){2.5}
  \ArrowArcn(0,44)(44,28,-28)
  \Vertex(37.5,22){2.5}
  \Vertex(37.5,66){2.5}
  \ArrowLine(50,88)(37.5,66)
  \ArrowLine(37.5,22)(50,0)
  \DashLine(25,44)(37.5,66){3}
  \DashLine(37.5,22)(25,44){3}
  \Text(33,75)[lb]{$\fbf$}
  \Text(-1,51)[lb]{$(\zb)$}
  \Text(17,53)[lb]{$\hkn$}
  \Text(48,44)[lc]{$\ff$}
  \Text(17,35)[lt]{$\hb$}
  \Text(33,13)[lt]{$\ff$}
  \Text(1,1)[lb]{}
\end{picture}
&+&
\begin{picture}(55,88)(0,41)
  \Photon(0,44)(25,44){3}{5}
  \Vertex(25,44){2.5}
  \ArrowLine(50,88)(25,44)
  \ArrowLine(25,44)(50,0)
  \Text(33,75)[lb]{$\fbf$}
  \Text(33,13)[lt]{$\ff$}
  \Text(1,1)[lb]{}
\SetScale{2.0}
  \Line(17.5,17.5)(7.5,27.5)
  \Line(7.5,17.5)(17.5,27.5)
\end{picture}
\nl 
\nl [15mm]
& &  & &  & &
\begin{picture}(75,20)(0,7.5)
\Text(12.5,13)[bc]{$\ff$}
\Text(62.5,13)[bc]{$\ff$}
\Text(37.5,26)[bc]{$\ff$}
\Text(37.5,-8)[tc]{$\hb$}
\Text(0,-7)[lt]{$$}
  \ArrowLine(0,10)(25,10)
  \ArrowArcn(37.5,10)(12.5,180,0)
  \DashCArc(37.5,10)(12.5,180,0){3}
  \Vertex(25,10){2.5}
  \Vertex(50,10){2.5}
  \ArrowLine(50,10)(75,10)
\end{picture}
\eaa
\]
\vspace*{2mm}
\caption[$\hb$ cluster.]
{$\hb$ cluster: the vertices and the counterterm. 
\label{fig:H_cluster}}
\end{figure}

The diagrams of \fig{fig:H_cluster} contribute to the $\hb$ cluster.
Separating UV poles, we have:
\begin{align}
\vvertil{\gamma{\sss H}}{\sss{Q}}{\sman} = &
  \cvertil{\gamma{\sss H}}{\sss{Q}}{\sman},
\nll[1mm]
\vvertil{\gamma{\sss H}}{\sss{D}}{\sman} = &
  \cvertil{\gamma{\sss H}}{\sss{D}}{\sman},
\nll[1mm]
 \vvertil{z{\sss H}}{\sss L}{\sman}   = &
  \frac{1}{4} \rtw \pole
+ \cvertil{z{\sss H}}{\sss L}{\sman},
\nll[1mm]
 \vvertil{z{\sss H}}{\sss{Q}}{\sman}  = & 
 \frac{1}{16} \frac{1}{\qfm} \rtw  \frac{1}{\stws} \pole
+ \cvertil{z{\sss H}}{\sss{Q}}{\sman},
\nll[2mm]
  \vvertil{z{\sss H}}{\sss{D}}{\sman} = &
  \cvertil{z{\sss H}}{\sss{D}}{\sman},
\label{higgs_formfactors}
\end{align}

\noindent
with the finite parts:
\begin{align}
{\cal F}^{\gamma{\sss H}}_{\sss Q}\lpar \sman \rpar =&
 \frac{1}{8} \ruw \biggl\{
   8 \mfs \cff{-\mfs}{-\mfs}{-s}{\mfl}{\mhl}{\mfl}
\nll[1mm] &
            + \fbff{0}{-s}{\mfl}{\mfl}+\Lmmt-2 
            - \fbff{d1}{-\mfs}{\mfl}{\mhl}
\nll[2mm] &
               -2 \lpar 1-4 \ruh \rpar \mhs \bff{0p}{-\mfs}{\mfl}{\mhl}
\nll &
               -2 \frac{\mhs}{\sdtit} \Longab{\mfl}{\mfl}{\mhl} \biggr\}, 
\\[3mm]
{\cal F}^{\gamma{\sss H}}_{\sss D}\lpar \sman \rpar =&
 - \frac{Q_t}{2\tcif} 
\frac{\rtw }{\sdtit} 
\biggl\{
     -6 \mhs 
 \cff{-\mfs}{-\mfs}{-s}{\mfl}{\mhl}{\mfl}
\nll[1mm] &
      +3 \fbff{0}{-\sman}{\mfl}{\mfl}-4 \fbff{0}{-\mfs}{\mfl}{\mhl}-\Lmmt    
\nll[1mm] &
      +2 + \fbff{d1}{-\mfs}{\mfl}{\mhl}
      +6 \frac{\mhs}{\sdtit} \Longab{\mfl}{\mfl}{\mhl}    
                                                \biggr\},
\end{align}
\newpage
\begin{align}
{\cal F}^{z{\sss H}}_{\sss L}\lpar \sman \rpar =&
 \frac{1}{4}\ruw \biggl\{
    4  \mfs  \cff{-\mfs}{-\mfs}{-s}{\mfl}{\mhl}{\mfl}
\nll[1mm] &
  +\Bigl[4\lpar 1-\ruz\rpar +(1-\rhz)^2 \Rz \Bigr]
\nll[-1mm] &\hspace*{3cm}\times
     \mzs \cff{-\mfs}{-\mfs}{-s}{\mhl}{\mfl}{\mzl}
\nll &
   +2 \fbff{0}{-s}{\mzl}{\mhl}-\frac{1}{2} \fbff{0}{-s}{\mfl}{\mfl}
   +\frac{1}{2} \Lmmt + 2 
\nll &
   -\frac{1}{2} \fbff{d1}{-\mfs}{\mfl}{\mhl}
   - (1-4 \ruh) {\mhs} \bff{0p}{-\mfs}{\mfl}{\mhl} 
\nll &
   +(1-\rhz) \Rz
   \Bigl[ \fbff{0}{-\mfs}{\mzl}{\mfl}-\fbff{0}{-\mfs}{\mfl}{\mhl} \Bigr]
\nll &
                +\frac{\mzs}{\sdtit}
  \biggl[
                 \lpar \rhz-8 \ruz \rpar \Longab{\mfl}{\mfl}{\mhl}
\nll &
  -2 \lpar 3 - \rhz + 4\rtz \rpar 
          \LongHi{\mfl}{\mhl}{\mzl} 
  \biggr]           \biggr\},
\\[3mm]
{\cal F}^{z {\sss H}}_{\sss Q}\lpar \sman \rpar =&
\ruw\biggl\{
  \mfs \cff{-\mfs}{-\mfs}{-s}{\mfl}{\mhl}{\mfl}
\nll[1mm] &
 +\Bigl[ 1 + \lpar 1-\rhz \rpar \Rz \Bigr] 
 \mzs \cff{-\mfs}{-\mfs}{-s}{\mhl}{\mfl}{\mzl}
\nll[1mm] &
 +\frac{1}{8} \Bigl[ \fbff{0}{-s}{\mfl}{\mfl}+\Lmmt-2 \Bigr]
\nll &
 +\Rz \Bigl[ \fbff{0}{-\mfs}{\mzl}{\mfl}-\fbff{0}{-\mfs}{\mfl}{\mhl} \Bigr]
\nll &
 -\frac{1}{8}\fbff{d1}{-\mfs}{\mfl}{\mhl}
 -\frac{1}{4}\lpar 1-4 \ruh \rpar\mhs \bff{0p}{-\mfs}{\mfl}{\mhl}
\nll &
 -\frac{1}{4} \frac{\mhs}{\sdtit} \Longab{\mfl}{\mfl}{\mhl} 
\nll &  
 +\frac{1}{4} \frac{\af}{\delta_f} 
\biggl(
   \Bigl[ 4 \rtz  + \lpar 3+\rhz \rpar (1-\rhz) \Rz \Bigr]
\nll[-1mm] &\hspace*{3cm}\times
   \mzs \cff{-\mfs}{-\mfs}{-s}{\mhl}{\mfl}{\mzl}
\nll[1mm] &
                + \fbff{0}{-s}{\mfl}{\mfl} - 2 \fbff{0}{-s}{\mzl}{\mhl}-3
\nll[1mm] &
+ \lpar 3+\rhz \rpar \Rz
\Bigl[\fbff{0}{-\mfs}{\mzl}{\mfl} -\fbff{0}{-\mfs}{\mfl}{\mhl} \Bigr]
\nll &
  - 2 \frac{\mzs}{\sdtit} 
         \bigg[ \lpar 1 - 4 \ruh \rpar \rhz \Longab{\mfl}{\mfl}{\mhl}
\nll &
  - \lpar 3 - \rhz + 4\rtz \rpar 
          \LongHi{\mfl}{\mhl}{\mzl} 
    \bigg] 
\biggr)
\biggr\},
\end{align}
\begin{align}
{\cal F}^{z{\sss H}}_{\sss D}\lpar \sman \rpar = &
-\frac{\vf}{2\tcif\ctws} 
\frac{1}{\sdtit}  \biggl\{\hspace{-1mm}
   -3 \rtz \mhs \cff{-\mfs}{-\mfs}{-s}{\mfl}{\mhl}{\mfl}
\nll &
   + 2 \biggl[
         2 \lpar \rhz-1 \rpar \frac{\mfs}{s}-\rhz+2 \rtz\biggr] 
\nll[-1.5mm] &\hspace{32mm}\times
 \mzs  \cff{-\mfs}{-\mfs}{-s}{\mhl}{\mfl}{\mzl}
\nll[-1mm] &
                -4 \frac{\mfs}{s} 
    \Bigl[ \fbff{0}{-\mfs}{\mzl}{\mfl}-\fbff{0}{-\mfs}{\mfl}{\mhl} \Bigr]
\nll &
 -2 \fbff{0}{-\sman}{\mzl}{\mhl}
 +2\fbff{0}{-\mfs}{\mzl}{\mfl}
\nll &
                +\frac{3}{2} \rtz 
 \Bigl[\fbff{0}{-\sman}{\mfl}{\mfl}
      -\fbff{0}{-\mfs}{\mfl}{\mhl} \Bigr]
\nll &
                +\frac{1}{2} \rhz \Bigl[
       \fbff{0}{-\mfs}{\mfl}{\mhl}+\Lmmh-1 \Bigr]
\nll &
 -\frac{1}{2} \rtz \Bigl[
       \fbff{0}{-\mfs}{\mfl}{\mhl}+\Lmmt-2 \Bigr]
\nll &
  +3 \rtz \frac{\mhs}{\sdtit} \Longab{\mfl}{\mfl}{\mhl}
             \biggr\}.\qquad
\end{align}
Again, the five (one does not exist) scalar form factors in \eqn{higgs_formfactors} 
are {\em separately} gauge-invariant. Note also that UV poles persisting
in the scalar form factors of the $\hb$ cluster cancel exactly the corresponding poles
of the $\zb$ cluster. In other words, the form factors of the `neutral sector'
cluster ($\zb+\hb$) are UV finite.

In total, we have 11 separately gauge-invariant {\em building blocks}
that originate from $\zb$ and $\hb$ clusters.
\subsubsection{Form factors of the $W$ cluster}
\begin{figure}[h]
\vspace{-18.5mm}
\[
\baa{cccccc}

\begin{picture}(55,88)(0,41)
  \Photon(0,44)(25,44){3}{5}
  \Vertex(25,44){2.5}
  \PhotonArc(0,44)(44,-28,28){3}{9}
  \Vertex(37.5,22){2.5}
  \Vertex(37.5,66){2.5}
  \ArrowLine(50,88)(37.5,66)
  \ArrowLine(37.5,66)(25,44)
  \ArrowLine(25,44)(37.5,22)
  \ArrowLine(37.5,22)(50,0)
  \Text(33,75)[lb]{$\fbf$}
  \Text(21,53)[lb]{$\fbfp$}
  \Text(47.5,44)[lc]{$\wb$}
  \Text(21,35)[lt]{$\ffp$}
  \Text(33,13)[lt]{$\ff$}
  \Text(1,1)[lb]{}
\end{picture}
&+&
\begin{picture}(55,88)(0,41)
  \Photon(0,44)(25,44){3}{5}
  \Vertex(25,44){2.5}
  \DashCArc(0,44)(44,-28,28){3}
  \Vertex(37.5,22){2.5}
  \Vertex(37.5,66){2.5}
  \ArrowLine(50,88)(37.5,66)
  \ArrowLine(37.5,66)(25,44)
  \ArrowLine(25,44)(37.5,22)
  \ArrowLine(37.5,22)(50,0)
  \Text(33,75)[lb]{$\fbf$}
  \Text(19,53)[lb]{$\fbfp$}
  \Text(47.5,44)[lc]{$\hkg$}
  \Text(19,35)[lt]{$\ffp$}
  \Text(33,13)[lt]{$\ff$}
  \Text(1,1)[lb]{}
\end{picture}
&+&
\begin{picture}(55,88)(0,41)
  \Photon(0,44)(25,44){3}{5}
  \Vertex(25,44){2.5}
  \ArrowLine(50,88)(25,44)
  \ArrowLine(25,44)(50,0)
  \Text(33,75)[lb]{$\fbf$}
  \Text(33,13)[lt]{$\ff$}
  \Text(1,1)[lb]{}
\SetScale{2.0}
  \Line(17.5,17.5)(7.5,27.5)
  \Line(7.5,17.5)(17.5,27.5)
\end{picture}
\nl \nl [14mm]
&& 
\begin{picture}(75,20)(0,7.5)
\Text(12.5,13)[bc]{$\ff$}
\Text(62.5,13)[bc]{$\ff$}
\Text(37.5,26)[bc]{$\ffp$}
\Text(37.5,-8)[tc]{$\hkg$}
\Text(0,-7)[lt]{$$}
  \ArrowLine(0,10)(25,10)
  \ArrowArcn(37.5,10)(12.5,180,0)
  \DashCArc(37.5,10)(12.5,180,0){3}
  \Vertex(25,10){2.5}
  \Vertex(50,10){2.5}
  \ArrowLine(50,10)(75,10)
\end{picture}
&\hspace*{-2.5mm}+\hspace*{-2.5mm}& 
\begin{picture}(75,20)(0,7.5)
\Text(12.5,13)[bc]{$\ff$}
\Text(62.5,13)[bc]{$\ff$}
\Text(37.5,26)[bc]{$\ffp$}
\Text(37.5,-8)[tc]{$\wb$}
\Text(0,-7)[lt]{$$}
  \ArrowLine(0,10)(25,10)
  \ArrowArcn(37.5,10)(12.5,180,0)
  \PhotonArc(37.5,10)(12.5,180,0){3}{7}
  \Vertex(25,10){2.5}
  \Vertex(50,10){2.5}
  \ArrowLine(50,10)(75,10)
\end{picture}
\nl[-7mm] 
\begin{picture}(55,88)(0,41)
  \Photon(0,44)(25,44){3}{5}
  \Vertex(25,44){2.5}
  \ArrowArcn(0,44)(44,28,-28)
  \Vertex(37.5,22){2.5}
  \Vertex(37.5,66){2.5}
  \ArrowLine(50,88)(37.5,66)
  \ArrowLine(37.5,22)(50,0)
  \Photon(25,44)(37.5,66){3}{5}
  \Photon(37.5,22)(25,44){3}{5}
  \Text(33,75)[lb]{$\fbf$}
  \Text(15,53)[lb]{$\wb$}
  \Text(48,44)[lc]{$\ffp$}
  \Text(4,35)[lt]{$\wb(\hkg)$}
  \Text(33,13)[lt]{$\ff$}
  \Text(1,1)[lb]{}
\end{picture}
&+&
\begin{picture}(55,88)(0,41)
  \Photon(0,44)(25,44){3}{5}
  \Vertex(25,44){2.5}
  \ArrowArcn(0,44)(44,28,-28)
  \Vertex(37.5,22){2.5}
  \Vertex(37.5,66){2.5}
  \ArrowLine(50,88)(37.5,66)
  \ArrowLine(37.5,22)(50,0)
  \DashLine(25,44)(37.5,66){3}
  \Photon(37.5,22)(25,44){3}{5}
  \Text(33,75)[lb]{$\fbf$}
  \Text(19,53)[lb]{$\hkg$}
  \Text(48,44)[lc]{$\ffp$}
  \Text(4,35)[lt]{$\wb(\hkg)$}
  \Text(33,13)[lt]{$\ff$}
  \Text(1,1)[lb]{}
\end{picture}
&+&
\begin{picture}(55,88)(0,41)
  \Photon(0,44)(25,44){3}{5}
  \Vertex(25,44){2.5}
  \ArrowLine(50,88)(25,44)
  \ArrowLine(25,44)(50,0)
  \Text(33,75)[lb]{$\fbf$}
  \Text(33,13)[lt]{$\ff$}
  \Text(1,1)[lb]{}
\SetScale{2.0}
  \Line(17.5,17.5)(7.5,27.5)
  \Line(7.5,17.5)(17.5,27.5)
\end{picture}
\eaa
\]
\vspace{9mm}
\caption
[$\wb$ cluster.]
{$\wb$ cluster: the first row shows the abelian diagrams of the cluster,
the last row the non-abelian diagrams; the second row shows diagrams that contribute
to both counterterm crosses (last diagrams in first and third rows).
\label{fig:W_cluster} }
\end{figure}
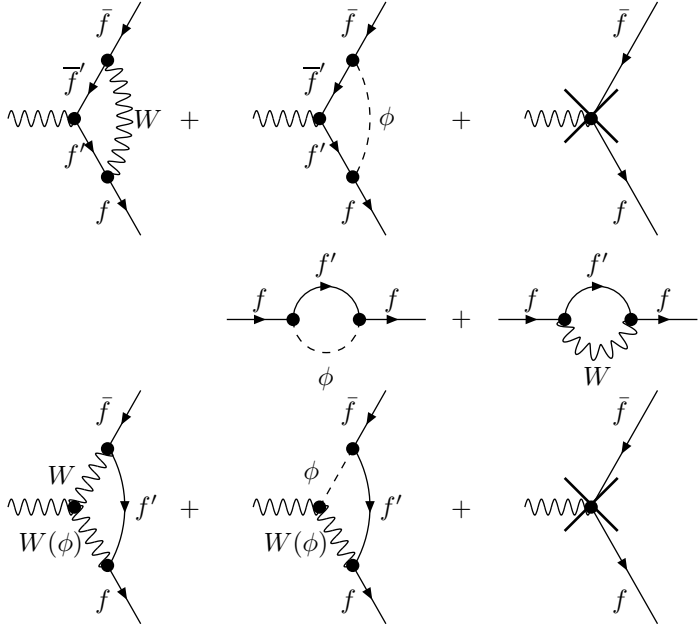

Finally, the $\wb$ cluster is made of the diagrams shown in 
Fig.~\ref{fig:W_cluster}.

In the formulae below, we present contributions to scalar form factors from all 
the diagrams of the $\wb$ cluster, not subdividing them into abelian and
non-abelian contributions. To some extent two sub-clusters are automatically 
marked by the type of arguments of $\scff{0}$ functions and typical coupling
constants. Separating poles, we have:
\begin{align}
\vvertil{\gamma{\sss W}}{\sss{L}}{\sman}  = &
2 \pole+ \cvertil{\gamma{\sss W}}{\sss{L}}{\sman},
\nll
  \vvertil{\gamma{\sss W}}{\sss{Q}}{\sman}= &
  \cvertil{\gamma{\sss W}}{\sss{Q}}{\sman},
\nll[1mm]
\vvertil{\gamma{\sss W}}{\sss{D}}{\sman}  = &
  \cvertil{\gamma{\sss W}}{\sss{D}}{\sman},
\nll
  \vvertil{z{\sss W}}{\sss{L}}{\sman}     = &
  2\ctws \pole
+ \cvertil{z{\sss W}}{\sss{L}}{\sman},
\nll
  \vvertil{z{\sss W}}{\sss{Q}}{\sman}     = &
  \cvertil{z{\sss W}}{\sss{Q}}{\sman},
\nll[1mm]
 \vvertil{z{\sss W}}{\sss{D}}{\sman}      = &  
 \cvertil{z{\sss W}}{\sss{D}}{\sman}.
\end{align}

\vspace*{-3mm}
Here the finite parts are:
\begin{align}
 {\cal F}^{\gamma W}_{\sss L} =& 
\frac{\qfp}{2\tcif}
\biggl\{ \Big[ 3+\lpar 1 + \rD \rpar^2+\frac{2}{\Rw} \Big]
              \mws \cff{-\mfs}{-\mfs}{-s}{\mfpl}{\mwl}{\mfpl} 
\nll [-1.5mm]&
      -\frac{1}{2} \lpar 6 + \rD \rpar 
            \Bigl[ \fbff{0}{-s}{\mfpl}{\mfpl}-\fbff{0}{-\mfs}{\mfpl}{\mwl} \Bigr]
\nll &
      +\frac{1}{2} \lpar 2-\rD\rpar \Bigl[\fbff{d\sss{W}}{-\mfs}{\mfpl}{\mwl}-1 \Bigr]
\nll &
      +\Big[ 4-\lpar 2+\rD \rpar \lpar 3 + 3\rtw + \rbw \rpar \Big]
       \frac{ \mws}{\sdtit} \Longab{\mfl}{\mfpl}{\mwl}
\biggr\}
\nll &
 -\rbw \lpar 2-\rD  \rpar \mws \cff{-\mfs}{-\mfs}{-s}{\mwl}{\mfpl}{\mwl}
\nll &
 +\frac{1}{2} \lpar  6-\rD \rpar 
     \Bigl[ \fbff{0}{-s}{\mwl}{\mwl}-\fbff{0}{-\mfs}{\mfpl}{\mwl} \Bigr]
\nll [-2mm]&
 + 2 \fbff{0}{-\mfs}{\mfpl}{\mwl}
 + \frac{1}{2} \lpar 2-3\rD \rpar \Bigl[ \fbff{d\sss{W}}{-\mfs}{\mfpl}{\mwl} + 1 \Bigr]
\nll [-1.5mm]&
 -\bigg(4-\Big[ 2 + \rtw \lpar 13+\rtw\rpar - \rbw \lpar 1+\rbw \rpar \Big]
      \frac{\mws}{\sdtit} \bigg) 
\nll[-1mm]&\hspace*{70mm}\times\Longna{\mfl}{\mfpl}{\mwl} 
\nll[-1mm]&
 - \frac{\qf}{4\tcif}
   \Bigl(\Delta_{{\cal F}_{\rm Im}}-3\Bigr) i\, {\rm Im} \fbff{0}{-\mfs}{\mwl}{\mfpl},
\label{gammaWL}
\end{align}
where
$
\ds{\Delta_{{\cal F}_{\rm Im}}}=
\ds{
\lpar 1-\rbw \rpar \frac{\lpar 2+\rbw\rpar}{\rtw}+\rtw}\,.
$

The last term in \eqn{gammaWL} is due to a non-cancellation of the imaginary part of the
function $\fbff{0}{-\mfs}{\mwl}{\mfpl}$ which appear in
real counterterms and complex-valued vertices.
\begin{align}
{\cal F}^{\gamma W}_{\sss Q} =&
   \frac{\rtw}{4\qf} \biggl\{
   \qfp \biggl[-4 \mws \cff{-\mfs}{-\mfs}{-s}{\mfpl}{\mwl}{\mfpl}
\nll &
   +\fbff{0}{-s}{\mfpl}{\mfpl}-\fbff{0}{-\mfs}{\mfpl}{\mwl}-1
\nll &
   -\frac{2+\rbw}{\rtw}  \fbff{d\sss{W}}{-\mfs}{\mfpl}{\mwl}
\nll &
   -\biggl( \frac{\lpar 1-\rbw\rpar \lpar 2+\rbw\rpar}{\rtw}
   -1 - \rtw + 2\rbw \biggr) 
\nll[-1mm]&\hspace{58mm}\times\mws \bff{0p}{-\mfs}{\mfpl}{\wml}
\nll &
   +2 \lpar 3+\rD \rpar\frac{\mws}{\sdtit}\Longab{\mfl}{\mfpl}{\mwl}\biggr]
\nll &
   -2 \tcif \bigg[ 2 \mfps \cff{-\mfs}{-\mfs}{-s}{\wml}{\mfpl}{\wml}
\nll &
   -\fbff{0}{-s}{\wml}{\wml}+\fbff{0}{-\mfs}{\mfpl}{\mwl}-1
\nll &
   +\frac{2+\rbw}{\rtw} \fbff{d\sss{W}}{-\mfs}{\mfpl}{\mwl}
\nll &
   +\biggl( \frac{ 2-\rbw \lpar 1+\rbw \rpar}{\rtw}
     -1-\rtw+2\rbw \biggr)
\nll &
\times \mws \bff{0p}{-\mfs}{\mfpl}{\wml}
   +2 \big( 7+\rP \big) \frac{ \mws}{\sdtit} \Longna{\mfl}{\mfpl}{\mwl}
 \bigg] \biggr\}
\nll &
 + \frac{1}{4} \Delta_{{\cal F}_{\rm Im}} i\,{\rm Im} \fbff{0}{-\mfs}{\mwl}{\mfpl}\,,
\\ 
{\cal F}^{\gamma{\sss W}}_{\sss D} =& 
-\frac{1}{\sdtit } \bigg\{
   \frac{\qfp}{2\tcif} \biggl( -2 \lrbr 8+ {\rD}^2-6 \rbw+\frac{2}{\Rw} \rrbr 
\nll  &
\times \mws \cff{-\mfs}{-\mfs}{-s}{\mfpl}{\mwl}{\mfpl}
\nll  &
 + \lpar 10+\rtw-3 \rbw \rpar 
         \Bigl[\fbff{0}{-s}{\mfpl}{\mfpl}-\fbff{0}{-\mfs}{\mfpl}{\mwl} \Bigr]
\nll &
 + \lpar 2+\rP \rpar
       \Bigl[ \fbff{d\sss{W}}{-\mfs}{\mfpl}{\mwl} + 1 \Bigr]
\nll &
 + 6 \Bigl[ \lpar 1+\rtw \rpar  \lpar 2+\rtw\rpar-\rbw \lpar 1+\rbw \rpar \Bigr]
             \frac{\mws}{\sdtit}  \Longab{\mfl}{\mfpl}{\mwl} \biggr)
\nll[-1mm] &
 + 2 \lrbr \lpar 1-\rP \rpar \lpar 2-\rtw \rpar-\rbw 
                    \lpar 1-4 \rbw-\frac{1}{\Rw} \rpar \rrbr
\nll  &
   \times \mws \cff{-\mfs}{-\mfs}{-s}{\mwl}{\mfpl}{\mwl}
\nonumber
\end{align}
\begin{align}
 & + \lpar 6-\rtw-5 \rbw \rpar \Bigl[ \fbff{0}{-s}{\mwl}{\mwl}
                                  - \fbff{0}{-\mfs}{\mfpl}{\mwl} \Bigr]
\nll &
 + \lpar 2+\rP \rpar \Bigl[ \fbff{d\sss{W}}{-\mfs}{\mfpl}{\mwl} - 1 \Bigr]
\nll &
 - 6 \bigg[ \lpar 1-\rtw \rpar \lpar 2+  \rtw      \rpar-\rbw 
                              \lpar 1+2 \rtw+\rbw \rpar \bigg]
\nll &
\times  \frac{\mws}{\sdtit} \Longna{\mfl}{\mfpl}{\mwl} \bigg\},
\\
{\cal F}^{z{\sss W}}_{\sss L} =& \frac{\vpa{\ffp}{}}{4 \tcif} 
\biggl\{
  \lrbr 3 + \lpar 1 + \rD \rpar^2 + \frac{2}{\Rw} \rrbr 
      \mws \cff{-\mfs}{-\mfs}{-s}{\mfpl}{\mwl}{\mfpl}
\nll &
  - \frac{1}{2}\lpar 6+\rD \rpar
     \Bigl[\fbff{0}{-s}{\mfpl}{\mfpl}-\fbff{0}{-\mfs}{\mfpl}{\mwl} \Bigr]
\nll &
  + \frac{1}{2}\lpar 2-\rD \rpar 
    \Bigl[ \fbff{d\sss{W}}{-\mfs}{\mfpl}{\mwl} - 1 \Bigr]
\nll &
  + \bigg[ 4-\lpar 2+\rD\rpar \lpar 3+3\rtw+\rbw \rpar \bigg]
      \frac{\mws}{\sdtit} \Longab{\mfl}{\mfpl}{\mwl} 
\biggr\}
\nll[1mm]&
  + \rbw\mws \cff{-\mfs}{-\mfs}{-s}{\mfpl}{\mwl}{\mfpl} 
\nll[2mm]&
  + \frac{1}{4} \biggl\{ \rD\fbff{0}{-s}{\mwl}{\wml}
               -\rtw\Bigl[\fbff{0}{-\mfs}{\mfpl}{\mwl} - 1\Bigr]
\nll &
               +\rbw\Bigl[\fbff{0}{-s}{\mfpl}{\mfpl} - 2 \Bigr]
 - \lpar 2+\rbw\rpar \fbff{d\sss{W}}{-\mfs}{\mfpl}{\mwl}
\nll &
 - \bigg[ \lpar 1-\rtw \rpar \lpar 2+\rtw \rpar
 - \rbw \lpar 1-2 \rtw+\rbw \rpar \bigg] 
\nll &
\times \mws \bff{0p}{-\mfs}{\mfpl}{\wml}
\nll &
  -2\rbw \lpar 1+\rD \rpar \frac{\mws}{\sdtit} \Longab{\mfl}{\mfpl}{\mwl} 
\biggr\}
\nll &
  - \ctws \biggl\{ \bigg( 4 \lpar 1-\rtw \rpar 
  + \frac{1}{2} \rbw \lrbr 4+\frac{\stws-\ctws}{\ctws} 
      \lpar 4+\rD \rpar \rrbr\bigg)
\nll [2mm] &
  \times \mws \cff{-\mfs}{-\mfs}{-s}{\mwl}{\mfpl}{\mwl}
\nll [2mm] &
 + \frac{1}{2} \lpar 2+\rD \rpar
   \Bigl[  \fbff{0}{-\mfs}{\mwl}{\mwl}-\fbff{0}{-\mfs}{\mfpl}{\mwl} \Bigr]
\nll &
 - \frac{1}{2} \lpar 2-\rD \rpar
   \Bigl[ \fbff{d\sss{W}}{-\mfs}{\mfpl}{\mwl} + 1 \Bigr]
 - 2 \fbff{0}{-\mfs}{\mfpl}{\mwl}
\nll &
 - \frac{1}{2} \biggl( 4 + 12 \rtw-\frac{\stws-\ctws}{\ctws} 
                                \rtw \lpar 7+\rtw \rpar 
\nll &
 - \rbw \lrbr 4 + \frac{\stws-\ctws}{\ctws}\lpar 1 - \rbw \rpar 
\rrbr \biggr)
    \frac{\mws}{\sdtit}  \Longna{\mfl}{\mfpl}{\mwl}\biggr\}
\nonumber
\end{align}
\begin{align}
\nll &
 + \frac{1}{8\tcif}
\bigg(3 \vpau-\vmau\Delta_{{\cal F}_{\rm Im}}  \bigg) i\, {\rm Im} \fbff{0}{-\mfs}{\mwl}{\mfpl}\,,
\\
{\cal F}^{z{\sss W}}_{\sss Q} =& - \frac{\rtw}{4}\frac{\vpa{\ffp}{}}{\vma{t}{}}
\bigg\{4 \mws 
\cff{-\mfs}{-\mfs}{-s}{\mfpl}{\mwl}{\mfpl}
\nll &
  -\fbff{0}{-s}{\mfpl}{\mfpl}+\fbff{0}{-\mfs}{\mfpl}{\mwl}+1
\nll &
  -2\lpar 3+\rD \rpar \frac{\mws}{\sdtit} \Longab{\mfl}{\mfpl}{\mwl}
\bigg\}
\nll &
           -\frac{1}{4}\lpar 2+\rbw\rpar\fbff{d\sss{W}}{-\mfs}{\mfpl}{\mwl}
\nll &
           -\frac{1}{4}
           \bigg[ \lpar 1-\rtw \rpar\lpar 2+\rD\rpar+\rbw \rD\bigg] 
      \mws \bff{0p}{-\mfs}{\mfpl}{\wml}
\nll &
      +\tcif\ctws \frac{\rtw}{\vma{t}{}} 
        \biggl\{\frac{\stws-\ctws}{\ctws} 
    \bigg(\rbw\mws \cff{-\mfs}{-\mfs}{-s}{\mwl}{\mfpl}{\mwl}
\nll &
   -\frac{1}{2} \Bigl[\fbff{0}{-s}{\mwl}{\mwl} - \fbff{0}{-\mfs}{\mfpl}{\mwl} + 1 
\Bigr] \bigg)
\nll &
 -\bigg[8-\frac{\stws-\ctws}{\ctws} \lpar 3+\rtw+\rbw \rpar \bigg] \frac{\mws}{\sdtit} 
 \Longna{\mfl}{\mfpl}{\mwl} \biggr\}
\nll &
 + \frac{1}{4}\Delta_{{\cal F}_{\rm Im}}i\,{\rm Im} \fbff{0}{-\mfs}{\mwl}{\mfpl},
\\[1mm]
{\cal F}^{z{\sss W}}_{\sss D} =& \frac{1}{\sdtit}\biggr\{
  \frac{\vpa{\ffp}{}}{2\tcif}  
\biggr( \lrbr  8 + \lpar{\rD}\rpar^2 - 6 \rbw + \frac{2}{\Rw} \rrbr
\nll &
\times \mws \cff{-\mfs}{-\mfs}{-s}{\mfpl}{\mwl}{\mfpl}
\nll &
   -\frac{1}{2}\lpar 10+\rtw-3 \rbw \rpar 
      \Bigl[\fbff{0}{-s}{\mfpl}{\mfpl}-\fbff{0}{-\mfs}{\mfpl}{\mwl}\Bigr]
\nll & 
   -\frac{1}{2}\lpar 2+\rP \rpar
               \Bigl[\fbff{d\sss{W}}{-\mfs}{\mfpl}{\mwl} + 1 \Bigr]
\nll &
   -3 \bigg[ \lpar 1+\rtw \rpar \lpar 2+\rtw \rpar
                               -\rbw \lpar 1+\rbw \rpar \bigg]
                   \frac{\mws}{\sdtit}  \Longab{\mfl}{\mfpl}{\mwl}
                           \biggr)
\nll &
  + \rbw \biggl(\mws \cff{-\mfs}{-\mfs}{-s}{\mfpl}{\mwl}{\mfpl}
\nll &
  +\frac{1}{2}\Bigl[\fbff{0}{-s}{\mfpl}{\mfpl}-\fbff{0}{-\mfs}{\mfpl}{\mwl}-1\Bigr]
\nll & 
 -\frac{1}{2}\fbff{d\sss{W}}{-\mfs}{\mfpl}{\mwl}
  -3 \lpar 1+\rD \rpar \frac{\mws}{\sdtit} \Longab{\mfl}{\mfpl}{\mwl}
                         \biggr)
\nll &
  -\ctws\biggl(
\bigg[ 4\rbw-\frac{\stws-\ctws}{\ctws}
             \biggl( \lpar 1-\rtw\rpar \lpar 2-\rtw \rpar
\nonumber
\end{align}
\begin{align}
&
   -\rbw  \Big( 5-\rtw-4\rbw-\frac{1}{\Rz} \Big) \biggr)
  \bigg]
\nll [2mm] &
\times  \mws \cff{-\mfs}{-\mfs}{-s}{\mwl}{\mfpl}{\mwl}
\nll [2mm] &
 +\frac{1}{2}\lrbr 4 - \frac{\stws-\ctws}{\ctws} \lpar 4 - \rtw
 - 5 \rbw \rpar  \rrbr
\nll &
 \times \Bigl[ \fbff{0}{-s}{\mwl}{\mwl}-\fbff{0}{-\mfs}{\mfpl}{\mwl} \Bigr]
\nll &
 +\frac{1}{2}\lpar 4 - \frac{\stws-\ctws}{\ctws} r_{tb}^+ \rpar
  \biggl[ \fbff{d\sss{W}}{-\mfs}{\mfpl}{\mwl} - 1 
\nll &
  -6 \lpar  1 - \rP \rpar  
     \frac{\mws}{\sdtit}  \Longna{\mfl}{\mfpl}{\mwl}\biggr] \biggl) \biggr\}.  
\end{align}
\vspace*{1mm}

Here we introduce more ratios, 
which were not given in ~\eqn{abbrev-old}:
\vspace*{-1mm}
\bqa 
 r^{\pm}=  \rtw\pm\rbw\,.
\label{moredefinitions}
\eqa
\vspace*{-5mm}

\noindent
Furthermore, we used one more `subtracted' function:
\bqa
\fbff{d\sss{W}}{-\mfs}{\mfpl}{\mwl}&=&\frac{1}{\rtw}
\biggl\{\lpar 1-\rbw\rpar\Bigl[\fbff{0}{-\mfs}{\mfpl}{\mwl}
\\ &&
+\Lmmw-1\Bigr]
-\rbw\Bigl[\Lmmb-\Lmmw\Bigr]\biggl\},
\nonumber
\eqa
and the three auxiliary functions:
\begin{align}
L_{ab}( M_1,M_2,M_3)  = &
\lpar M^2_3 + M^2_1 - M^2_2 \rpar 
        \cff{-\mfs}{-\mfs}{-\sman}{M_2}{M_3}{M_2}
\nll &
 -\fbff{0}{-\sman}{M_2}{M_2} + \fbff{0}{-\mfs }{M_2}{M_3},
\vspace*{-5mm}
\\
L_{na}(M_1,M_2,M_3)   = &
 \lpar M^2_3 - M^2_1 - M^2_2 \rpar \cff{-\mfs}{-\mfs}{-\sman}{M_3}{M_2}{M_3}
\nll &
 +\fbff{0}{-\sman}{M_3}{M_3} - \fbff{0}{-\mfs}{M_3}{M_2},
\vspace*{-5mm}
\\
\LongHi{M_1}{M_2}{M_3} =&
     \lrbr \frac{1}{2} \lpar M_2^2+M_3^2 \rpar
                           -2 M_1^2  \rrbr
\\ &
 \times   \cff{-\mfs}{-\mfs}{-\sman}{M_2}{M_1}{M_3}
        +           \fbff{0}{-\sman}{M_3}{M_2}
\nll&
        -\frac{1}{2}\fbff{0}{-\mfs}{M_1}{M_2} 
        -\frac{1}{2}\fbff{0}{-\mfs}{M_3}{M_1}.
\nonumber
\end{align}
Four scalar form factors, ${\cal F}^{\gamma(z){\sss W}}_{\sss{Q,D}}$, as 
follows from calculations, are both gauge-invariant and finite,
thus enlarging the number of gauge-invariant {\em building blocks} to 15.
On the contrary, two form factors, ${\cal F}^{\gamma(z){\sss W}}_{\sss{L}}$,
are neither gauge-invariant nor finite. Gauge dependence on $\gpar$, 
as well as UV poles, of $L$ form factors cancel in the sum with the $\wb\wb$ box 
and the self-energy contributions. 
\subsection{Library of scalar form factors in the limit $\mfl=0$}
In the limit $\mfl=0$ only nine scalar form factors of vertex 
clusters give non-zero contributions:
$F^{\gamma {\sss A}}_{\sss Q},\; F^{z {\sss A}}_{\sss L,Q},\;
 F^{\gamma {\sss Z}}_{\sss L,Q},\; F^{z {\sss Z}}_{\sss L,Q}$ 
and $F^{\gamma,z {\sss W}}_{\sss L}$, which, in turn, may be expressed in terms of
three auxiliary functions $F^{\sss A}_{\rm lim},\;F^{\sss Z}_{\rm lim}$ and $F^{\sss W}_{\rm lim}$.
These nine form factors are:
\begin{align}
{F^{\gamma {\sss A}}_{\sss Q}}_{\hspace*{-2mm}\rm lim} =&
{F^{z {\sss A}}_{\sss L}}_{\hspace*{-2mm}\rm lim} =
{F^{z {\sss A}}_{\sss Q}}_{\hspace*{-2mm}\rm lim} = \qf^2\stws F^{\sss A}_{\rm lim}\,,
\nll[.75mm]
{F^{\gamma {\sss Z}}_{\sss L}}_{\hspace*{-2mm}\rm lim} =& 
  \frac{1}{\ctws} 2 \qu \vf \af F^{\sss Z}_{\rm lim}\,,
\nll[.75mm]
{F^{\gamma {\sss Z}}_{\sss Q}}_{\hspace*{-2mm} \rm lim}=& 
{F^{z{\sss Z}}_{\sss Q}}_{\hspace*{-2mm} \rm lim} = 
  \frac{1}{ 4\ctws} \delta_f^2 F^{\sss Z}_{\rm lim}\,,
\nll[.75mm]
{F^{z{\sss Z}}_{\sss L}}_{\hspace*{-2mm} \rm lim} =& 
\frac{1}{4 \tciu \ctws} \lpar 3 \vf^2+\af^2 \rpar \af F^{\sss Z}_{\rm lim}\,,
\nll[.75mm]
{F^{\gamma {\sss W}}_{\sss L}}_{\hspace*{-2mm} \rm lim} =& \frac{\qfp}{2\tciu} 
F^{\sss W}_{\rm lim}
          - \biggl[ \rbw (2+\rbw)+4\lrbw + \lrbw^2 (2+\rbw) \Rw \biggr]
\nll &
        \times   \wms \coswdw
\nll &
          - \biggl[1 - \frac{1}{2}\rbw + \lrbw \lpar 2+\rbw \rpar \Rw \biggr] 
\nll &
        \times \Big[ \bofsww+\Lndm-1 \Big] 
\nll &
          -\biggl[ \frac{3}{2 \lrbw^2}+\frac{1}{2\lrbw}+1+\lpar 2+\rbw \rpar \Rw \biggr]
                                                               \Big[ \Lndm - \Lnwm \Big]
\nll &
          - 2\Lndm+\frac{3}{2\lrbw}+2+\frac{\rbw}{4}\,,
\nll[.75mm]
{F^{z{\sss W}}_{\sss L}}_{\hspace*{-2mm}\rm lim} =& \frac{\vpa{\ffp}{}}{4\tciu}F^{\sss W}_{\rm lim}
\nll &
   + \frac{\rbw}{2}\biggl\{ \big( 2 + \lrbw^2 \Rw \big)\wms \cusdwd
\nll &
          -\frac{1}{2}  \Big[ \bofsww - \bofsdd + 2 \Big]
\nll &
          - \Rw         \Big[ \bofsdd + \Lnwm - 1 
\nll &
          - \rbw \big( \bofsdd + \Lndm - 1 \big) \Big]
                       \biggr\}
\nll &
        - \ctws\biggl\{            
         \bigg[4 + \frac{\rbw}{2} \big[4+\lpar 4-\rbw\rpar t_{\sss W} \big]
            +\biggl( 2-\frac{\rbw}{2} \big( 4+\lrbw  t_{\sss W} \big) \biggr)
\nll &
    \times   \lrbw\Rw \bigg] 
\wms \cff{-\mfs}{-\mfs}{-s}{\wml}{\dml}{\wml}
\nn
\end{align}
\begin{align}
&
          +\frac{1}{2} \biggl( 2-\rbw
             +\big[4-\rbw \big(4+\lrbw t_{\sss W} \big) \big] \Rw \biggr)
\nll &
       \times  \Big[ \bofsww+\Lnwm-1 \Big]
\nll &     
-\biggl(\frac{3}{2 \lrbw^2} + \frac{1}{2\lrbw} - 2
        +\frac{1}{2}\rbw \big[ 1 + \big( 4-\rbw t_{\sss W} \big) \Rw \big] \biggr)
\nll &
      \times   \Big[ \Lndm-\Lnwm \Big]
\nll &
          +2\Lnwm-\frac{3}{2\lrbw}-2-\frac{1}{4} \rbw
\bigg\},
\end{align}
while the three auxiliary functions read:
\begin{align}
F^{\sss A}_{\rm lim} =& 2\sman\cff{-\mfs}{-\mfs}{-\sman}{\mfl}{\lambda}{\mfl}
\nll &
-3\fbff{0}{-s}{0}{0}-3\Lmmt+2 L_{\lambda}(\mfs)+2\,,
\nll[1mm]
F^{\sss Z}_{\rm lim} =& \frac{2 \lpar 1+\Rz \rpar^2}{\Rz} \mzs \cosozo
\nll &
          -\lpar 3+2 \Rz \rpar \Big[ \bofsoo + \Lnzm - 1 \Big]-\frac{1}{2},
\nll[1mm]
F^{\sss W}_{\rm lim} =& 
     \biggl( \frac{2}{\Rw}+3+\lrbw^2 \big[ 1+\lpar 2+\rbw \rpar \Rw \big] \biggr)
\nll &
      \times  \wms \cff{0}{0}{-s}{\dml}{\wml}{\dml} 
\nll &
          -\frac{1}{2} \Big[ 6-\rbw+2 \lrbw \lpar 2+\rbw \rpar \Rw\Big]  
\nll &                   
                   \times  \Big[ \bofsdd + \Lndm - 1 \Big]
\nll &
          +\frac{1}{2} \biggl[ \frac{3}{\lrbw^2}+\frac{1}{\lrbw}+2+2 \lpar 2+\rbw \rpar \Rw\biggr] 
 \Big[ \Lndm - \Lnwm \Big]
\nll &
          +\frac{3}{ 2 \lrbw}-2-\frac{3}{4}\rbw\,,
\end{align}
where
\bqa
L_{\lambda}(\mfs)&=&\ln\frac{\mfs}{\lambda^2}\,,
\nll[1mm]
\lrbw&=&1-\rbw\,,
\eqa
with $\lambda$ being the photon mass.

The formulae, presented in this subsection, are useful to compute the one-loop corrections 
for the channels with massless or light final state fermions $\ff$ and heavy virtual fermions 
$\ffp$.
\clearpage
\subsection{Library of scalar form factors for electron vertex}
Besides $B\ft\ft$ clusters, we need also $Bee$ clusters, which can, in principle,
be taken from \cite{Bardin:1999yd} or derived from the formulae of previous subsection 
in the limit $\mfpl\to0$ and $\mfl=\mel$. Here we simply list the results:
\bqa
\cvetril{\gamma ee}{\sss{L}}{\sman} &=& 
 -\frac{2}{\cows} \qe v_e a_e \cvertil{{\sss Z},e}{}{\sman}  
 +\cvertil{{\sss{Wna}},e}{}{\sman},
\nll
\cvetril{\gamma ee}{\sss{Q}}{\sman} &=& {\cal F}^{{\sss A},e} \lpar\sman\rpar
 +\frac{1}{4 \cows} \delta_e^2 \cvertil{{\sss{Z}},e}{}{\sman},
\nll
\cvetril{zee}{\sss{L}}{\sman}  &=& {\cal F}^{{\sss A},e} \lpar\sman\rpar
 -\frac{1}{2 \cows} \lpar 3 v_e^2 + a_e^2 \rpar a_e
  \cvertil{{\sss Z},e}{}{\sman}
 +\cvertil{{\sss W},e}{}{\sman},
\nll
\cvetril{zee}{\sss{Q}}{\sman}  &=& {\cal F}^{{\sss A},e} \lpar\sman\rpar
 +\frac{1}{4 \cows} \delta_e^2
    \cvertil{{\sss Z},e}{}{\sman},
\label{Beeffs}
\eqa
with
\begin{equation}
  \cvertil{{\sss{W}},e}{}{\sman}=
 -\cvertil{{\sss{Wab}},e}{}{\sman}
+ \cows
  \cvertil{{\sss{Wna}},e}{}{\sman}.
\end{equation}
In \eqn{Beeffs} we use four more auxiliary functions:
\begin{align}
{\cal F}^{{\sss A},e}\lpar\sman\rpar =& Q_e^2 \stws \Big[ 
2\sman\cff{-\mes}{-\mes}{-\sman}{\mel}{\lambda}{\mel}
\nll &
-3\fbff{0}{-s}{0}{0}-3\Lmme+2 L_{\lambda}(\mes)+2 \Big],
\\[.5mm]
{\cal F}^{{\sss Z},e}\lpar\sman\rpar
=&
 2 \frac{\lpar 1+\Rz \rpar^2}{\Rz} \mzs \cff{0}{0}{-s}{0}{\mzl}{0}
\nll &
 -3 \Bigl[ \fbff{0}{-s}{0}{0} + \Lmmz \Bigr]
\nll &
+\frac{5}{2} 
  -2 \Rz \Bigl[ \fbff{0}{-s}{0}{0} + \Lmmz - 1 \Bigr],
\\[.5mm]
{\cal F}^{{\sss{Wna}},e}\lpar\sman\rpar
=&
 -2 \Rw 
\Bigl[\mws \cff{0}{0}{-s}{\mwl}{0}{\mwl}
\nll &
         +\fbff{0}{-s}{\mwl}{\mwl}+\Lmmw-1 \Bigr]  
\nll &
  -4 \mws \cff{0}{0}{-s}{\mwl}{0}{\mwl}
\nll &
      -\fbff{0}{-s}{\mwl}{\mwl}-3\Lmmw+\frac{9}{2}\,,\qquad
\\[.5mm]
{\cal F}^{{\sss{Wab}},e}\lpar\sman\rpar
=&  \sigma_\nu
 \biggl\{   
 \frac{\lpar 1 + \Rw \rpar^2}{\Rw} \mws \cff{0}{0}{-s}{0}{\mwl}{0}
\nll &
      -\frac{3}{2} \Bigl[ \fbff{0}{-s}{0}{0} + \Lmmw \Bigr]
\nll &
+\frac{5}{4}
      -\Rw \Bigl[ \fbff{0}{-s}{0}{0} + \Lmmw - 1 \Bigr] \biggr\},
\end{align}
where $L_{\lambda}(\mes)=\ln(\mes/\lambda^2)$.

\subsection{Amplitudes of boxes}
The contributions of QED $AA$ and $ZA$ boxes form gauge-invariant and UV-finite subsets.
In terms of six structures  $(L,R) \otimes (L,R,D)$ they read:
\begin{align}
\bigg({\cal B}^{\sss  LK} \bigg)^{d+c}\hspace*{-5.5mm}  = 
\label{anyboxamplitude}
k^{\sss LK}\frac{g^4}{s}
&\biggl[ \;\;\,
    \lrbr  \gdmu \gdp \otimes  \gdmu \gdp \rrbr
{\cal F}^{\sss  LK}_{\sss LL} \lpar s,t,u \rpar
\nll &
  + \lrbr  \gdmu \gdp \otimes  \gdmu \gdm  \rrbr
{\cal F}^{\sss  LK}_{\sss LR}\lpar s,t,u \rpar 
\nll[3mm] & 
  + \lrbr  \gdmu \gdm \otimes  \gdmu \gdp \rrbr 
{\cal F}^{\sss  LK}_{\sss RL}\lpar s,t,u \rpar
\nll[3mm] &
  + \lrbr  \gdmu \gdm   \otimes  \gdmu \gdm \rrbr
{\cal F}^{\sss  LK}_{\sss RR} \lpar s,t,u \rpar
\\[3mm] & 
  + \lrbr \gdmu \gdp  \otimes \lpar -i \mfl I D_\mu \rpar \rrbr
{\cal F}^{\sss  LK}_{\sss LD}\lpar s,t,u \rpar
\nll &
  + \lrbr \gdmu \gdm  \otimes \lpar -i \mfl I D_\mu \rpar \rrbr 
{\cal F}^{\sss  LK}_{\sss RD}\lpar s,t,u \rpar
\biggr],
\nonumber
\end{align}
where $ LK = AA, ZA, ZZ $ and for shortening of presentation we factorize out normalization factors: 
\begin{equation}
k^{\sss AA} = \stwf \qes\qfs\,,
\qquad
k^{\sss ZA} = \frac{\stws\qe\qf}{\ctws}\,,
\qquad
k^{\sss ZZ} = \frac{1}{32\ctwf}\,.
\end{equation}
\subsubsection{$AA$ box contribution}
 There are only two $AA$ diagrams, {\em direct} and {\em crossed}:

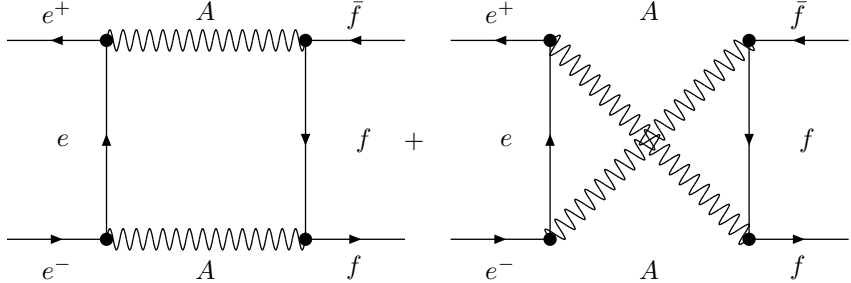
\begin{figure}[!h]
\vspace*{-20mm}
\[
\hspace*{-2cm}
{
\begin{array}{ccc}
\begin{picture}(100,105)(0,49.5)
  \ArrowLine(0,15)(37.5,15)
  \ArrowLine(37.5,15)(37.5,90)
  \ArrowLine(37.5,90)(0,90)
  \Photon(37.5,15)(112.5,15){4}{15}
  \Photon(37.5,90)(112.5,90){4}{15}
  \Vertex(37.5,15){2.5}
  \Vertex(37.5,90){2.5}
  \Vertex(112.5,15){2.5}
  \Vertex(112.5,90){2.5}
  \ArrowLine(150,90)(112.5,90)
  \ArrowLine(112.5,90)(112.5,15)
  \ArrowLine(112.5,15)(150,15)
\Text(18.75,105)[tc]{$e^+$}
  \Text(75,105)[tc]{$A$}
  \Text(131.25,105)[tc]{$\bar f$}
 \Text(18.75,52.5)[lc]{$e$}
  \Text(131.8,52.5)[lc]{$\ff$}
  \Text(18.75,0)[cb]{$e^-$}
  \Text(75,0)[bc]{$A$}
  \Text(131.25,0)[cb]{$\ff$}
\end{picture}
\qquad \qquad 
&+&
\begin{picture}(100,105)(0,49.5)
  \ArrowLine(0,15)(37.5,15)
  \ArrowLine(37.5,15)(37.5,90)
  \ArrowLine(37.5,90)(0,90)
  \Photon(37.5,15)(112.5,90){4}{20}
  \Photon(37.5,90)(112.5,15){4}{20}
  \Vertex(37.5,15){2.5}
  \Vertex(37.5,90){2.5}
  \Vertex(112.5,15){2.5}
  \Vertex(112.5,90){2.5}
  \ArrowLine(150,90)(112.5,90)
  \ArrowLine(112.5,90)(112.5,15)
  \ArrowLine(112.5,15)(150,15)
\Text(18.75,105)[tc]{$e^+$}
  \Text(75,105)[tc]{$A$}
  \Text(131.25,105)[tc]{$\bar f$}
  \Text(18.75,52.5)[lc]{$e$}
  \Text(131.8,52.5)[lc]{$\ff$}
  \Text(18.75,0)[cb]{$e^-$}
  \Text(75,0)[bc]{$A$}
  \Text(131.25,0)[cb]{$\ff$}
\end{picture}
\end{array}
}
\]
\vspace{10mm}
\caption{Direct and crossed $AA$ boxes. \label{AA_box_dc} }
\end{figure}
\vspace{2mm}
The six form factors of $AA$ boxes might be expressed in terms of
four auxiliary functions ${\cal F}_1$ and ${\cal H}_{1,2,3}$:
\begin{align}
 \Faa{LL}\lpar s,t,u \rpar =&  \Faa{RR}\lpar s,t,u \rpar
                   = {\cal H}_1 \lpar s,t \rpar - {\cal H}_1 \lpar s,u \rpar 
            + {\cal H}_2 \lpar s,t \rpar + {\cal H}_3 \lpar s,u \rpar,
\nll 
 \Faa{LR}\lpar s,t,u \rpar =& \Faa{RL}\lpar s,t,u \rpar
                   = {\cal H}_1 \lpar s,t \rpar - {\cal H}_1 \lpar s,u \rpar 
            - {\cal H}_2 \lpar s,u \rpar - {\cal H}_3 \lpar s,t \rpar,
\nll
 \Faa{LD}\lpar s,t,u \rpar =& \Faa{LD}\lpar s,t,u \rpar
                   = {\cal F}_1 \lpar s,t \rpar - {\cal F}_1\lpar s,u \rpar.
\end{align}
The auxiliary functions are rather short:
\begin{align}
{\cal F}_1 \lpar s,t \rpar =& -\frac{1}{2} \frac{s }{\sdfit} 
                \biggl\{\frac{1}{\sdfit} \Big[ - \tmi^3  \jaat           
\nll &
               +t s^2  \cesoeo  \Big]
\nll &
 + t \biggl[\frac{4 \mfs}{\sdtit}+\frac{s\lpar s - 2  \mfs \rpar}{\sdfit} \biggr]\ctsoto 
\nll &
       - 2\frac{t}{\sdtit}\Bigl[ \bofsoo - \boftto \Bigr]
\nll &
       + 2  \frac{t}{\tmi}\Bigl[ \boftet - \boftto \Bigr]  \biggr\}\,,
\\[4mm]
{\cal H}_1 \lpar s,t \rpar =&  
- \frac{\tmi}{4}\biggl[2 - \lpar t+\frac{\tpl\tmi^2}{\sdfit}\rpar\frac{s}{\sdfit}\biggr]\jaat
\nll &
+\tmi \cateot
\\ &
      +\frac{s \mfs}{2 \sdfit}\lpar 1 -2  \frac{t}{\tmi}\rpar \Bigl[ \boftet-\boftto \Bigr],
\nll[4mm]
{\cal H}_2 \lpar s,t \rpar =& 
\frac{ s}{4\sdfit}  \biggl\{
         \lpar s + 2\tmi \rpar \lpar 1 - \tpl \frac{s}{\sdfit}\rpar  s \, \cesoeo
\nll & 
   - \biggl[  2  \mfs s - \lpar s - 4 \mfs \rpar
    \bigg(s + 2\tmi- \lpar  s\tpl + 2 t\tmi \rpar \frac{s}{\sdfit}\bigg) \biggr] 
\nll &\hspace{4cm}
\times \ctsoto
\nll &
       + 2  \tmi  \Bigl[ \bofsoo - \boftet \Bigr] 
\nll &
       - 4  \mfs  \Bigl[ \boftet - \boftto \Bigr]  \biggr\} \, ,
\\[4mm]
{\cal H}_3 \lpar s,t \rpar =& \frac{s}{4 \sdfit^2} \lpar s + 2 \tmi \rpar \tmi^3
\jaat\,.
\end{align}
\vspace*{2mm}
Here
\begin{equation}
\sdfit=-\tman\uman + \mfq\,,
\end{equation}
and $J_{\sss AA}\lpar Q^2,P^2;M_1,M_2 \rpar$ is due to a procedure of 
disentangling
of the infrared divergences from $\sdff{0}$.
Its explicit expression reads $(P^2 > 0, Q^2 < 0$, and $M_1$ is ignored 
everywhere but for the arguments of $\ln$):
\newpage
\begin{align} 
&\hspace{-2mm}
J_{\sss  AA}\lpar Q^2,P^2;M_1,M_2 \rpar \hspace*{-.5mm} =
\frac{1}{\Delta_{\sss P}}
         \Biggl\{
       \ln\frac{\Delta_{\sss P}^2}{-Q^2 P^2}\ln\lpar \frac{P^2}{-Q^2}\rpar
-\frac{1}{2}\ln^2\lpar\frac{M_1^2}{-Q^2}\rpar
\\ &
 - \frac{1}{2}\ln^2\lpar\frac{M_2^2}{-Q^2}\rpar
 + \ln^2\lpar 1+\frac{M^2_2}{P^2} \rpar
 -  2 \Litwo \lpar\frac{P^2}{\Delta_{\sss P}} \rpar 
 + i \pi \ln\lrbr \frac{\Delta_{\sss P}^2}{M_1^2 M_2^2}\rrbr
        \Biggr\}.
\label{JAA}
\nonumber
\end{align}
Furthermore, the relevant infrared divergent $\scff{0}$ function 
(again $P^2 > 0$), is
\begin{align} 
&\hspace{-5mm}
C_0^{\sss {\rm IR}}
\lpar -M_1^2, -M_2^2, P^2; M_1, \lambda, M_2 \rpar
 =
\frac{1}{2\Delta_{\sss P}}
         \Bigg\{
       \ln\lrbr\frac{\Delta_{\sss P}^2}{ M_1^2 M_2^2}\rrbr
             \ln \frac{ P^2}{\tHlas} 
\\ &
      - 2\Litwo\lpar\frac{P^2}{\Delta_{\sss P}}\rpar
      -\frac{1}{2} \ln^2\lpar \frac{M_1^2}{P^2}\rpar
      -\frac{1}{2} \ln^2\lpar \frac{M_2^2}{P^2}\rpar 
     + \ln^2\lpar 1+\frac{M_2^2}{P^2}\rpar
           \Bigg\},
\nonumber
\end{align} 
where $\Delta_{\sss P}=P^2+M^2$.
\subsubsection{$ZA$ box contribution}
In $\Rxi$ gauge there are eight $ZA$ boxes; however, since the electron mass is 
ignored,
only four diagrams without $\phi_0$ contribute, see \fig{ZA_box_dc}.

The six relevant scalar form factors are conveniently presentable in the form
of differences of $\ff$- and $u$- dependent functions:
\bqa
\Fza{IJ}\lpar s,t,u \rpar  &=& \Fza{IJ}(s,t)-\Fza{IJ}(s,u),
\label{FzaIJ}
\eqa
where the index ${IJ}$ is any pair of $L,R\otimes L,R,D$. The 12 $\Fza{IJ}$ functions depend on
6 auxiliary functions by means of equations where the coupling constants are factored out:
\begin{align}
\Fza{LL}\lpar s,t \rpar =&  \vpa{e}{} \vpa{\ff}{} {\cal G}_1\lpar s,t \rpar 
                + \vpa{e}{} \vma{\ff}{} {\cal G}_2\lpar s,t \rpar,
\nll 
\Fza{LL}\lpar s,u \rpar =&  \vpa{e}{} \vma{\ff}{} {\cal H}_1\lpar s,u \rpar 
                + \vpa{e}{} \vpa{\ff}{} {\cal H}_2\lpar s,u \rpar,
\nll 
\Fza{RR}\lpar s,t \rpar =&   \vma{e}{} \vma{\ff}{} {\cal G}_1\lpar s,t \rpar 
                + \vma{e}{} \vpa{\ff}{} {\cal G}_2\lpar s,t \rpar,
\nll 
\Fza{RR}\lpar s,u \rpar =&   \vma{e}{} \vpa{\ff}{} {\cal H}_1\lpar s,u \rpar 
                + \vma{e}{} \vma{\ff}{} {\cal H}_2\lpar s,u \rpar,
\nll 
\Fza{LR}\lpar s,t \rpar =&   \vpa{e}{} \vpa{\ff}{} {\cal H}_1\lpar s,t \rpar 
                + \vpa{e}{} \vma{\ff}{} {\cal H}_2\lpar s,t \rpar,
\nll 
\Fza{LR}\lpar s,u \rpar =&   \vpa{e}{} \vma{\ff}{} {\cal G}_1\lpar s,u \rpar 
                + \vpa{e}{} \vpa{\ff}{} {\cal G}_2\lpar s,u \rpar,
\nll 
\Fza{RL}\lpar s,t \rpar =&   \vma{e}{} \vma{\ff}{} {\cal H}_1\lpar s,t \rpar 
                + \vma{e}{} \vpa{\ff}{} {\cal H}_2\lpar s,t \rpar,
\nll 
\Fza{RL}\lpar s,u \rpar =&   \vma{e}{} \vpa{\ff}{} {\cal G}_1\lpar s,u \rpar 
                + \vma{e}{} \vma{\ff}{} {\cal G}_2\lpar s,u \rpar,
\nll
\Fza{LD}\lpar s,t \rpar =&   \vpa{e}{} \vpa{\ff}{} {\cal F}_1\lpar s,t \rpar 
                + \vpa{e}{} \vma{\ff}{} {\cal F}_2\lpar s,t \rpar,
\nll 
\Fza{LD}\lpar s,u \rpar =&   \vpa{e}{} \vma{\ff}{} {\cal F}_1\lpar s,u \rpar 
                + \vpa{e}{} \vpa{\ff}{} {\cal F}_2\lpar s,u \rpar,
\nll 
\Fza{RD}\lpar s,t \rpar =&   \vma{e}{} \vma{\ff}{} {\cal F}_1\lpar s,t \rpar 
                + \vma{e}{} \vpa{\ff}{} {\cal F}_2\lpar s,t \rpar,
\nll 
\Fza{RD}\lpar s,u \rpar =&   \vma{e}{} \vpa{\ff}{} {\cal F}_1\lpar s,u \rpar 
                + \vma{e}{} \vma{\ff}{} {\cal F}_2\lpar s,u \rpar.
\end{align}

\begin{figure}[!h]
\vspace{-13mm}
\[
\hspace*{-2cm}
{
\begin{array}{ccc}
\begin{picture}(100,105)(0,49.5)
  \ArrowLine(0,15)(37.5,15)
  \ArrowLine(37.5,15)(37.5,90)
  \ArrowLine(37.5,90)(0,90)
  \Photon(37.5,15)(112.5,15){4}{15}
  \Photon(37.5,90)(112.5,90){4}{10}
  \Vertex(37.5,15){2.5}
  \Vertex(37.5,90){2.5}
  \Vertex(112.5,15){2.5}
  \Vertex(112.5,90){2.5}
  \ArrowLine(150,90)(112.5,90)
  \ArrowLine(112.5,90)(112.5,15)
  \ArrowLine(112.5,15)(150,15)
\Text(18.75,105)[tc]{$e^+$}
  \Text(75,108.75)[tc]{$Z$}
  \Text(131.25,112.5)[tc]{$\bar f$}
  \Text(18.75,52.5)[lc]{$e$}
  \Text(131.8,52.5)[lc]{$\ff$}
  \Text(18.75,0)[cb]{$e^-$}
  \Text(75,-3.75)[bc]{$A$}
  \Text(131.25,-3)[cb]{$\ff$}
\end{picture}
\qquad \qquad \quad
&+& 
\hspace*{-0.5mm}
\begin{picture}(100,105)(0,49.5)
  \ArrowLine(0,15)(37.5,15)
  \ArrowLine(37.5,15)(37.5,90)
  \ArrowLine(37.5,90)(0,90)
  \Photon(37.5,15)(112.5,15){4}{10}
  \Photon(37.5,90)(112.5,90){4}{15}
  \Vertex(37.5,15){2.5}
  \Vertex(37.5,90){2.5}
  \Vertex(112.5,15){2.5}
  \Vertex(112.5,90){2.5}
  \ArrowLine(150,90)(112.5,90)
  \ArrowLine(112.5,90)(112.5,15)
  \ArrowLine(112.5,15)(150,15)
\Text(18.75,105)[tc]{$e^+$}
  \Text(75,108.75)[tc]{$A$}
  \Text(131.25,112.5)[tc]{$\bar f$}
  \Text(18.75,52.5)[lc]{$e$}
  \Text(131.8,52.5)[lc]{$\ff$}
  \Text(18.75,0)[cb]{$e^-$}
  \Text(75,-3.75)[bc]{$Z$}
  \Text(131.25,-3)[cb]{$\ff$}
\end{picture}
\\ \\ 
\begin{picture}(100,105)(0,49.5)
  \ArrowLine(0,15)(37.5,15)
  \ArrowLine(37.5,15)(37.5,90)
  \ArrowLine(37.5,90)(0,90)
  \Photon(37.5,15)(112.5,15){4}{15}
\Line(37.5,90)(42.5,90)
\Line(47.5,90)(52.5,90)
\Line(57.5,90)(62.5,90)
\Line(67.5,90)(72.5,90)
\Line(77.5,90)(82.5,90)
\Line(87.5,90)(92.5,90)
\Line(97.5,90)(102.5,90)
\Line(107.5,90)(112.5,90)
  \Vertex(37.5,15){2.5}
  \Vertex(37.5,90){2.5}
  \Vertex(112.5,15){2.5}
  \Vertex(112.5,90){2.5}
  \ArrowLine(150,90)(112.5,90)
  \ArrowLine(112.5,90)(112.5,15)
  \ArrowLine(112.5,15)(150,15)
\Text(18.75,105)[tc]{$e^+$}
  \Text(75,108.75)[tc]{$\phi^0$}
  \Text(131.25,112.5)[tc]{$\bar f$}
  \Text(18.75,52.5)[lc]{$e$}
  \Text(131.8,52.5)[lc]{$\ff$}
  \Text(18.75,0)[cb]{$e^-$}
  \Text(75,-3.75)[bc]{$A$}
  \Text(131.25,-3)[cb]{$\ff$}
\end{picture}
\qquad \qquad \quad
&+&
\hspace*{-0.5mm}
\begin{picture}(100,105)(0,49.5)
  \ArrowLine(0,15)(37.5,15)
  \ArrowLine(37.5,15)(37.5,90)
  \ArrowLine(37.5,90)(0,90)
\Line(37.5,15)(42.5,15)
\Line(47.5,15)(52.5,15)
\Line(57.5,15)(62.5,15)
\Line(67.5,15)(72.5,15)
\Line(77.5,15)(82.5,15)
\Line(87.5,15)(92.5,15)
\Line(97.5,15)(102.5,15)
\Line(107.5,15)(112.5,15)
  \Photon(37.5,90)(112.5,90){4}{15}
  \Vertex(37.5,15){2.5}
  \Vertex(37.5,90){2.5}
  \Vertex(112.5,15){2.5}
  \Vertex(112.5,90){2.5}
  \ArrowLine(150,90)(112.5,90)
  \ArrowLine(112.5,90)(112.5,15)
  \ArrowLine(112.5,15)(150,15)
  \Text(18.75,105)[tc]{$e^+$}
  \Text(75,108.75)[tc]{$A$}
  \Text(131.25,112.5)[tc]{$\bar f$}
  \Text(18.75,52.5)[lc]{$e$}
  \Text(131.8,52.5)[lc]{$\ff$}
  \Text(18.75,0)[cb]{$e^-$}
  \Text(75,-3.75)[bc]{$\phi^0$}
  \Text(131.25,-3)[cb]{$\ff$}
\end{picture}
\\  \\  
\begin{picture}(100,105)(0,49.5)
  \ArrowLine(0,15)(37.5,15)
  \ArrowLine(37.5,15)(37.5,90)
  \ArrowLine(37.5,90)(0,90)
  \Photon(37.5,15)(112.5,90){4}{20}
  \Photon(37.5,90)(112.5,15){4}{10}
  \Vertex(37.5,15){2.5}
  \Vertex(37.5,90){2.5}
  \Vertex(112.5,15){2.5}
  \Vertex(112.5,90){2.5}
  \ArrowLine(150,90)(112.5,90)
  \ArrowLine(112.5,90)(112.5,15)
  \ArrowLine(112.5,15)(150,15)
\Text(18.75,105)[tc]{$e^+$}
  \Text(75,108.75)[tc]{$Z$}
  \Text(131.25,112.5)[tc]{$\bar f$}
  \Text(18.75,52.5)[lc]{$e$}
  \Text(131.8,52.5)[lc]{$\ff$}
  \Text(18.75,0)[cb]{$e^-$}
  \Text(75,-3.75)[bc]{$A$}
  \Text(131.25,-3)[cb]{$\ff$}
\end{picture}
\qquad \qquad \quad
&+& 
\hspace*{-0.5mm}
\begin{picture}(100,105)(0,49.5)
  \ArrowLine(0,15)(37.5,15)
  \ArrowLine(37.5,15)(37.5,90)
  \ArrowLine(37.5,90)(0,90)
  \Photon(37.5,15)(112.5,90){4}{10}
  \Photon(37.5,90)(112.5,15){4}{20}
  \Vertex(37.5,15){2.5}
  \Vertex(37.5,90){2.5}
  \Vertex(112.5,15){2.5}
  \Vertex(112.5,90){2.5}
  \ArrowLine(150,90)(112.5,90)
  \ArrowLine(112.5,90)(112.5,15)
  \ArrowLine(112.5,15)(150,15)
\Text(18.75,105)[tc]{$e^+$}
  \Text(75,108.75)[tc]{$A$}
  \Text(131.25,112.5)[tc]{$\bar f$}
  \Text(18.75,52.5)[lc]{$e$}
  \Text(131.8,52.5)[lc]{$\ff$}
  \Text(18.75,0)[cb]{$e^-$}
  \Text(75,-3.75)[bc]{$Z$}
  \Text(131.25,-3)[cb]{$\ff$}
\end{picture}
\\  \\  
\begin{picture}(100,105)(0,49.5)
  \ArrowLine(0,15)(37.5,15)
  \ArrowLine(37.5,15)(37.5,90)
  \ArrowLine(37.5,90)(0,90)
  \Photon(37.5,15)(112.5,90){4}{20}
 \Line(37.5,90)(41.5,88)
 \Line(45.5,84)(49.5,80)
 \Line(53.5,76)(57.5,72)
 \Line(61.5,68)(65.5,64)
 \Line(69.5,60)(73.5,56)
 \Line(77.5,52)(81.5,48)
 \Line(85.5,44)(89.5,40)
 \Line(93.5,36)(97.5,32)
 \Line(101.5,28)(105.5,24)
 \Line(109.5,19)(112.5,15)
  \Vertex(37.5,15){2.5}
  \Vertex(37.5,90){2.5}
  \Vertex(112.5,15){2.5}
  \Vertex(112.5,90){2.5}
  \ArrowLine(150,90)(112.5,90)
  \ArrowLine(112.5,90)(112.5,15)
  \ArrowLine(112.5,15)(150,15)
\Text(18.75,105)[tc]{$e^+$}
  \Text(75,108.75)[tc]{$\phi^0$}
  \Text(131.25,112.5)[tc]{$\bar f$}
  \Text(18.75,52.5)[lc]{$e$}
  \Text(131.8,52.5)[lc]{$\ff$}
  \Text(18.75,0)[cb]{$e^-$}
  \Text(75,-3.75)[bc]{$A$}
  \Text(131.25,-3)[cb]{$\ff$}
\end{picture}
\qquad \qquad  \quad
&+& 
\hspace*{-0.5mm}
\begin{picture}(100,105)(0,49.5)
  \ArrowLine(0,15)(37.5,15)
  \ArrowLine(37.5,15)(37.5,90)
  \ArrowLine(37.5,90)(0,90)
  \Photon(37.5,90)(112.5,15){4}{20}
 \Line(37.5,15)(41.5,19 )
 \Line(45.5,24)(49.5,28)
 \Line(53.5,32)(57.5,36)
 \Line(61.5,40)(65.5,44)
 \Line(69.5,48)(73.5,52)
 \Line(77.5,56)(81.5,60)
 \Line(85.5,64)(89.5,68)
 \Line(93.5,72)(97.5,76)
 \Line(101.5,80)(105.5,84)
 \Line(109.5,88)(112.5,90)
  \Vertex(37.5,15){2.5}
  \Vertex(37.5,90){2.5}
  \Vertex(112.5,15){2.5} 
  \Vertex(112.5,90){2.5}
  \ArrowLine(150,90)(112.5,90)
  \ArrowLine(112.5,90)(112.5,15)
  \ArrowLine(112.5,15)(150,15)
\Text(18.75,105)[tc]{$e^+$}
  \Text(75,108.75)[tc]{$A$}
  \Text(131.25,112.5)[tc]{$\bar f$}
  \Text(18.75,52.5)[lc]{$e$}
  \Text(131.8,52.5)[lc]{$\ff$}
  \Text(18.75,0)[cb]{$e^-$}
  \Text(75,-3.75)[bc]{$\phi^0 $}
  \Text(131.25,-3)[cb]{$\ff$}
\end{picture}
\end{array}
}
\]
\vspace{14mm}
\caption{Direct and crossed $ZA$ boxes. \label{ZA_box_dc} }
\vspace*{-10mm}
\end{figure}
\clearpage

Finally, we present these 6 auxiliary functions:
\begin{align}
{\cal F}_1 \lpar s,t \rpar = & -\frac{1}{8}\frac{s}{\sdfit} \bigg\{            
     \tmi \bigg(   \Rz+\frac{\tmi}{s}-2                  
                + 2t\frac{\szmi}{\sdfit} \bigg)  \jazt
\nll &
   -\bigg(\szpl+2\frac{\szmi\tmi^2}{\sdfit}\bigg) \ceszeo
\nll &
   - 2\bigg( t+2\tmi+\frac{t \mzs}{\tmi}-2\frac{\szmi t \tmi}{\sdfit} \bigg)
\nll[-1mm]&\hspace*{4cm}  \times \cattze
\nll[-1mm]&
   -\bigg[  2\frac{\tpl}{\sdtit}\lpar \mzs - 4 \mfs \rpar  
           + \szmi \lpar 1+2\frac{\tmi\tpl}{\sdfit} \rpar
    \bigg] 
\nll&\hspace*{4cm}\times \ctszto     
\nll[-2mm]&
    + 2\frac{t}{\tmi} \bigg[ 2\boftet 
\nll[-3.5mm]&\hspace*{1.25cm} 
- \boftzt - \boftto \bigg]
\nll[-2mm] &               
   - 2\frac{\tpl}{\sdtit} \bigg[ 2\bofszo
\nll[-3.5mm]&\hspace*{1.25cm} 
- \boftzt-\boftto \bigg]
                  \bigg\},
\nll
{\cal F}_2 \lpar s,t \rpar =& \frac{1}{8}\frac{s}{\sdfit} 
                  \, \bigg[\frac{ \tmi^2}{s}\jazt
\nll &
-2 t \cattze\bigg],
\\
{\cal H}_1 \lpar s,t \rpar = & 
     - \frac{s \mfs}{8} \, \bigg\{ 
 \bigg(  \frac{1}{s}
  - \lrbr t - \tmi \lpar \Rz-2+\frac{\szmi\tpl}{\sdfit} \rpar \rrbr\frac{1}{\sdfit} \bigg) 
\nll&\hspace*{4cm}\times \jazt
\nll &
 -2 \bigg(\frac{\Rz}{\tmi} + 
      \bigg[t- \tmi \lpar \Rz-2+\frac{\szmi\tpl}{\sdfit}\rpar \bigg]\frac{1}{\sdfit} \bigg)
     \nll&\hspace*{4cm}\times      \cattze 
\nll &
    - \frac{1}{\sdfit} \bigg(
      \bigg[\szpl+\szmi \lpar s+2\tmi \rpar \frac{\tmi}{\sdfit} \bigg]   \ceszeo 
\nll &  
    + \bigg[ \szmi - 2 \tpl 
          - \szmi\lpar s-4\mfs \rpar \frac{\tmi}{\sdfit} \bigg] 
\nll[-1mm]&\hspace*{4cm}\times
\ctszto
\nll &
    + 2 \Bigl[ \bofszo - \boftet \Bigr]
\nll &
    - 2 \frac{\mfs}{\tmi} \Big[2\boftet
\nll[-3.5mm]&\hspace*{1.25cm}
-\boftzt-\boftto \Big] \bigg) \bigg\},
\vspace*{-15mm}
\end{align}
\clearpage

\begin{align}
{\cal H}_2 \lpar s,t \rpar = & 
\frac{s}{4} \,  \bigg\{ 
\biggl[-  \frac{\tmi}{\szmi} +\frac{\mfs}{2}\bigg( \frac{1}{s} -\frac{\mfs}{\sdfit}\bigg)\biggr]
 \jazt
\nll &
+  \frac{\tmi}{ \szmi} \cateot
\nll &
-\biggl(\frac{\tmi}{ \szmi}+\frac{m^4_t}{\sdfit}\biggr) \cattze 
                     \bigg\},
\nll
{\cal G}_1 \lpar s,t \rpar = & 
\frac{s}{4} \, \bigg\{\bigg(\hspace*{-.5mm}
  - \frac{\tmi}{\szmi} -\frac{\mfs}{2 s} 
  + \frac{1}{2 \sdfit} \bigg[\mfq + \tmi \bigg( 2\tpl-\mfs\Rz \bigg) 
              - \frac{\szmi t \tmi\tpl}{\sdfit}\bigg]  \bigg)
\nll &
  \times \jazt
           + \frac{\tmi}{\szmi}   \cateot
\nll & - \bigg[
    \frac{\tmi}{\szmi } 
  - \frac{1   }{\sdfit}
\bigg(\mfq + 2\tmi\tpl  + \frac{\mfs \mzs t}{\tmi}- \frac{\szmi t \tmi\tpl}{\sdfit} \bigg)\bigg]  
\nll &\hspace*{3cm}\times   \cattze 
\nll &
   + \frac{1}{2 \sdfit} \bigg[ 
       \bigg( t \szpl + \frac{\szmi\tmi^2\tpl}{\sdfit} \bigg)  \ceszeo 
\nll &
-\bigg(  2\mfs\tmi
            - t\lpar\szpl-4\mfs \rpar - \frac{\szmi\tmi\tpl^2}{\sdfit} \bigg)
\nll &\hspace*{3cm}\times \ctszto 
\nll &
      + 2t \bigg( \bofszo -  \boftet  
\nll &
           - \frac{\mfs }{\tmi}\Big[  2\boftet
\nll[-3.5mm]&\hspace*{1.25cm}
-\boftzt-\boftto \Big] \bigg)\bigg]  \bigg\},
\\
{\cal G}_2 \lpar s,t \rpar =& \frac{ s \mfs}{8} \, 
\bigg[ \lpar\frac{1}{s}-\frac{t}{\sdfit}\rpar\jazt
\nll &
-2\frac{t}{\sdfit} \cattze  \bigg],
\end{align}
where a new notation were introduced for the invariants:
\begin{equation}
s_{\pm}= s \pm \mzs\,,
\qquad
t_{\pm}= t \pm \mfs\,,
\end{equation}
and for the new function $J_{\sss {IJ}}\lpar Q^2,P^2;M_1,M_2 \rpar$
:
\begin{align}
\hspace*{-3mm}
J_{\sss {AZ}}\lpar Q^2,P^2;M_1,M_2 \rpar
=& \frac{1}{P^2+M^2_2} 
    \ln\bigg(\frac{ Q^2 + \mzs}{\mzs}\bigg)
    \ln\bigg[\frac{M^2_1 M^2_2}{\lpar P^2+M^2_2 \rpar^2} \bigg].
\end{align}
\subsubsection{The $\wb\wb$ box}
\begin{figure}[!h]
\vspace{-20mm}
\[
\hspace*{-14mm}
\begin{array}{ccc}
\begin{picture}(100,105)(0,49.5)
  \ArrowLine(0,15)(37.5,15)
  \ArrowLine(37.5,15)(37.5,90)
  \ArrowLine(37.5,90)(0,90)
  \Photon(37.5,15)(112.5,90){3}{10}
  \Photon(37.5,90)(112.5,15){3}{10}
  \Vertex(37.5,15){2.5}
  \Vertex(37.5,90){2.5}
  \Vertex(112.5,15){2.5}
  \Vertex(112.5,90){2.5}
  \ArrowLine(150,90)(112.5,90)
  \ArrowLine(112.5,90)(112.5,15)
  \ArrowLine(112.5,15)(150,15)
\Text(18.75,105)[tc]{$e^+$}
  \Text(75,102)[tc]{$W$}
  \Text(131.25,105)[tc]{$\bar t$}
  \Text(18.75,52.5)[lc]{$\nu_e$}
  \Text(120,52.5)[lc]{$b(s,d)$}
  \Text(18.75,0)[cb]{$e^-$}
  \Text(75,0)[bc]{$W$}
  \Text(131.25,0)[cb]{$\ft$}
\end{picture}
\qquad \qquad\quad
&+& 
\hspace*{-0.5mm}
\begin{picture}(100,105)(0,49.5)
  \ArrowLine(0,15)(37.5,15)
  \ArrowLine(37.5,15)(37.5,90)
  \ArrowLine(37.5,90)(0,90)
  \Photon(37.5,15)(112.5,15){3}{10}
  \Photon(37.5,90)(112.5,90){3}{10}
  \Vertex(37.5,15){2.5}
  \Vertex(37.5,90){2.5}
  \Vertex(112.5,15){2.5}
  \Vertex(112.5,90){2.5}
  \ArrowLine(150,90)(112.5,90)
  \ArrowLine(112.5,90)(112.5,15)
  \ArrowLine(112.5,15)(150,15)
\Text(18.75,105)[tc]{$e^+$}
  \Text(75,102)[tc]{$W$}
  \Text(131.25,105)[tc]{$\bar b$}
  \Text(18.75,52.5)[lc]{$e$}
  \Text(120,52.5)[lc]{$(u,c)t$}
  \Text(18.75,0)[cb]{$e^-$}
  \Text(75,0)[bc]{$W$}
  \Text(131.25,0)[cb]{$b$}
\end{picture}
\end{array}
\]
\vspace{12mm}
\caption{Crossed and direct $WW$ boxes. \label{WW_box_dc} }
\end{figure}
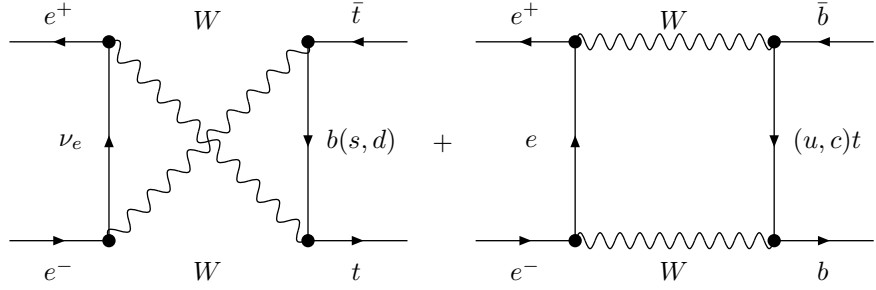

There is only one, {\it direct} or {\it crossed}, $\wb\wb$ diagram contributing to our 
process, see \fig{WW_box_dc}.

\vspace{0.4cm}

\noindent
 Here we give the contribution of this diagram
to the scalar form factor $LL$:
\begin{equation}
\Big( {\cal B}^{\sss WW} \Big)^{c} = (2\pi)^4\ib\frac{g^4}{16\pi^2}\frac{1}{\sman}
\gadi{\mu}\gap\otimes\gadi{\mu}\gap
{\cal F}_{\sss {LL}}^{\sss WW}\lpar\sman,\uman \rpar,
\label{WW_box_def}
\end{equation}
where
\begin{align}
{\cal F}_{\sss {LL}}^{\sss WW}\lpar \sman, \uman \rpar =&
 \frac{s}{\tcit}\bigg[ \lpar\uman-\mbs\rpar
   \dffp{0}{0}{-\mts}{-\mts}{-\sman}{-\uman} \dffm{\mwl}{0}{\mwl}{\mbl} 
\nll &
\hspace*{-1.6mm}+\hspace*{-0.6mm}
\cff{-\mts}{-\mts}{-\sman}{\mwl}{0}{\mwl}
\hspace*{-0.6mm}+\hspace*{-0.6mm}\cff{0}{0}{-\sman}{\mwl}{0}{\mwl}
             \bigg],\quad
\label{WW_box_c}
\end{align}
\vspace*{-5mm}

\noindent
with $\sman$, $\tman$, and $\uman$ being the usual Mandelstamm variables
satisfying
\begin{equation}
\sman+\tman+\uman=2\mfs\,,\qquad\ff=\ft\mbox{ or }\fb\,.
\end{equation}
For the processes, where the direct box contributes, we have:
\begin{align}
{\cal F}^{\sss WW}_{\sss {LL}} =& \frac{\sman}{2\tcib} \bigg\{
   \biggl[ - \lpar 2 - K_2 \rpar \tmip
 + \biggl( 2 \wmf \tpl 
+ K_1\lpar \mds \tmip-\mts\tpl \rpar \frac{\tmi}{t} \biggr) \frac{1}{\sdfit}
\nll &
 - K_1^2 \tpl \tmi^2 \frac{1}{\sdfit^2}
   \biggr]
 \dffp{0}{0}{-\mbs}{-\mds}{-\sman}{-\tman} \dffm{\mwl}{0}{\mwl}{\mul} 
\nll & 
-\biggl( K_2 + \biggl[ K_2\tpl\tmi - K_1 \biggl( \frac{\mdf}{t}+t \biggr) \biggr] \frac{1}{\sdfit}
     + K_1\frac{\tpl^2\tmi^2}{t}\frac{1}{\sdfit^2} 
  \biggr)
\nll &
                     \times \cff{-\mds}{-\mds}{-t}{\wml}{\uml}{\wml}
\nll &
    +2\bigg( K_2 t + K_1 \tpl\tmi\frac{1}{\sdfit} \bigg) \tmi \frac{1}{\sdfit}
        \cff{0}{-\mds}{-t}{0}{\wml}{\uml}
\nn
\end{align}
\newpage
$\,$\vspace*{-6mm}
\begin{align}
&
   - \bigg[ 2 - K_2 + \lpar K_2 t \tmi - K_1 \tpl \rpar \frac{\tmi}{t} \frac{1}{\sdfit}
       + K_1 \frac{\tpl\tmi^3}{t}\frac{1}{\sdfit^2}
     \bigg]
\nll[-1mm]&\hspace*{5cm}
                    \times         \cff{0}{0}{-s}{\wml}{0}{\wml}
\nll[-1mm] &
   +2 \tmi \frac{1}{\sdfit} \Big[ \fbff{0}{-s}{\wml}{\wml} - \fbff{0}{-\mds}{\wml}{\uml} \Big]
\nll &
   -2 \tpl \frac{1}{\sdfit} \Big[ \fbff{0}{-t}{0}{\uml} - \fbff{0}{-\mds}{\wml}{\uml} \Big]
                              \bigg\}, 
\end{align}
\vspace*{-3mm}
\begin{align}
{\cal F}^{\sss WW}_{\sss {LQ}} =&
 \frac{\sman\mds}{4 \tcie\stws\qb} 
 \biggl\{
   -\biggl(
\mts \frac{\tmip}{t^2}
    - \frac{1}{2} \biggl[ (\tmip)^2-2 K_1 \tpl-\biggl( \mds\frac{\tmip}{t}-K_1\biggr)^2 \biggr]
              \frac{1}{\sdfit}
\nll &
     +  K_1^2 \tpl \tmi \frac{1}{\sdfit^2} 
    \biggr) \dffp{0}{0}{-\mds}{-\mds}{-\sman}{-\tman} \dffm{\wml}{0}{\wml}{\uml}  
\nll &
   +\biggl[
     \frac{\mts}{t^2}-\bigg( K_2 \tpl+K_1\frac{\mds}{t} \bigg) \frac{1}{\sdfit} 
     - K_1 \frac{\tpl^2\tmi}{t}  \frac{1}{\sdfit^2}   
    \biggr] 
\nll[-1mm]&\hspace*{5cm}
        \times
\cff{-\mds}{-\mds}{-s}{\wml}{\uml}{\wml}
\nll &
   +2 \biggl[\lpar K_2 + \frac{K_1}{\tmi}\rpar t + K_1\tpl\tmi\frac{1}{\sdfit}
      \biggr] \frac{1}{\sdfit} \cff{0}{-\mds}{-t}{0}{\wml}{\uml} 
\nll &
   -\biggl[
    \frac{\mts}{t^2}+\biggl( K_2 \tmi - K_1 \frac{\mds}{t} \biggr) \frac{1}{\sdfit}
     + K_1 \frac{\tpl\tmi^2}{t}\frac{1}{\sdfit^2}
    \biggr]
\nll[-1mm]&\hspace*{5cm}
        \times                     \cff{0}{0}{-s}{\wml}{0}{\wml}
\nll[-1mm] &
   + 2 \frac{1}{\sdfit} \Bigl[\fbff{0}{-s}{\wml}{\wml} - \fbff{0}{-\mds}{\wml}{\uml} \Bigr]
\nll &
   -  \frac{2 \tpl}{\tmi}  \frac{1}{\sdfit} 
                        \Bigl[\fbff{0}{-t}{0}{\uml} - \fbff{0}{-\mds}{\wml}{\uml}    \Bigr]
                                                       \biggr\},
\end{align}
\vspace*{-3mm}
\begin{align}
{\cal F}^{\sss WW}_{\sss {LD}}
          =& -\frac{\sman}{\tcib}\frac{1}{\sdfit}
\biggl\{
   \bigg[2\mts\tmi + K_3 \tplp+ K_1^2\tmi t\frac{1}{\sdfit}   \bigg]
\nll &
 \times \dffp{0}{0}{-\mds}{-\mds}{-\sman}{-\tman} \dffm{\mwl}{0}{\mwl}{\mul} 
\nll &
   +\bigg[K_3 + 2 \bigg( \frac{\mds \tplp}{t} - K_3 \bigg) \tpl\frac{1}{\sdfit}
     + K_1\tmi\tpl\frac{1}{\sdfit} \bigg] \frac{1}{\sdfit} 
\nll &
 \times \cff{-\mds}{-\mds}{-s}{\wml}{\uml}{\wml}
\nll &
 - t \bigg[\frac{\tplp}{t} + \frac{K_1}{\tmi} + 2 K_1\tmi\frac{1}{\sdfit}\bigg] 
  \cff{0}{-\mds}{-t}{0}{\wml}{\uml} 
\nll &
   -\bigg(K_3 - K_1\tmi^2\frac{1}{\sdfit}\bigg) \cff{0}{0}{-s}{\wml}{0}{\wml}
\nll &
 -2\tpl\frac{1}{\sdtit}\Big[\fbff{0}{-s}{\wml}{\wml} - \fbff{0}{-\mds}{\wml}{\uml} \Big]
\nll &
 +2\frac{t}{\tmi}\Big[\fbff{0}{-t}{0}{\uml} - \fbff{0}{-\mds}{\wml}{\uml} \Big]
                              \biggr\},
\vspace*{-10mm}
\end{align}
\newpage
\noindent
where
\bqa
 K_1&=& 2\mws+\frac{\tmip\tmi}{\tman}\,,
\nll
 K_2&=& \frac{\mfs\mfps}{\tman^2}+1\,
\nll
 K_3&=& 2\mws-\mfps\frac{\tmi}{\tman}\,,
\nll
\tmip &=& t-\mus,\qquad\tplp= t+\mus,
\nll
\tmi  &=& t-\mds,\qquad\tpl = t+\mds.
\eqa

In the limit $ m_b \longrightarrow 0$, for $ F^{\sss WW}_{\sss {LL}}$ one has:
\begin{align}
F^{\sss WW}_{{\sss {LL}} \,\,{\rm lim}} =&\frac{\sman}{2\tcib}\Biggl\{ 
       - \bigg[\bigg(\frac{1}{2} \frac{\tmip}{u} + 1 \bigg)\tmip 
  - K^{\rm lim}_1 \frac{t}{u} 
  + \frac{ (K^{\rm lim}_1)^2}{u} \bigg(\frac{t}{u} + \frac{1}{2}\bigg) \bigg]
\nll &
 \times \dffp{0}{0}{0}{0}{-\sman}{-\tman} \dffm{\wml}{0}{\wml}{\uml} 
\nll &
       +\bigg[ \frac{t}{u}-1 - \frac{K^{\rm lim}_1}{u} \bigg(\frac{t}{u} + 1 \bigg) \bigg]
\nll & 
\times \Big[\cff{0}{0}{-s}{\wml}{0}{\wml} + \cff{0}{0}{-s}{\wml}{\uml}{\wml} \Big]
\nll &
       - 2 \frac{t}{u} \bigg( 1-\frac{K^{\rm lim}_1}{u} \bigg) \cff{0}{0}{-t}{0}{\wml}{\uml}
\nll &
       - \frac{2}{u} \Big[\fbff{0}{-s}{\wml}{\wml}  - \fbff{0}{-t}{0}{\uml} \Big],
\end{align}
where
\bqa
 K^{\rm lim}_1&=& 2\mws+\tmip\,.
\eqa
\subsubsection{ The $\zb\zb$ box}
\begin{figure}[!h]
\vspace{-20mm}
\[
\hspace*{-14mm}
\begin{array}{ccc}
\begin{picture}(100,105)(0,49.5)
  \ArrowLine(0,15)(37.5,15)
  \ArrowLine(37.5,15)(37.5,90)
  \ArrowLine(37.5,90)(0,90)
  \Photon(37.5,15)(112.5,90){3}{10}
  \Photon(37.5,90)(112.5,15){3}{10}
  \Vertex(37.5,15){2.5}
  \Vertex(37.5,90){2.5}
  \Vertex(112.5,15){2.5}
  \Vertex(112.5,90){2.5}
  \ArrowLine(150,90)(112.5,90)
  \ArrowLine(112.5,90)(112.5,15)
  \ArrowLine(112.5,15)(150,15)
\Text(18.75,105)[tc]{$e^+$}
  \Text(75,102)[tc]{$Z$}
  \Text(131.25,105)[tc]{$\bar f$}
  \Text(18.75,52.5)[lc]{$e$}
  \Text(131.8,52.5)[lc]{$\ff$}
  \Text(18.75,0)[cb]{$e^-$}
  \Text(75,0)[bc]{$Z$}
  \Text(131.25,0)[cb]{$\ff$}
\end{picture}
\qquad \qquad\quad
&+& 
\hspace*{-0.5mm}
\begin{picture}(100,105)(0,49.5)
  \ArrowLine(0,15)(37.5,15)
  \ArrowLine(37.5,15)(37.5,90)
  \ArrowLine(37.5,90)(0,90)
  \Photon(37.5,15)(112.5,15){3}{10}
  \Photon(37.5,90)(112.5,90){3}{10}
  \Vertex(37.5,15){2.5}
  \Vertex(37.5,90){2.5}
  \Vertex(112.5,15){2.5}
  \Vertex(112.5,90){2.5}
  \ArrowLine(150,90)(112.5,90)
  \ArrowLine(112.5,90)(112.5,15)
  \ArrowLine(112.5,15)(150,15)
\Text(18.75,105)[tc]{$e^+$}
  \Text(75,102)[tc]{$Z$}
  \Text(131.25,105)[tc]{$\bar f$}
  \Text(18.75,52.5)[lc]{$e$}
  \Text(131.8,52.5)[lc]{$\ff$}
  \Text(18.75,0)[cb]{$e^-$}
  \Text(75,0)[bc]{$Z$}
  \Text(131.25,0)[cb]{$\ff$}
\end{picture}
\end{array}
\]
\vspace{12mm}
\caption{Crossed and direct $ZZ$ boxes. \label{ZZ_box_dc} }
\end{figure}
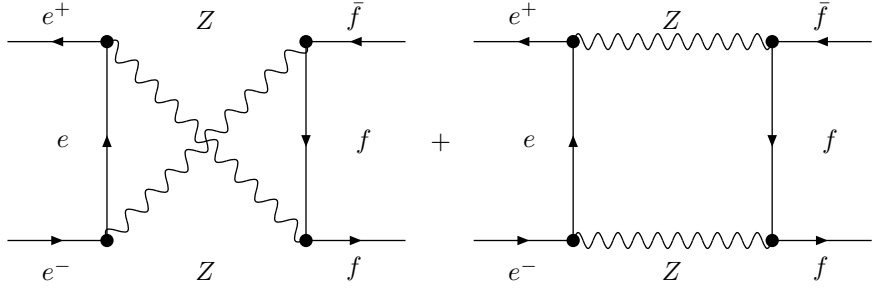

There are four $\zb\zb$ box diagrams, which form a gauge-invariant and UV finite
subset, see \fig{ZZ_box_dc}.
\vspace{0.4cm}

First, we use three auxiliary functions ${\cal{F,H,G}}$:
\bqa
{\cal F}^{\sss ZZ}_{\sss LD}\lpar s,t,u \rpar  &=&
 {\cal F}\lpar \vpae,\vpau,\vmau,s,t \rpar  
-{\cal F}\lpar \vpae,\vmau,\vpau,s,u \rpar, 
\nll
{\cal F}^{\sss ZZ}_{\sss RD} \lpar s,t,u \rpar &=&
 {\cal F}\lpar \vmae,\vmau,\vpau,s,t \rpar
-{\cal F}\lpar \vmae,\vpau,\vmau,s,u \rpar,   
\nll
{\cal F}^{\sss ZZ}_{\sss LR}\lpar s,t,u \rpar  &=&
 {\cal H}\lpar \vpae,\vpau,\vmau,s,t \rpar
-{\cal G}\lpar \vpae,\vmau,\vpau,s,u \rpar,  
\nll
{\cal F}^{\sss ZZ}_{\sss RL}\lpar s,t,u \rpar  &=&
 {\cal H}\lpar \vmae,\vmau,\vpau,s,t \rpar 
-{\cal G}\lpar \vmae,\vpau,\vmau,s,u \rpar,  
\nll
{\cal F}^{\sss ZZ}_{\sss LL} \lpar s,t,u \rpar &=&
  {\cal G}\lpar \vpae,\vpau,\vmau,s,t \rpar 
- {\cal H}\lpar \vpae,\vmau,\vpau,s,u \rpar, 
\nll
{\cal F}^{\sss ZZ}_{\sss RR}\lpar s,t,u \rpar  &=&
  {\cal G}\lpar \vmae,\vmau,\vpau,s,t \rpar 
- {\cal H}\lpar \vmae,\vpau,\vmau,s,u \rpar. 
\eqa

Separating out $\zb$ fermion coupling constants and some common factors,
we introduce more auxiliary functions. For ${\cal F}\lpar \vpae,\vpau,\vmau,s,t \rpar$
defined as
\bqa
{\cal F}\lpar \vpae,\vpau,\vmau,s,t \rpar &=&
 \frac{s}{\sdfit} 
 \biggl[   \vpae^2 \vpau^2     {\cal F}_1\lpar s,t \rpar 
       +  \vpae^2 \vpau \vmau {\cal F}_2\lpar s,t \rpar   \biggr],
\eqa 
there are two
\begin{align}
{\cal F}_1(s,t) = \hspace{1mm} &
  \bigg(   \tpl \mzs+\tmis 
           +\lrbr \lpar-\sz+2 \mzs \rpar \tmis + 4 t \mzf \rrbr
            \frac{\tmi}{ \sdfit}
  \bigg) 
\nll &
\times 
 \dffp{0}{0}{-\mfs}{-\mfs}{-s}{-t}
 \dffm{\mzl}{0}{\mzl}{\mfl}
\nll &
  -\bigg[ 
      \mfs - \mzs 
 - \frac{ \tmi \lpar \tmi t + 2 \mzs \tpl \rpar + s \mfq}{\sdfit}
                 +\frac{ 2  \tpl \lpar \mzs-2 \mfs \rpar}{\sdtit}    
   \bigg]
\nll&\hspace*{4cm}\times \cff{-\mfs}{-\mfs}{s}{\mzl}{\mfl}{\mzl}
\nll &
  -  2\bigg[
                   t \lpar 1+\frac{\mzs}{\tmi}\rpar
      +\frac{ t^3 + 2 t \tmi \mzs 
      -\mfs \lpar 3 t\tmi + \mfq \rpar }{\sdfit}
     \bigg]
\nll&\hspace*{4cm}\times \,\cff{0}{-\mfs}{t}{0}{\mzl}{\mfl}
\nll&
  -  \bigg[
      \mzs- \mfs  + 
       \frac{ \lpar t-2 \tmi-2 \mzs \rpar \tmis + s \mfq} 
            { \sdfit}
 \bigg]
\cff{0}{0}{s}{\mzl}{0}{\mzl} 
\nll&
-\lpar 1 + \frac{s + 2 \tmi}{ \sdtit} \rpar
\Bigl[
 \bff{0}{-s}{\mzl}{\mzl}
-\bff{0}{-\mfs}{\mzl}{\mfl}
\Bigr]
\nll &
  + \frac{2 t}{\tmi}
\Bigl[ \bff{0}{t}{\mfl}{0} - \bff{0}{-\mfs}{\mzl}{\mfl} \Bigr],
\end{align}
and
\begin{align}
{\cal F}_2\lpar s,t \rpar = &
  -\lpar \tmis + 2 \mzs  t  \rpar
 \dffp{0}{0}{-\mfs}{-\mfs}{-s}{-t}
 \dffm{\mzl}{0}{\mzl}{\mfl}
\nll [1mm] &
   - \tpl \cff{-\mfs}{-\mfs}{s}{\mzl}{\mfl}{\mzl}
\nll[2mm] &
   + 2 t   \cff{0}{-\mfs}{t}{0}{\mzl}{\mfl}
   - \tmi   \cff{0}{0}{s}{\mzl}{0}{\mzl}.   
\end{align}
For ${\cal H}$ written out as 
\begin{align}
{\cal H}\lpar \vpae,\vpau,\vmau,s,t \rpar = &
\frac{ \mfs s}{ \sdfit}
\biggl[ 
   \vpae^2 \vpau^2       {\cal  H}_1 \lpar s,t \rpar 
+  \vpae^2 \vpau \vmau   {\cal  H}_2 \lpar s,t \rpar \biggr]
+  \vpae^2 \vmau^2       {\cal  H}_3 \lpar s,t \rpar,  
\end{align}
we need three auxiliary subfunctions:
\begin{align}
{\cal H}_1\lpar s,t \rpar 
 = & 
  \bigg[
 \frac{ s\tmi }{2}-(\tmi+\mzs)^2-2 \mzs t-\frac{\szs \tmi \tpl}{2\sdfit}
 \bigg]
\nll &  
\times 
\dffp{0}{0}{-\mfs}{-\mfs}{-s}{-t}
 \dffm{\mzl}{0}{\mzl}{\mfl}
\nll &
 - \bigg(
  \tpl-\frac{\sz \lrbr \tpls+\sdtit \mfs \rrbr}{2\sdfit}
   \bigg) 
\cff{-\mfs}{-\mfs}{s}{\mzl}{\mfl}{\mzl}
\nll &
 + \bigg( 
 2 t+\tmi+\frac{2\mzs t}{\tmi}-\frac{\sz \tmi \tpl }{\sdfit}
  \bigg)
\cff{0}{0}{s}{\mzl}{0}{\mzl}
\nll &
 -  \bigg[
\tmi +  \frac{\sz \lpar s \mfs-\tmis \rpar}{2 \sdfit}
    \bigg] 
\cff{0}{-\mfs}{t}{0}{\mzl}{\mfl}
\nll &
 -2 \frac{\mfs}{\tmi} 
 \biggl( \bff{0}{t}{\mfl}{0}-\bff{0}{-\mfs}{\mzl}{\mfl} \biggr)
\nll &
 - \bff{0}{t}{\mfl}{0}+\bff{0}{s}{\mzl}{\mzl},
\\[2mm]
{\cal H}_2\lpar s,t \rpar =&
  \lrbr \tmi \lpar s\hspace*{-1mm} +\hspace*{-1mm} \tmi \rpar + 2 \mfs \mzs \rrbr
\dffp{0}{0}{-\mfs}{-\mfs}{-s}{-t}
\dffm{\mzl}{0}{\mzl}{\mfl}
\nll[2mm]  &
 - \lpar s + \tmi - 2 \mfs\rpar \cff{-\mfs}{-\mfs}{s}{\mzl}{\mfl}{\mzl} 
\nll[2mm]    &
   - 2 \mfs\cff{-\mfs}{-\mfs}{s}{0}{\mzl}{\mfl}
\nll[2mm]  &
   + \lpar s+\tmi \rpar\cff{0}{0}{s}{\mzl}{0}{\mzl}, 
\end{align}
and
\begin{align}
{\cal H}_3\lpar s,t \rpar = & - s\biggl[  \tmi 
\dffp{0}{0}{-\mfs}{-\mfs}{-s}{-t}
\dffm{\mzl}{0}{\mzl}{\mfl}
\\ &
   + \cff{-\mfs}{-\mfs}{s}{\mzl}{\mfl}{\mzl} 
   + \cff{0}{-\mfs}{t}{\mzl}{0}{\mzl} \biggr].      
\nonumber
\end{align}
Finally, ${\cal G}$ also, defined as follows:
\begin{equation}
{\cal G} \lpar \vpae,\vpau,\vmau,s,t \rpar = \frac{s}{\sdfit}
\biggl[
  \vpae^2 \vpau^2 {\cal G}_1\lpar s,t \rpar  
+ \vpae^2 \vpau \vmau {\cal G}_2\lpar s,t \rpar  \biggr],
\end{equation}
needs only two additional functions:
\begin{align}
{\cal G}_1\lpar s,t \rpar
=&-
   \biggl[
     \tmi \tpl \bigg( 2 \mzs + \frac{\szs t}{2\sdfit}\bigg) 
      -\tmi   \bigg( \frac{s \mfs}{2}-\tmi t\bigg) 
      + t\mzs \lpar 2 \mfs-\mzs \rpar
   \biggr]
\nll &
 \times 
\dffp{0}{0}{-\mfs}{-\mfs}{-s}{-t}
\dffm{\mzl}{0}{\mzl}{\mfl}
\nll &
 + \bigg[    
\tpl t+ \frac{\sz}{2} \bigg( t - \frac{\tmi\tpls}{\sdfit} \bigg)
   \bigg]
 \cff{-\mfs}{-\mfs}{s}{\mzl}{\mfl}{\mzl}
\nll &
 -  \bigg[
  \tpl \lpar t+\tmi \rpar + 2 \mzs \mfs \frac{t}{\tmi}
 -\frac{\tpl \tmi t \sz}{\sdfit}
    \bigg]
 \cff{0}{-\mfs}{-t}{0}{\mzl}{\mfl}
\nll &
 - \bigg[ \tmi t + \frac{\sz}{2} 
            \bigg( t - \frac{ \tpl \tmis}{\sdfit} \bigg)
   \bigg] 
\cff{0}{0}{s}{\mzl}{0}{\mzl} 
\nll &
 + 2 \mfs \bigg( 1 + \frac{\mfs}{\tmi} \bigg)
 \bigg[ \bff{0}{t}{\mfl}{0} -\bff{0}{-\mfs}{\mzl}{\mfl} \bigg]
\nll [1mm] &
 -  t \bigg[ \bff{0}{s}{\mzl}{\mzl}-\bff{0}{t}{\mfl}{0} \bigg]\,,
\end{align}
and
\begin{align}
{\cal G}_2\lpar s,t \rpar = \hspace{1mm} &
 \mfs \bigg[     \bigg( \tmis + 2 \mzs t \bigg) 
\dffp{0}{0}{-\mfs}{-\mfs}{-s}{-t}\dffm{\mzl}{0}{\mzl}{\mfl} 
\nll [1mm] &
                  + \tpl \cff{-\mfs}{-\mfs}{s}{\mzl}{\mfl}{\mzl}
                  - 2 t     \cff{0}{-\mfs}{t}{0}{\mzl}{\mfl}
\nll [1mm] &  
                + \tmi  \cff{0}{0}{s}{\zml}{0}{\mzl}\biggr].
\end{align}
We recall that 
$\sdfit = - t u + \mfq\,$
together with $\sdtit$ of \eqn{moredefinitions}
denotes remnants of Gram determinants that remained after cancellation
of factors $\sman$ and $4$, leading to a simplification of the expressions.

\subsubsection{Transition to the $\it L,Q,D$ basis}
Since the $\zb\zb$ box contribution is given in the ${\it L,R,D}$ basis, while all the rest
is in the ${\it L,Q,D}$ basis, we should transfer one of them to a chosen basis.
At this phase of the calculations there is not much difference which basis is 
chosen. For definiteness we choose the ${\it L,Q,D}$ basis and transfer the $\zb\zb$ box 
contribution to it.The transition formulae are simple:
\bqa
{\tilde{\cal F}}^{\sss ZZ}_{\sss LL}\lpar s,t,u \rpar &=&   
 {\cal F}^{\sss ZZ}_{\sss LL}\lpar s,t,u \rpar
+{\cal F}^{\sss ZZ}_{\sss RR}\lpar s,t,u \rpar
-{\cal F}^{\sss ZZ}_{\sss LR}\lpar s,t,u \rpar
-{\cal F}^{\sss ZZ}_{\sss RL}\lpar s,t,u \rpar,
\nll
{\tilde{\cal F}}^{\sss ZZ}_{\sss QL} \lpar s,t,u \rpar &=& 
 2\Bigl[ {\cal F}^{\sss ZZ}_{\sss RL}\lpar s,t,u \rpar
       -{\cal F}^{\sss ZZ}_{\sss RR}\lpar s,t,u \rpar \Bigr], 
\nll
{\tilde{\cal F}}^{\sss ZZ}_{\sss LQ} \lpar s,t,u \rpar  &=&  
 2\Bigl[{\cal F}^{\sss ZZ}_{\sss LR}\lpar s,t,u \rpar
             -{\cal F}^{\sss ZZ}_{\sss RR}\lpar s,t,u \rpar  \Bigr],
\nonumber
\eqa
\bqa
{\tilde{\cal F}}^{\sss ZZ}_{\sss QQ} \lpar s,t,u \rpar &=&  
 4{\cal F}^{\sss ZZ}_{\sss RR}\lpar s,t,u \rpar,
\nll[1mm]
{\tilde{\cal F}}^{\sss ZZ}_{\sss LD} \lpar s,t,u \rpar &=&   
        {\cal F}^{\sss ZZ}_{\sss LD}\lpar s,t,u \rpar
       -{\cal F}^{\sss ZZ}_{\sss RD}\lpar s,t,u \rpar,
\nll[1mm]
{\tilde{\cal F}}^{\sss ZZ}_{\sss QD} \lpar s,t,u \rpar &=& 
 2{\cal F}^{\sss ZZ}_{\sss RD}\lpar s,t,u \rpar.   
\eqa
\subsubsection{Box--Born interferences}
Any box, described by the amplitude in~\eqn{anyboxamplitude}, interfering with $\gamma$ and $\zb$ 
exchange tree-level amplitudes, gives rise to two contributions to the differential cross-sections,
which are useful for internal cross-checks:
\begin{align}
  \sigma_{{\sss{\rm BOX}}\otimes{\sss {\rm BORN_\gamma}}} \propto 
\hspace{1mm}& 8 \qe \qt 
 {\rm Re} \biggl\{
    \bigg(
         \Big[ \lpar s+\tmi \rpar^2+s\mfs \Big] 
                  \lpar {\cal F}_{\sss LL}+{\cal F}_{\sss RR} \rpar
\label{BoxBorngamma}
\\ &
  \hspace*{-1mm}     +\hspace*{-1mm} \lpar s\mfs+\tmi^2 \rpar 
                  \lpar {\cal F}_{\sss LR}+{\cal F}_{\sss RL} \rpar
  \hspace*{-1mm}     -2\mfs \lpar s t+\tmi^2 \rpar 
          \lpar {\cal F}_{\sss LD}+{\cal F}_{\sss RD} \rpar \bigg) \biggr\},
\nll
  \sigma_{{\sss{\rm BOX}}\otimes{{\sss {\rm BORN}}_{\sss Z}}} \propto\hspace{1mm}  &    
\label{BoxBornZ}
  8 {\rm Re} \biggl\{\bigg(
  \Big[\lpar s+\tmi \rpar^2+s\mfs\Big] \vma{\ff}{}
    \lpar \vpae {\cal F}_{\sss LL}+\vma{e}{} {\cal F}_{\sss RR} \rpar
\nll &\hspace*{-1mm}
       +2 \lpar s+\tmi \rpar^2 \af            \vpae
                                  {\cal F}_{\sss LL}
       +2\tmi^2\af                \vma{e}{} {\cal F}_{\sss RL}
\nll[2mm] &
  \hspace*{-1mm}     + \lpar s \mfs+\tmi^2 \rpar \vma{\ff}{}     
                 \lpar\vpae {\cal F}_{\sss LR}+\vma{e}{}{\cal F}_{\sss RL}\rpar
\hspace*{-1mm}
       +\hspace*{-1mm}2 s \mfs \af 
            \lpar \vpae {\cal F}_{\sss LR}+\vma{e}{} {\cal F}_{\sss RR}\rpar
\nll & \hspace*{-1mm}
      -2 \mfs \lpar s t+\tmi^2 \rpar \vf     
              \lpar \vpae {\cal F}_{\sss LD}+\vma{e}{}{\cal F}_{\sss RD}\rpar           
                            \bigg) {\chi_{\sss Z}}^* \biggr\}.
\end{align}

\section{Scalar form factors for electroweak amplitude}
Having all the building blocks, it is time to construct
{\it complete} electroweak scalar form factors.
\subsection{Vertices scalar form factors}
We begin with two vertex contributions:
\begin{figure}[th]
\vspace*{-20mm}
\[
\begin{array}{ccc}
\begin{picture}(125,86)(0,40)
    \Photon(25,43)(100,43){3}{15}
    \Vertex(25,43){12.5}
    \ArrowLine(0,0)(25,43)
    \ArrowLine(25,43)(0,86)
      \ArrowLine(125,86)(100,43)
      \Vertex(100,43){2.5}
      \ArrowLine(100,43)(125,0)
\Text(14,74)[lb]{$\fbe$}
\Text(108,74)[lb]{$\bar f$}
\Text(62.5,50)[bc]{$(\gamma,\zb)$}
\Text(14,12)[lt]{$\fe$}
\Text(108,12)[lt]{$\ff$}
\end{picture}
\qquad
&&
\qquad
\begin{picture}(125,86)(0,40)
 \Photon(25,43)(100,43){3}{15}
 \Vertex(100,43){12.5}
 \ArrowLine(125,86)(100,43)
 \ArrowLine(100,43)(125,0)
   \ArrowLine(0,0)(25,43)
   \Vertex(25,43){2.5}
   \ArrowLine(25,43)(0,86)
\Text(14,74)[lb]{$\fbe$}
\Text(108,74)[lb]{$\bar f$}
\Text(62.5,50)[bc]{$(\gamma,\zb)$}
\Text(14,12)[lt]{$\fe$}
\Text(108,12)[lt]{$\ff$}
\end{picture}
\end{array}
\]
\vspace{8mm}
\caption[Electron and final fermion vertices 
in $\fe\fbe\to(\gamma,\zb)\to\ff\fbf$.]
{\it
Electron (a) and final fermion (b) vertices 
in $\fe\fbe\to(\gamma,\zb)\to\ff\fbf$.
\label{zavert6}}
\vspace*{-15mm}
\end{figure}
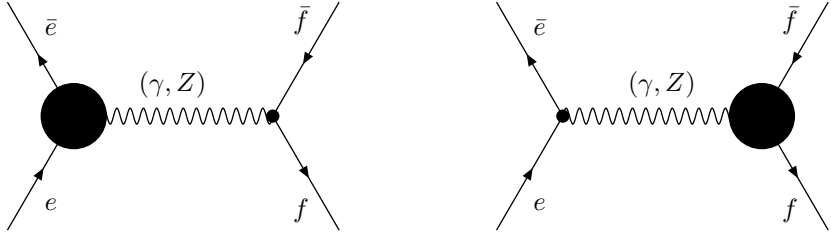
\clearpage

 In the same way as described in~\cite{Bardin:1999yd}, we reduce two vertex 
contributions to our six form factors:
\bqa
\vvertil{}{\sss{LL}}{\sman}&=&
    \vvertil{zee}{\sss{L}}{\sman} 
 +  \vvertil{zff}{\sss{L}}{\sman} - 4\ctws \Delta(\mwl),  
\nll [2mm]
\vvertil{}{\sss{QL}}{\sman}&=&
    \vvertil{zee}{\sss{Q}}{\sman}
 +  \vvertil{zff}{\sss{L}}{\sman} - 2\ctws \Delta(\mwl)            
 +  k
 \lrbr \vvertil{\gamma ff}{\sss{L}}{\sman}- 2\Delta(\mwl) \rrbr,
\nll [2mm]
\vvertil{}{\sss{LQ}}{\sman}&=&
    \vvertil{zee}{\sss{L}}{\sman} - 2\ctws \Delta(\mwl)
  + \vvertil{zff}{\sss{Q}}{\sman}    
 +  k
 \lrbr \vvertil{\gamma ee}{\sss{L}}{\sman}- 2\Delta(\mwl) \rrbr,
\nll [1mm]
\vvertil{}{\sss{QQ}}{\sman}&=&
    \vvertil{zee}{\sss{Q}}{\sman}
 +  \vvertil{zff}{\sss{Q}}{\sman}                   
 -  \frac{k}{\stws} 
    \lrbr \vvertil{\gamma ee}{\sss{Q}}{\sman}
 +  \vvertil{\gamma ff}{\sss{Q}}{\sman} \rrbr,
\nll [1mm]
\vvertil{}{\sss{LD}}{\sman}&=&
    \vvertil{zff}{\sss{D}}{\sman},
\nll [3mm]
\vvertil{}{\sss{QD}}{\sman}&=&
    \vvertil{zff}{\sss{D}}{\sman}
 +  k
    \vvertil{\gamma ff}{\sss{D}}{\sman},
\eqa
where 
\bqa
k = \cows \lpar \Rz - 1 \rpar.
\label{koeff_k}
\eqa

With the term containing $\Delta(\mwl)$,
\bqa
\Delta(\mwl) = \pole - \ln \frac{\mws}{\mu^2}\,,
\eqa    
we explicitly show the contribution of the  
so-called {\em special} vertices \cite{Passarino:1991b}.
Note that they accompany every $L$ form factor.
The poles $1/{\bar{\varepsilon}}$ originating from special vertices
will be canceled in the sum of all contributions, including self-energies 
and boxes.
\subsection{Bosonic self-energies and bosonic counterterms}
 The contributions to form factors from bosonic self-energy diagrams and 
counterterms, originating from bosonic self-energy diagrams,
come from four classes of diagrams; their sum is depicted by a black
circle in \fig{zavert8}.
Their contribution to the four scalar form factors is derived
in~\cite{Bardin:1999ak}, ~\cite{Bardin:1999yd}:
\begin{align}
\vvertil{ct}{\sss{LL}}{\sman}  = & 
      \Dz{}{\sman}-\stws\Pgg^{}(0)
     +\frac{\cows-\siws}{\siws} \lpar \delrho{} + \bdelrho{\bos} \rpar,   
\label{Fct_b_LL}
\\[2mm]
\vvertil{ct}{\QL(\LQ)}{\sman} = & 
      \Dz{}{\sman}-
      \lpar \Pzg^{}(\sman) + \bPzga{\bos}{\sman} \rpar
     -\stws \Pgg^{}(0)-\lpar\delrho{}+\bdelrho{\bos} \rpar, 
\label{Fct_b_LQ}
\\
\vvertil{ct,\bos}{\sss{QQ}}{\sman} = &
      \Dz{\bos}{\sman}-2\lpar \Pzg^{\bos}(\sman) + \bPzga{\bos}{\sman} \rpar
     + k \Bigl[\Pgg^{\bos}\lpar s \rpar 
     - \Pgg^{\bos}\lpar 0 \rpar \Bigr]                  
\nll&
-\stws \Pgg^{\bos}\lpar 0 \rpar 
     -\frac{1}{\siws} \lpar \delrho{\bos} + \bdelrho{\bos} \rpar,    
\label{Fct_b_QQ}
\end{align}
\begin{align}
\vvertil{ct,\fer}{\sss{QQ}}{\sman} = &
      \Dz{\fer}{\sman}-2\Pzg^{\fer}(\sman)
     -\stws \Pgg^{\fer}\lpar 0 \rpar 
     -\frac{1}{\siws} \delrho{\fer}.
\label{Fct_f_QQ}
\end{align}

\begin{figure}[h]
\vspace{-20mm}
\[
\begin{array}{ccc}
\begin{picture}(125,86)(0,40)
    \Photon(25,43)(50,43){3}{5}
    \Vertex(62.5,43){12.5}
    \Photon(75,43)(100,43){3}{5}
  \ArrowLine(125,86)(100,43)
  \Vertex(100,43){2.5}
  \ArrowLine(100,43)(125,0)
    \ArrowLine(0,0)(25,43)
    \Vertex(25,43){2.5}
    \ArrowLine(25,43)(0,86)
\Text(14,74)[lb]{$\fbe$}
\Text(108,74)[lb]{$\fbf$}
\Text(37.5,50)[bc]{$(\ph,\zb)$}
\Text(87.5,50)[bc]{$(\ph,\zb)$}
\Text(14,12)[lt]{$\fe$}
\Text(108,12)[lt]{$\ff$}
\end{picture}
\qquad
 &=&
\nll  \nll
\begin{picture}(125,86)(0,40)
    \Photon(25,43)(50,43){3}{5}
    \GCirc(62.5,43){12.5}{0.5}
    \Photon(75,43)(100,43){3}{5}
  \ArrowLine(125,86)(100,43)
  \Vertex(100,43){2.5}
  \ArrowLine(100,43)(125,0)
    \ArrowLine(0,0)(25,43)
    \Vertex(25,43){2.5}
    \ArrowLine(25,43)(0,86)
\Text(14,74)[lb]{$\fbe$}
\Text(108,74)[lb]{$\fbf$}
\Text(37.5,50)[bc]{$(\ph,\zb)$}
\Text(87.5,50)[bc]{$(\ph,\zb)$}
\Text(14,12)[lt]{$\fe$}
\Text(108,12)[lt]{$\ff$}
\end{picture}
\qquad
&+&
\qquad
\begin{picture}(125,86)(0,40)
     \Photon(25,43)(100,43){3}{15}
\SetScale{2.0}
     \Line(26.5,16.75)(28,18.25)
     \Line(29,19.25)(34,24.25)
     \Line(35,25.25)(36.5,26.75)
      \Line(26.5,26.75)(28,25.25)
      \Line(29,24.25)(34,19.25)
      \Line(35,18.25)(36.5,16.75)
\SetScale{1.0}
     \ArrowLine(125,86)(100,43)
     \Vertex(100,43){2.5}
     \ArrowLine(100,43)(125,0)
  \ArrowLine(0,0)(25,43)
  \Vertex(25,43){2.5}
  \ArrowLine(25,43)(0,86)
\Text(14,74)[lb]{$\fbe$}
\Text(108,74)[lb]{$\fbf$}
\Text(37.5,50)[bc]{$(\ph,\zb)$}
\Text(87.5,50)[bc]{$(\ph,\zb)$}
\Text(14,12)[lt]{$\fe$}
\Text(108,12)[lt]{$\ff$}
\end{picture} 
\nll \nll
\begin{picture}(125,86)(0,40)
  \Photon(25,43)(100,43){3}{15}
\SetScale{2.0}
    \Line(45,16.5)(46.5,18)
    \Line(47.5,19)(52.5,24)
    \Line(53.5,25)(55,26.5)
    \Line(45,26.5)(46.5,25)
    \Line(47.5,24)(52.5,19)
    \Line(53.5,18)(55,16.5)
\SetScale{1.0}
  \ArrowLine(125,86)(100,43)
  \Vertex(100,43){2.5}
  \ArrowLine(100,43)(125,0)
    \ArrowLine(0,0)(25,43)
    \Vertex(25,43){2.5}
    \ArrowLine(25,43)(0,86)
\Text(14,74)[lb]{$\fbe$}
\Text(108,74)[lb]{$\fbf$}
\Text(62.5,50)[bc]{$(\ph,\zb)$}
\Text(14,12)[lt]{$\fe$}
\Text(108,12)[lt]{$\ff$}
\end{picture}
\qquad
&+&
\qquad
\begin{picture}(125,86)(0,40)
  \Photon(25,43)(100,43){3}{15}
\SetScale{2.0}
    \Line(7.5,16.5)(9,18)
    \Line(10,19)(15,24)
    \Line(16,25)(17.5,26.5)
    \Line(7.5,26.5)(9,25)
    \Line(10,24)(15,19)
    \Line(16,18)(17.5,16.5)
\SetScale{1.0}
    \Vertex(25,43){2.5}
    \ArrowLine(0,0)(25,43)
    \ArrowLine(25,43)(0,86)
 \ArrowLine(125,86)(100,43)
 \Vertex(100,43){2.5}
 \ArrowLine(100,43)(125,0)
\Text(14,74)[lb]{$\fbe$}
\Text(108,74)[lb]{$\fbf$}
\Text(62.5,50)[bc]{$(\ph,\zb)$}
\Text(14,12)[lt]{$\fe$}
\Text(108,12)[lt]{$\ff$}
\end{picture}
\end{array}
\]
\vspace{12mm}
\caption[Bosonic self-energies
and counterterms for $\fe\fbe\to(\zb,\ph)\to\ff\fbf$.]
{\it
Bosonic self-energies
and counterterms for $\fe\fbe\to(\zb,\ph)\to\ff\fbf$.
\label{zavert8}} 
\end{figure}

 We note that the term $ k \bigl[\Pgg^{\fer}\lpar\sman\rpar
-\Pzg^{\fer}\lpar\sman\rpar\bigr]$ is conventionally extracted from 
$\vvertil{ct,\fer}{\sss{QQ}}{\sman}$. 
This contribution is shifted to $A_{\gamma}^{\sss{\rm{IBA}}}$, \eqn{Born_modulo-old}.

 In \eqns{Fct_b_LL}{Fct_f_QQ} $\bdelrho{\bos}$ and $\bPzga{\bos}{\sman}$
stand for shifts of bosonic self-energies. They have the same origin as 
special vertices and they are equal to:
\bqa
\bdelrho{\bos}      &=& 4\stws\Delta(\mwl),
\\[2mm]
\bPzga{\bos}{\sman} &=& -2\Rw \Delta(\mwl),
\eqa
see Eqs.~(6.137) and ~(6.139) of ~\cite{Bardin:1999ak}.
These poles also cancel in the sum of all contributions.
\subsection{Total scalar form factors of the one-loop amplitude}

 Adding all contributions together, we observe the cancellation of all poles.
The ultraviolet-finite results for six scalar form factors are:
\begin{align}
\vvertil{}{\sss{LL}}{s,t,u} = &
    \cvetril{zee}{\sss{L}}{s}
 +  \cvetril{zff}{\sss{L}}{s}  
 +  \cvertil{ct}{\sss{LL}}{s}  
 +  16k\cvetril{\sss {\rm BOX}}{\sss LL}{s,t,u},
\nll [3mm]
\vvertil{}{\sss{QL}}{s,t,u} = &
    \cvetril{zee}{\sss{Q}}{s}
 +  \cvetril{zff}{\sss{L}}{s}             
 + k~\cvetril{\gamma ff}{\sss{L}}{s}
 +  \cvertil{ct}{\sss{QL}}{s}
 +  16k\cvetril{\sss {\rm BOX}}{\sss QL}{s,t,u},
\nll [3mm]
\vvertil{}{\sss{LQ}}{s,t,u} = &
    \cvetril{zee}{\sss{L}}{s}
 +  \cvetril{zff}{\sss{Q}}{s}    
 +  k~\cvetril{\gamma ee}{\sss{L}}{s} 
 +  \cvertil{ct}{\sss{LQ}}{s}
 +  16k \cvetril{\sss {\rm BOX}}{\sss LQ}{s,t,u},
\nll [3mm]
\vvertil{}{\sss{QQ}}{s,t,u} = &
    \cvetril{zee}{\sss{Q}}{s}
 +  \cvetril{zff}{\sss{Q}}{s}                   
 - \frac{k}{\stws}~\Big[ \cvetril{\gamma ee}{\sss{Q}}{s}
 +  \cvetril{\gamma ff}{\sss{Q}}{s} \Big]
 +  \cvertil{ct}{\sss{QQ}}{s}
\nll &\hspace*{6.5cm}
 +  16k\cvetril{\sss {\rm BOX}}{\sss QQ}{s,t,u},
\nonumber
\end{align}
\begin{align}
 \vvertil{}{\sss{LD}}{s,t,u} = &
    \cvetril{zff}{\sss{D}}{s}
 +  16k\cvetril{\sss {\rm BOX}}{\sss LD}{s,t,u},
\nll [4mm]
\vvertil{}{\sss{QD}}{s,t,u} = &
    \cvetril{zff}{\sss{D}}{s}
+ k~\cvetril{\gamma ff}{\sss{D}}{s}
 +  16k\cvetril{\sss {\rm BOX}}{\sss QD}{s,t,u}.
\label{soular_ff}
\end{align}
For $IJ=LQ,QL,QQ,LD,QD$ components of a box contribution we have: 
\begin{align}
 \cvetril{\sss {\rm BOX}}{\sss IJ}{s,t,u}  = &
  k^{\sss  AA} \cvetril{\sss  AA}{\sss IJ}{s,t,u}
+ k^{\sss  ZA} \cvetril{\sss  ZA}{\sss IJ}{s,t,u} 
+ k^{\sss  ZZ} \cvetril{\sss  ZZ}{\sss IJ}{s,t,u}
\nll[1mm] &\hspace*{4cm}
+ k^{\sss  WW} \cvetril{\sss  WW}{\sss IJ}{s,t,u}\,,
\label{com_box}
\end{align}
where
\begin{equation}
k^{\sss  WW} = \frac{1}{16}\,.
\end{equation}
Moreover,
\bqa
\cvetril{\gamma(z)ff}{\sss{L,Q,D}}{s}=
\sum_{\sss B=A,Z,H,W}
\cvetril{\gamma(z){\sss B}}{\sss{L,Q,D}}{s}\,,    
\eqa
except for $\cvetril{\gamma {\sss A}}{\sss{L}}{s} =0$ and 
$\cvetril{\gamma {\sss H}}{\sss{L}}{s} =0$.
\section{Improved Born Approximation cross-section}

\subsection{Improved Born Approximation cross-section}
In this section we give the improved  Born approximation (IBA) differential in the
scattering angle cross-section. It is derived by simple squaring the $(\ph+\zb)$ exchange
IBA amplitude, \eqns{Born_modulo-old}{structures-old}, and accounting for proper 
normalization factors. We simply give the result:
\begin{equation}
\frac{d\sigma^{\sss{\rm{IBA}}}}{d\cos\vartheta}=\frac{\pi\alpha^2}{s^3}\beta_{\ft} N_{c}
\bigl(\sigma_{\gamma\gamma}^{\sss IBA}
     +\sigma_{\gamma{\sss{Z}}  }^{\sss IBA}
     +\sigma_{\sss ZZ          }^{\sss IBA}
\bigr),
\label{differential-cs}
\end{equation}
where $\beta_{\ft}=\sqrt{1-{4\mfs}/{\sman}}$ and 
\begin{equation}
\sigma_{\gamma\gamma}^{\sss IBA} = 
      \qf^2 \lpar s^2+ 2st+2\tmis \rpar \Bigl| \alpha\lpar s \rpar \Bigr|^2, 
\end{equation}
\begin{align}
\sigma_{\gamma{\sss{Z}}}^{\sss IBA}  =& 2 \qf{\rm Re} \biggl\{ \chi \biggl(
            2\lrbr \lpar s+\tmi\rpar^2+s \mfs \rrbr\FLLt 
\nll &
            +\lpar s^2+2st+2\tmis\rpar\biggl[
 \FQLt + \FLQt  + \FQQt   \biggr]
\nll &
           - 4 \mfs \lpar st+\tmis \rpar \biggl[   \FLDt   
 +  \FQDt   \biggr] \biggr)
\alpha^*\lpar s \rpar \biggr\},
\end{align}
\begin{align}
\sigma_{\sss ZZ}^{\sss IBA} = &  |\chi|^2 {\rm Re} \biggl\{
   8 \lpar s+\tmi\rpar^2 \bigg[ \Bigl| \FLLt \Bigr|^2+ \FLLt \FQLtc \bigg]
\nll[2mm] &
 + 2 \lrbr \lpar s+\tmi \rpar^2+\tmis \rrbr \Bigl| \FQLt \Bigr|^2
\nll[2mm] &
 + 4 \lrbr\lpar s+\tmi\rpar^2+s\mfs \rrbr\bigg[ 2  \FLLt \FLQtc 
\nll &
 +  \FLLt \FQQtc + \FQLt \FLQtc  \bigg]
\nll &
 + \lrbr s^2+2 \lpar s t+ \tmis \rpar\rrbr \biggl[ 2 \Bigl| \FLQt \Bigl|^2 
                          +\Bigl| \FQQt \Bigr|^2
\nll &
 + 2 \bigg( \FQLt + \FLQt \bigg) \FQQtc  \biggr]
\nll & 
 - 8 \mfs \lpar st+\tmis \rpar \bigg[ \bigg( 2\FLDt + \FQDt \bigg) \FLLtc 
\nll & 
 + \bigg( \FLDt + \FQDt \bigg) \FQLtc 
\nll & 
 + \bigg( 2 \FLDt + \FQDt \bigg) \FLQtc 
\nll & 
 + \bigg( \FLDt + \FQDt \bigg)\FQQtc  \bigg] 
\nll &
 - 2\mfs \lpar s t + \tmis \rpar \sdtit 
       \bigg[ 2 \Bigl| \FLDt \Bigr|^2 
\nll &
 + 2 \FLDt \FQDtc  + \Bigl| \FQDt \Bigr|^2 \bigg] \biggr\}.
\end{align}
\subsection{The $e^+e^-\to\ff\bar{f}$ process in the helicity amplitudes}
\eqnzero

  According to the analysis of the EW part and QED parts, we have the {\bf complete} answer for 
the amplitude of the process $e^+e^-\to\ft\bar{t}$.
  
  The aim of this section is to adapt the helicity amplitude techniques for
the description of our process. 
  We produced an alternative analytic answer for the same amplitude
using the method suggested in Ref.~\cite{hel_VegaWudka}.

  In general, there are 16 helicity amplitude for any $2 \ff \to 2 \ff $ process. 
For the unpolarized case,
and when the electron mass is ignored, we are left with only six helicity amplitudes,
which depend on kinematical variables and our six scalar form factors: 
\begin{align}
{\cal A}_{++++}  =& 0,\qquad
{\cal A}_{+++-}   =  0,\qquad
{\cal A}_{++-+}   =  0,\qquad 
{\cal A}_{++--}   =  0,
\nonumber
\end{align}
\begin{align}
{\cal A}_{+-+-}  =& s\lpar 1-\cos\vartheta\rpar 
 \bigg( \qe \qf F_{\sss GG}
        +{\chi_{\sss Z}}\delta_e \Big[\lpar 1+\beta_f\rpar \tcit F_{\sss QL}
                              +\delta_f F_{\sss QQ}\Big]\bigg),
\nll
{\cal A}_{+--+}  
                  =& s \lpar 1+\cos\vartheta\rpar 
 \bigg(  \qe\qf F_{\sss GG}
        +{\chi_{\sss Z}}\delta_e \Big[(1-\beta_f)\tcit  F_{\sss QL}
        +\delta_f F_{\sss QQ}    \Big] \bigg),
\nll
{\cal A}_{+---} =& {\cal A}_{+-++}
                   =  2\sqs\mfl \sin\vartheta
 \bigg(\qe \qf F_{\sss GG}
\nll &
       +{\chi_{\sss Z}}\delta_e \Big[ \tcit F_{\sss QL}+\delta_f  F_{\sss QQ}
     +\frac{1}{2} s {\beta_f}^2  \tcit  F_{\sss QD}\Big] \bigg),
\nll
{\cal A}_{-+++} =& {\cal A}_{-+--} 
                   =  -2\sqs\mfl \sin\vartheta
 \bigg(\qe\qf F_{\sss GG}
\nll &
 +{\chi_{\sss Z}} \bigg[2 \tcie \tcit F_{\sss LL}+2 \tcie  \delta_f F_{\sss LQ}
 +\delta_e \tcit F_{\sss QL}
                                 +\delta_e \delta_f F_{\sss QQ}
\nll &
  +\frac{1}{2} s {\beta_f}^2 \tcit
       \Big( 2\tcie F_{\sss LD}+\delta_e F_{\sss QD}\Big) \bigg] \bigg),
\nll
{\cal A}_{-++-} =&  s \lpar 1+\cos\vartheta \rpar 
  \bigg( \qe\qf F_{\sss GG}
         +{\chi_{\sss Z}} \Big[ \lpar 1+\beta_f \rpar 
    \Big( 2\tcie \tcit F_{\sss LL}+\delta_e \tcit  F_{\sss QL} \Big)
\nll &
+\delta_f\lpar 2 \tcie F_{\sss LQ}+\delta_e  F_{\sss QQ} \rpar \Big] \bigg),
\nll
{\cal A}_{-+-+} = &  s \lpar 1-\cos\vartheta \rpar 
   \bigg(
  \qe\qf F_{\sss GG}
+{\chi_{\sss Z}} \Big[\lpar 1-\beta_f\rpar \tcit
        \Big( 2\tcie    F_{\sss LL}+\delta_e F_{\sss QL}\Big)
\nll &
             +\delta_f \Big( 2 \tcie F_{\sss LQ}
             +\delta_e  F_{\sss QQ} \Big) \Big] \bigg),
\nll
{\cal A}_{--++}  = & 0,\qquad
{\cal A}_{--+-}  =   0,\qquad
{\cal A}_{---+}  =   0,\qquad
{\cal A}_{----}  =   0.
\end{align}

Here
\bqa
\cos\vartheta &=& \lpar t-\mfs+\frac{s}{2} \rpar \frac{2}{s\beta_f}\,,
\eqa
and  for the amplitude 
$ {\cal A}_{ \lambda_{i} \lambda_{j} \lambda_{k} \lambda_{l} }  $
each index $ \lambda_{(i,j,k,l)}$ 
takes two values ($\pm=\pm 1$) meaning 2 times the projection of spins 
$e^+, e^-, t, \bar{t}$ onto their corresponding momentum.
The differential cross-section for the unpolarized case is:
\bqa
\frac{d\sigma}{d\cos\vartheta} = \,\frac{\pi \alpha^2}{s^3}\beta_f N_c \,
 \sum_{ \lambda_{i} \lambda_{j} \lambda_{k} \lambda_{l}}
\left|
{\cal A}_{ \lambda_{i} \lambda_{j} \lambda_{k} \lambda_{l} }  
\right|^2.
\label{dsigma_A}
\eqa

We checked that this expression is analytically identical to \eqn{differential-cs}.
Both expressions, \eqn{differential-cs} and \eqn{dsigma_A}, contains, however, spurious
contributions of the two-loop order (squares of one-loop terms), which should be 
suppressed, since we would like to have 
a complete one-loop result. 

 This may be achieved with a simple trick.
 First of all, let us note that if all form factors are: $ F_{\sss {IJ}} = 1 $ 
for ${IJ} = LL,\,LQ,\,QL,\,QQ$ and $F_{\sss {IJ}} = 0$ for ${IJ} = LD,\,QD$, 
we have the tree level.
 At the one-loop level $LL,\,LQ,\,QL,\,QQ$ form factors may be represented as:
\bqa 
{\bf F}_{\sss {IJ}} 
= 1 + \frac{\alpha}{4 \pi \siw^2} F_{\sss {IJ}}\,, 
\label{one_ll}
\eqa
and
\bqa 
{ \bf F}_{\sss {IJ}} 
= \frac{\alpha}{4 \pi \siw^2} F_{\sss {IJ}}\,, 
\label{one_llD}
\eqa
for ${IJ} = LD,\,QD$.

Instead of \eqn{one_ll} for the four form factors we write 
\bqa 
{ \bf F}_{\sss {IJ}} 
= Z + \frac{\alpha}{4 \pi \siw^2} F_{\sss {IJ}}\,,
\label{Pl_z}
\eqa
and note that the cross section is a function of six form factors.
 
 Then the one-loop result is apparently equal:
\bqa
\frac{{d\sigma}^{(1)}}{d\cos\vartheta} = 
 \frac{d\sigma}{d\cos\vartheta}[Z=1] - \frac{d\sigma}{d\cos\vartheta}[Z=0]. 
\label{oloop}
\eqa

\section{Annex}
\eqnzero

\subsection{QED vertices and soft-photon contributions}
Here we present virtual corrections due to QED vertices, a factorized part due to QED boxes
and soft-photon contributions. 
The expressions in this subsection can be also cast from \cite{Bardin:1999ak}.

The formal structure of factorized virtual and soft contributions is as follows:
\bqa
\delta^{{\rm virt+soft}} = 
\label{dlt_vs} 
\frac{\alpha}{\pi} 
 \bigg[ \qe^2 \delta^{{\rm virt+soft}}_{\sss{\rm ISR}} 
      + \qe\qf\delta^{{\rm virt+soft}}_{\sss{\rm IFI}} 
      + \qf^2 \delta^{{\rm virt+soft}}_{\sss{\rm FSR}} \bigg].
\eqa
There are thus three types of contributions: ISR, FSR and IFI.
\subsubsection{Initial-state radiation (ISR)}
The contributions of the initial-state QED $e^+e^-\gamma$ vertex and ISR soft photon
contribution are short, since the electron mass is ignored:
\bqa
\delta^{\rm virt}_{\sss{\rm ISR}} &=& 
            - \ln \frac{\mes}{\tHlas} 
                    \bigg( \lelog   - 1 \bigg)
            - \frac{1}{2}  \lelog^2 
            + \frac{3}{2}  \lelog   - 2 + 4 \Litwo\lpar 1 \rpar,
\nll
\delta^{\rm soft}_{\sss{\rm ISR}}    &=& 
\ln\lpar \frac{4 \omega^2}{s} \frac{\mes}{\tHlas} \rpar 
             \bigg(  \lelog - 1 \bigg)
       + \frac{1}{2} \lelog^2  -  2 \Litwo\lpar 1 \rpar,
\label{isr_virt}
\eqa
where
\begin{equation}
l_e  = \ln \lpar \frac{s}{\mes} \rpar.
\end{equation}
\subsubsection{Initial--final state interference (IFI)} 
This originates from contributions of QED boxes: \,$\gamma\gamma, \zb\gamma$ and 
initial--final state soft photons interference:
\bqa      
\delta^{\rm virt}_{\sss{\rm IFI}} 
\label{dltIFI}
&=& - 2 \ln \frac{s}{\tHlas} \ln \frac{\tmi}{\umi},
\\
\delta^{\rm soft}_{\sss{\rm IFI}} &=& 2 \ln \frac{4\omega^2}{\tHlas} \ln\frac{\tmi}{\umi}
\label{dltsft}
                  +\lrbr F^{\rm soft} \lpar s,t \rpar - F^{\rm soft}\lpar s,u \rpar \rrbr,
\eqa  
with
\begin{align}
F^{\rm soft} \lpar s,t \rpar =& 
         - \frac{1}{2} \lelog^2
         - \frac{1}{2} \ln^2\eta
         +2  \ln\eta
             \ln  \lpar 1+\frac{2 \mfs}{\betatp\tmi}\rpar
\\&
         -   \ln^2\lpar 1+\frac{2 \mfs}{\betatp\tmi}\rpar
         +   \ln^2\lpar-\frac{s t}{\tmis}\rpar
         +2  \ln\lpar-\frac{s t}{\tmis}\rpar \ln \lpar 1+\frac{\tmis}{s t}\rpar
\nll &
         +2  \Litwo\lpar 1 - \frac{2 t}{\tmi\betatp} \rpar
         -2  \Litwo\lpar\frac{-\betatm\tmi}{\betatp\tmi+2\mfs} \rpar
\nll &
         -2  \Litwo\lpar-\frac{\tmis}{ s t}\rpar
         -2  \Litwo\lpar 1 \rpar,
\nonumber
\end{align}
where we introduce the notation:
\bqa
\beta \equiv \beta_f&=&\sqrt{1-\frac{4 \mfs}{s}}\;,
\nll
\betatp&=& 1+\beta\,,
\nll
\betatm&=& 1-\beta\,,
\nll
\qquad\eta &=&\frac{\betatm}{\betatp}\;.
\eqa
\subsubsection{Final-state radiation (FSR)}
The contributions of one-loop QED $ f \bar f \gamma$ vertex and 
final-state soft photon radiation are: 
\begin{align}
\delta^{\rm virt}_{\rm{\sss FSR}}
        =&-\ln\frac{\mfs}{\tHlas} \bigg[-\bebeta \ln\eta-1\bigg]
                           -\frac{3}{2}\beta\ln\eta-2
\nll &
       +\bebeta \bigg[-\frac{1}{2} \ln^2\eta
                + 2 \ln\eta \ln\lpar 1-\eta\rpar
                + 2 \Litwo \lpar\eta\rpar + 4 \Litwo\lpar 1 \rpar \bigg],
\nll
\delta^{\rm soft}_{\rm{\sss FSR}}
    =& \ln\frac{4\omega^2}{\tHlas}\bigg[-\bebeta\ln\eta-1 \bigg]
        -\frac{1}{\beta} \ln\eta
\label{penka}
\nll &
       +\bebeta \bigg[-\frac{1}{2} \ln^2\eta
                + 2 \ln\eta \ln\lpar 1-\eta\rpar
               + 2 \Litwo\lpar\eta\rpar-2 \Litwo\lpar 1 \rpar\bigg].
\end{align}
 The contribution of the ISR given in~\eqn{isr_virt} may be obtaineded 
from these expressions
in the limit $\mfl=m_e \to 0$.

\subsubsection{Non-factorized final-state vertex `anomalous' contributions}

 To present this contribution let us introduce the definition
\bqa
 L_n=\ln\frac{\beta-1}{\beta+1}\,.
\label{Leta}
\eqa
The `anomalous' part of the QED vertex contribution to the differential cross-section
reads:
\begin{align}
\frac{d\sigma^a}{d\cos\vartheta} 
 = \hspace*{1mm}&
  4 \alpha^3 N_c \frac{\mfs}{s^4} \qfs \Big[
          \Big( \qes \qfs + 2 \qe\qf \ve \vf {\rm Re}({\chi_{\sss Z}})
       + \lpar v^2_e + a^2_e \rpar v^2_t  |{\chi_{\sss Z}}|^2 \Big)
\nll &
  \times \lpar s t+\tmi^2 \rpar {\rm Re}(L_n)
       + \qe \qf \aee \af s \lpar s + 2 \tmi \rpar {\rm Re}(L_n {\chi_{\sss Z}})
\nll &
       + \Big(
         \lpar v^2_e + a^2_e  \rpar  a^2_t
         \Big[  s \lpar s-4\mfs \rpar 
 +2\lpar  s t+\tmi^2 \rpar \Big]
\nll &
 + 2 \ve \aee \vf \af s \lpar s + 2 \tmi\rpar  \Big) 
|{\chi_{\sss Z}}|^2 {\rm Re}(L_n) \Big].
\end{align}
\subsection{An alternative form of the cross-section for QED boxes}
Here we present some useful formulae, which are not in the main stream of our approach
(described in the previous sections), but were used for internal cross-checks of the calculations
of the QED part of the process under consideration.

The QED boxes, \eqnsc{BoxBorngamma}{BoxBornZ}, may be greatly simplified purely algebraically
if the cross-sections are calculated analytically.
For the sum of $AA$ and $ZA$ boxes one may easily derive the cross-section:
\begin{align}
\label{alernativeQED}
\frac{d\sigma^{\sss{\rm BOX}} }{d\cos\vartheta} 
= \hspace*{1mm} & \frac{2\alpha^3}{s}\beta \qe\qf N_c \, {\rm Re} \,
  \biggl\{
                \qe^2 \qf^2 \, {\cal F}_{\sss V}                           
\\ &
  +\qe \qf \, {\chi_{\sss Z}} \bigg[\ve \vf \left({\cal F}^*_{\sss V} +{\cal H}_{\sss V} \right) 
  +                      \aee \af \left({\cal F}^*_{\sss A} +{\cal G}_{\sss V} \right) \bigg]  
\nll & 
  +|{\chi_{\sss Z}}|^2    \bigg[
\left( \ve^2 + \aee^2 \right) \lpar \,\vf^2 {\cal H}_{\sss V} + \af^2  {\cal H}_{\sss A} \rpar  
  +2 \aee \ve \af \vf \, \lpar {\cal G}_{\sss V} + {\cal G}_{\sss A} \rpar 
                          \bigg] \biggr\},
\nonumber
\end{align}
where $\chi_{\sss{Z}}(\sman)$ is defined by \eqn{propagators} and the six {\em cross-section
form factors} are:
\begin{align}
{\cal F}_{\sss V} =& {\cal F}_{\sss V} \lpar t \rpar - {\cal F}_{\sss V} \lpar u \rpar,
\nll
{\cal F}_{\sss A} =& {\cal F}_{\sss A} \lpar t \rpar + {\cal F}_{\sss A} \lpar u \rpar,
\nll 
{\cal H}_{\sss V} =& {\cal H}_{\sss V} \lpar t \rpar - {\cal H}_{\sss V} \lpar u \rpar,
\nll
{\cal H}_{\sss A} =& {\cal H}_{\sss A} \lpar t \rpar - {\cal H}_{\sss A} \lpar u \rpar,
\nll 
{\cal G}_{\sss V} =& {\cal G}_{\sss V} \lpar t \rpar + {\cal G}_{\sss V} \lpar u \rpar,
\nll
{\cal G}_{\sss A} =& {\cal G}_{\sss A} \lpar t \rpar + {\cal G}_{\sss A} \lpar u \rpar,
\end{align}
with
\begin{align}
{\cal F}_{\sss V} \lpar t \rpar =& 
\frac{1}{s} \biggl\{
        \frac{\tmi}{4} \bigg[ 2 \mfs+\lpar s + 2\tmi \rpar\bigg] \jaat
\nll &
+\frac{st}{2}\cesoeo 
\nll & 
+\tman\biggl(\frac{s-4\mfs}{2} +\frac{4\mtq}{\sdtit}\biggr)\ctsoto 
\nll &
   +\frac{2\tman\mfs}{\sdtit} \bigg[\boftto-\bofsoo\bigg]  
\nll &
      - \frac{ s \mfs}{\tmi}  \bigg[\boftet-\boftto\bigg]
\nll &
      - \frac{\lpar s+\tmi \rpar }{2} \bigg[\boftet-\bofsoo\bigg] 
    \bigg\},
\\[1mm]
{\cal F}_{\sss A} \lpar t \rpar =& \frac{1}{s} \bigg\{ 
        \frac{ s + 2\tmi}{4} \tmi \jaat 
\nll &
- \mfs  \bigg(\frac{1}{2} s \ctsoto
\nll &
  +\lpar \frac{s}{\tmi}+1 \rpar \bigg[ \boftet-\boftto \bigg] \bigg)
\nll &
      -\frac{\lpar s+\tmi \rpar}{2} \bigg[ \boftet-\bofsoo \bigg] \bigg\},
\end{align}
\begin{align}
{\cal H}_0 \lpar t \rpar =& 
\frac{1}{s^2}\bigg\{ 
   \lpar \tmi^2+\lpar s+\tmi \rpar^2 \rpar \frac{\tmi}{2}\jaat
\nll &
 + \tmi\biggl[\frac{1}{2}\szmi \lpar \szpl + 2 \tman \rpar
 -  \lpar \tmi^2+\lpar s+\tmi \rpar^2 \rpar \biggr] \jazt 
\nll &
 + \biggl( \frac{s \szmi \mfs}{\tmi}\mzs 
 + \tmi\Bigl[\szmi\lpar \szpl + 2 t \rpar 
 -  \lpar \tmi^2+\lpar s+\tmi \rpar^2 \rpar\Bigr]\biggr)
\nll[-1mm] &\hspace{5cm}\times\cattze
\nll &
 +\szmi\biggl( s t\, \ceszeo 
\nll &
       + s t\, \ctszto 
\nll &
       - \frac{s \mfs}{\tmi}\Big[ 2 \boftet - \boftto 
\nll &\hspace{1.7cm} - \boftzt \Big]
\nll &
       -  \lpar s+\tmi \rpar \Big[\boftet - \bofszo \Big] \biggr)
\bigg\},
\end{align}
\begin{align}
{\cal H}_{\sss V} \lpar t \rpar =& {\cal H}_0 \lpar t \rpar 
 -{\ds \frac{2\tmi\mfs}{\sman}}
\bigg[\jazt - {\ds\frac{1}{2}}\jaat
\nll &\hspace{9mm}
 + \cattze \bigg]
\nll &\hspace{9mm}
 -{\ds \frac{2\mfs\szmi t}{s^2}}\biggl(
  \biggl( 1- {\ds\frac{ \szmi }{\sdtit}} \biggr) \ctszto
\nll &\hspace{9mm}
 -{\ds\frac{1}{\sdtit}} \Big[ 2 \bofszo   - \boftto 
\nll &\hspace{9mm}
 -\boftzt\Big] \bigg)
\bigg\},
\\[1mm]
{\cal H}_{\sss A} \lpar t \rpar =& {\cal H}_0 \lpar t \rpar 
 +{\ds\frac{2 \mfs}{s^2}} 
\bigg\{ \tmi \bigg[\mzs\jazt
\nll &\hspace{9mm}
         - \sman {\ds\frac{1}{2}}\jaat\bigg]     
\nll &\hspace{9mm}
+ \bigl(\szmi \szpl +\tmi\sman \bigr) \cattze
\nll &\hspace{9mm}
         + \szmi \bigg[-t \ctszto 
\nll &\hspace{9mm}
         - \boftet+\bofszo\bigg] 
\bigg\},
\end{align}
\begin{align}
{\cal G}_{\sss V} \lpar t \rpar =&
- \frac{1}{s} \bigg\{- \frac{1}{2}\lpar s + 2\tmi \rpar \tmi \jaat 
\nll &
+ \tmi \biggl[\sman + 2\tmi   
- \frac{\szmi}{2s} \lpar \szpl +  2\tmi \rpar\biggr]\jazt
\nll &
+ \biggl[ \lpar s + 2\tmi \rpar \tmi 
- \frac{\szmi\tmi}{s} \lpar \szpl +  2\tmi \rpar 
 -\szmi \mfs \bigg(2+ \frac{\mzs }{\tmi} \bigg)\biggr]
\nll &\hspace{4cm}\times\cattze
\nll &
- \szmi 
\bigg(  \frac{1}{2} \mzs \bigg[ \ceszeo
\nll &\hspace{4cm}
+\ctszto \bigg]
\nll &
+ \frac{\mfs}{\tmi} \bigg[ -2 \boftet + \boftto
\nll &\hspace{1.2cm}
 +\boftzt\bigg]
\nll &
        - \frac{\lpar s+\tmi \rpar}{s} \bigg[\boftet-\bofszo\bigg]
\bigg)
\bigg\},
\end{align}
\begin{align}
\label{lffn}
{\cal G}_{\sss A} \lpar t \rpar =& {\cal G}_{\sss V} \lpar t \rpar -
        \mfs \frac{\szmi}{s^2}    
  \bigg[ 2 \szmi \cattze 
\\ &
+ \szpl \ctszto 
\nll &
+ 2 \boftet - \boftzt - \boftto \bigg].
\nonumber
\end{align}
Note the cancellation of $\sdfit$, leading to a great simplification of 
the expressions.
Equations (\ref{alernativeQED}--\ref{lffn}) were coded as a separate branch 
of {\tt eeffLib} and, together
with the vertex QED contributions described in the previous subsections, 
were used for an internal cross-check of the QED part of the calculations.

 Some factorized part of the $AA$ and $ZA$ boxes contribution is not included 
in~\eqn{alernativeQED}.
It has the form
\bqa
\frac{d\sigma}{d\cos\vartheta} \frac{\alpha}{\pi}\qe\qf \delta^{\rm virt}_{\sss{\rm IFI}},
\eqa
where $\delta^{\rm virt}_{\sss{\rm IFI}} $ is given by ~\eqn{dltIFI}.

 The whole QED contribution can be written as follows
\bqa
 \frac{d\sigma^{\sss {\rm QED}}}{d\cos\vartheta}=
 \frac{d\sigma^{\sss{\rm BORN}}}{d\cos\vartheta}\delta^{\rm virt+soft}
+\frac{d\sigma^a}               {d\cos\vartheta}
+\frac{d\sigma^{\sss{\rm BOX}} }{d\cos\vartheta}\,, 
\eqa
where $\delta^{\rm virt+soft} $ is defined by \eqns{dlt_vs}{penka}. 

\section{Numerical results and discussion}
\eqnzero

All the formulae derived in this article are realized in the {\tt FORTRAN} code with 
the tentative name {\tt eeffLib}. Numbers presented in this section are produced with 
the February 2002 version of the code. 
As compared to the December 2000 version, used to produce numbers for Ref.~\cite{eett_subm}, 
the current version contains full QED corrections together with the soft-photon contribution 
to the angular distribution $d\sigma/d\cos\vartheta$. 
In this section we present several examples of numerical results.
In particular, we will show
a comparison of the electroweak form factors (EWFF) {\em including} QED corrections between 
{\tt eeffLib} and another {\tt FORTRAN} code, which was automatically generated 
from {\tt form} log files with the aid of system {\tt s2n.f} ({\it symbols to numbers}), 
producing a {\tt FORTRAN} source code --- a part of our {\tt SANC} system. 
This comparison provides a powerful internal cross-check of our numerics that practically excludes 
the appearance of bugs in numerical results.

We begin with showing several examples of comparison with 
{\tt ZFITTER v6.30}~\cite{zfitterv6.30:2000}. In the present realization,
{\tt eeffLib} does not calculate $\mwl$ from $\mu$ decay and does not precompute either
Sirlin's parameter $\Delta r$ or the total $\zb$ width, which enters the $\zb$ boson propagator.
For this reason, the three parameters: $\mwl\,,\;\Delta r\,,\;\gz$ were being taken from
{\tt ZFITTER} and used as {\tt INPUT} for {\tt eeffLib}. Moreover, present {\tt eeffLib}
is a purely one-loop code, while  it was not foreseen in {\tt ZFITTER}, to access just one-loop
form factors with the aid of users flags. To accomplish the goals of comparison at the one-loop 
level, we had to modify the {\tt DIZET} electroweak library. The most important change 
was an addition to the {\tt SUBROUTINE ROKANC}:

\begin{verbatim}
*
* For eett
*  
      FLL=(XROK(1)-1D0+DR )*R1/AL4PI
      FQL=FLL+(XROK(2)-1D0)*R1/AL4PI
      FLQ=FLL+(XROK(3)-1D0)*R1/AL4PI
      FQQ=FLL+(XROK(4)-1D0)*R1/AL4PI 
\end{verbatim}
with the aid of which we reconstruct four scalar form factors from {\tt ZFITTER}'s effective
couplings $\rho$ and $\kappa$'s ($F_{\sss{LD}}$ and $F_{\sss{QD}}$ do not contribute 
in the massless approximation).
\subsection{Flags of {\tt eeffLib}\label{t-flaggs}}
Here we give a description of flags (user options) of {\tt eeffLib}. While creating 
the code, we followed the  principle to preserve as often as possible the meaning of flags as 
described in the {\tt ZFITTER} description~\cite{Bardin:1999yd}. In the list below, 
the comment `{\tt as in ZFD}' means that the flag has exactly the same meaning as in
~\cite{Bardin:1999yd}. Here we describe the set of flags of the February 2002 version
of {\tt eeffLib}.
\begin{itemize}
\item \verb+ALEM=3  ! as in ZFD +
\vspace*{-2.5mm}

\item \verb+ALE2=3  ! as in ZFD +
\vspace*{-2.5mm}

\item \verb+VPOL=0  ! =0 + $\alpha$(0); =1,=2 as in ZFD; =3 is reserved for later use \\
Note that the flag is extended to {\tt VPOL=0} to allow calculations `without running of 
$\alpha$'.
\vspace*{-2.5mm}

\item \verb+QCDC=0  ! as in ZFD +
\vspace*{-2.5mm}

\item \verb+ITOP=1  ! as in DIZET (internal flag) +
\vspace*{-2.5mm}

\item \verb+GAMS=1  ! as in ZFD +
\vspace*{-2.5mm}

\item \verb+WEAK=1  ! as in ZFD + (use {\tt WEAK=2} in v6.30 to drop out some higher-order 
terms)
\vspace*{-2.5mm}

\item \verb+IMOMS=1 ! =0 +  $\alpha$-scheme; =1 GFermi-scheme  \\
New meaning of an old flag: switches between two renormalization schemes.
\vspace*{-2.5mm}

\item \verb+BOXD=6 + \\
Together with {\tt WEAK=0} is used for an internal comparison 
of separate boxes and QED contributions: \\
      \verb+BOXD    ! =1 + with $AA$ boxes \\
      \verb+        ! =2 + with $ZA$ boxes \\
      \verb+        ! =3 + with $AA$ and $ZA$ boxes \\ 
      \verb+        ! =4 + with all QED contributions       \\
Together with {\tt WEAK=1} (working option), it has somewhat different meanings: \\
      \verb+BOXD    ! =0 + without any boxes   \\
      \verb+        ! =1 + with $AA$ boxes \\
      \verb+        ! =2 + with $ZA$ boxes \\
      \verb+        ! =3 + with $AA$ and $ZA$ boxes \\ 
      \verb+        ! =4 + with $WW$ boxes \\
      \verb+        ! =5 + with $WW$ and $ZZ$ boxes \\
      \verb+        ! =6 + with all QED and EW boxes 
\vspace*{+2mm}

{\bf `Treatment' options.}

\item \verb+GAMZTR=1+ treatment of $\Gamma_{\sss{Z}}$. \\
The option is implemented for the sake of comparison with 
{\tt FeynArts}: \\
      \verb+GAMZTR=0 + $\Gamma_{\sss{Z}} =  0 $ \\
      \verb+GAMZTR=1 + $\Gamma_{\sss{Z}}\ne 0 $
\vspace*{-2.5mm}

\item \verb+EWFFTR=0+ treatment of EW form factors. \\
Switches between form factors and 
effective {\tt ZFITTER} couplings $\rho$ and $\kappa$'s. The option is implemented for
comparison with {\tt ZFITTER}: \\
      \verb+EWFFTR=0 + electroweak form factors \\
      \verb+EWFFTR=1 + effective couplings $\rho$ and $\kappa$'s
\vspace*{-2.5mm}

\item \verb+FERMTR=1+ treatment of fermionic masses\\
Switches between three different
sets of `effective quark masses':\\
      \verb+FERMTR=1 + a `standard' set of fermions masses\\
      \verb+FERMTR=2,3+  `modified' sets 
\vspace*{-2.5mm}

\item \verb+VPOLTR=1+ treatment of photonic vacuum polarization \\ 
Switches between lowest-order expression $\alpha(s)=\alpha\lrbr 1+\Delta\alpha(s)\rrbr$ and its 
`resummed' version, $\alpha(s)=\alpha/\lrbr 1-\Delta\alpha(s)\rrbr$:\\
      \verb+VPOLTR=0 + lowest order\\
      \verb+VPOLTR=1 + resummed
\vspace*{-2.5mm}
 
\item \verb+EWRCTR=2+ treatment of electroweak radiative corrections\\ 
Switches between three variants for vertex corrections:\\
      \verb+EWRCTR=0 + electroweak form factors contain only QED additions  \\
      \verb+EWRCTR=1 + electroweak form factors do not contain QED additions\\ 
      \verb+EWRCTR=2 + electroweak form factors contain both QED and EW additions
\vspace*{-2.5mm}

\item \verb+EMASTR=0+ treatment of terms with $\ln(s/\mes)$ in $AA$ and $ZA$ boxes,
which are present in various functions but cancel in sum:\\
      \verb+EMASTR=0 + these terms are suppressed in all functions they enter \\
      \verb+EMASTR=1 + these terms are retained in all functions, which results in loss
of computer precision owing to numerical cancellation; results for \verb+EMASTR=0+ and 
\verb+EMASTR=1+ are equal 
\vspace*{-2.5mm}

\item \verb+EWWFFV=1+ treatment of vertex and box diagrams with virtual $\wb$ boson,
switches between two variants:\\
      \verb+EWWFFV=0 + variant of formulae without $b$-quark mass\\
      \verb+EWWFFV=1 + variant of formulae with finite $b$-quark mass
\vspace*{+2mm}

{\bf Options affecting QED contributions.}

\item \verb+IQED=4 + variants of inclusion virtual and soft photon QED contributions:\\
      \verb+IQED=1 + only initial-state radiation (ISR)     \\
      \verb+IQED=2 + only initial--final interference (IFI) \\
      \verb+IQED=3 + only ~final-state radiation (FSR)     \\
      \verb+IQED=4 + all QED contributions are included     
\vspace*{-2.5mm}

\item \verb+IBOX=4 + is active only if {\tt IQED}=2 or 4 and affects only Eq.~(5.11): \\
      \verb+IBOX=0 + $AA$ boxes interfering with $\ph$ exchange BORN \\
      \verb+IBOX=1 + $AA$ boxes \\
      \verb+IBOX=2 + $ZA$ boxes \\
      \verb+IBOX=3 or 4 + $AA$+$ZA$ boxes 
\end{itemize}

\subsection{{\tt eeffLib}--{\tt ZFITTER} comparison of scalar form factors}
First of all we discuss the results of a computation of the complete EW part of the
four scalar form factors (i.e. with $WW$ and $ZZ$ boxes),
\bqa
\vvertil{}{\sss{LL}}{s,t},\quad
\vvertil{}{\sss{QL}}{s,t},\quad
\vvertil{}{\sss{LQ}}{s,t},\quad
\vvertil{}{\sss{QQ}}{s,t},
\eqa
for three channels: $\fep\fem\to u\bar{u},\;d\bar{d}$ and $b\bar{b}$
for light final fermion masses (we set $m_u=m_d=0.1$ GeV) and for $b\bar{b}$ channel we use
the formulae in the limit $m_f=0$). 
We remind that {\tt ZFITTER} is able to deliver only massless results.


\begin{table}[h]
\caption[EWFF for the process $\fep\fem\to\ff\bar{f}$. {\tt eeffLib}--{\tt ZFITTER} comparison.]
{EWFF for the process $\fep\fem\to\ff\bar{f}$. {\tt eeffLib}--{\tt ZFITTER} comparison.
\label{table1}}
\vspace*{2mm}
\begin{tabular} {||c||l|l|l||}
\hline
{$\sqrt s$}&~~~~~~~~100 GeV   &~~~~~~~~200 GeV  &~~~~~~~~300 GeV \\
\hline
FF& & & \\
\hline
\hline
\multicolumn{4}{||c||}{$u\bar{u}$ channel, $m_u =0.1$ GeV, $m_d=0$}             \\
\hline
$F_{\sss{LL}}$&$12.89583~-~i1.84786$&$~8.24737~-i10.64653$&$~8.98371~-i12.88466$\\
{\tt ZF}      &$12.89583~-~i1.84786$&$~8.24736~-i10.64651$&$~8.98370~-i12.88466$\\
\hline
$F_{\sss{QL}}$&$29.30446~+~i3.67330$&$29.38217~+~i2.27613$&$31.59711~+~i1.59304$\\
{\tt ZF}      &$29.30445~+~i3.67330$&$29.38216~+~i2.27613$&$31.59710~+~i1.59304$\\
\hline
$F_{\sss{LQ}}$&$29.10831~+~i3.26973$&$29.48511~+~i0.92311$&$31.65835~-~i0.89711$\\
{\tt ZF}      &$29.10832~+~i3.26973$&$29.48512~+~i0.92312$&$31.65835~-~i0.89711$\\
\hline
$F_{\sss{QQ}}$&$44.88228~+~i8.85688$&$43.31854~+~i9.48286$&$44.18773~+i10.25197$\\
{\tt ZF}      &$44.88228~+~i8.85688$&$43.31854~+~i9.48286$&$44.18773~+i10.25196$\\
\hline
\hline
\multicolumn{4}{||c||}{$d\bar{d}$ channel, $m_d =0.1$ GeV, $m_u=0$}             \\
\hline
$F_{\sss{LL}}$&$13.70781~-~i1.51002$&$15.18630~-~i3.93706$&$~8.86000~-~i1.80409$\\
{\tt ZF}      &$13.70781~-~i1.51002$&$15.18629~-i~3.93706$&$~8.86000~-~i1.80409$\\
\hline
$F_{\sss{QL}}$&$29.64340~+~i4.12394$&$31.96819~+~i6.97877$&$31.69945~+~i8.03876$\\
{\tt ZF}      &$29.64340~+~i4.12394$&$31.96818~+~i6.97877$&$31.69944~+~i8.03876$\\
\hline
$F_{\sss{LQ}}$&$29.12112~+~i3.22780$&$30.28990~+~i1.73736$&$31.69321~-~i0.07000$\\
{\tt ZF}      &$29.12113~+~i3.22780$&$30.28990~+~i1.73733$&$31.69323~-~i0.07001$\\
\hline
$F_{\sss{QQ}}$&$44.87608~+~i8.79014$&$43.94906~+~i9.86523$&$44.19463~+i10.38253$\\
{\tt ZF}      &$44.87609~+~i8.79014$&$43.94905~+~i9.48286$&$44.19462~+i10.38251$\\
\hline
\hline
\multicolumn{4}{||c||}{$b\bar{b}$ channel, $m_b =0$, $m_t=173.8$ GeV}           \\
\hline
$F_{\sss{LL}}$&$11.16367~-~i0.65244$&$14.68726~-~i1.85834$&$11.26568~-~i3.38026$\\
{\tt ZF}      &$11.16367~-~i0.65244$&$14.68726~-~i1.85834$&$11.26568~-~i3.38026$\\
\hline
$F_{\sss{QL}}$&$26.88634~+~i4.76865$&$28.08780~+~i5.27877$&$31.02060~+~i3.91962$\\
{\tt ZF}      &$26.88634~+~i4.76865$&$28.08780~+~i5.27877$&$31.02060~+~i3.91962$\\
\hline
$F_{\sss{LQ}}$&$29.12113~+~i3.22780$&$30.28990~+~i1.73733$&$31.69323~-~i0.07001$\\
{\tt ZF}      &$29.12113~+~i3.22780$&$30.28990~+~i1.73733$&$31.69323~-~i0.07001$\\
\hline
$F_{\sss{QQ}}$&$44.87609~+~i8.79014$&$43.94905~+~i9.86523$&$44.19462~+i10.38251$\\
{\tt ZF}      &$44.87609~+~i8.79014$&$43.94905~+~i9.86523$&$44.19462~+i10.38251$\\
\hline
\hline
\end{tabular}
\end{table}

In this comparison we use flags as in subsection \ref{t-flaggs} and, moreover,
\bqa
\mwl    &=&80.4514958\;\mbox{GeV},
\nl
\Delta r&=&0.0284190602\,,
\nl
\Gamma_{\sss Z}&=&2.499 776\;\mbox{GeV}.
\label{firstPOs}
\eqa
The form factors are shown as complex numbers for the three c.m.s. energies 
(for $\tman = \mts-\sman/2$) and for $\tHs=\wml$\footnote{In the preprints~\cite{eett_subm}
and~\cite{part2:2001} we presented this and the following tables for the three values of 
scale $\tHs=\wml/10,\;\wml,\;10\wml$ and demonstrated the scale independence of  {\tt eeffLib}
numbers.}.
\tbn{table1} shows very good agreement with {\tt ZFITTER} results (up to 6 or 7 digits agree).
It should be stressed that total agreement with {\tt ZFITTER} is not expected because 
in the {\tt eeffLib} for $\fep\fem\to u\bar{u}$ and $d\bar{d}$ channels
we use massive expressions to compute the nearly massless case.
Certain numerical cancellations leading to losing some numerical precision are expected.  
We should conclude that the agreement is very good and demonstrates that our 
formulae have the correct $\mtl\to 0$ limit. Note, that there is an all digits agreement for 
$\fep\fem\to b\bar{b}$ channel since in both cases one uses the formulae in the limit $\mf=0$. 
\subsection{{\tt eeffLib}--{\tt ZFITTER} comparison of IBA cross-section}
As the next step of the comparison of {\tt eeffLib} with calculations from the 
literature, we present a comparison of the IBA cross-section.

In \tbn{table4} we show the differential cross-section \eqn{differential-cs} in pb
for four leptonic channels and in \tbn{table4p} for three quarkonic channes ($b\bar{b}$ channel
being shown twice for massless and massive $b$ quarks)
for three values of $\cos\vartheta=-0.9,\,0,\,+0.9$ with running e.m. coupling 
$\alpha\lpar\sman\rpar$.
Since the flag setting {\tt VPOL=1}, which is relevant to this case, affects the {\tt ZFITTER}
numbers, we now use, instead of \eqn{firstPOs}, the new {\tt INPUT} set:
\bqa
\mwl    &=&80.4467671\;\mbox{GeV},
\nl
\Delta r&=&0.0284495385\,,
\nl
\Gamma_{\sss Z}&=&2.499 538\;\mbox{GeV}.
\label{secondPOs}
\eqa
The numbers shown in first two rows of \tbnsc{table4}{table4p} 
exhibit a very good level of agreement with {\tt ZFITTER}.

Finally, in \tbn{table5}, we give a comparison of the cross-section integrated within  
the angular interval $|\cos\vartheta| \leq 0.999$. (Flag setting is the same as for Table 4.)

A typical deviation between {\tt eeffLib} and {\tt ZFITTER} is of the order $\sim 10^{-6}$, i.e. 
of the order of the required precision of the numerical integration over $\cos\vartheta$.
Examples of numbers obtained with {\tt eeffLib}, which were shown in this section, 
demonstrate that {\tt ZFITTER} numbers are recovered for light $\mtl$.
\clearpage


\begin{table}[t]
\caption[{\tt eeffLib}--{\tt ZFITTER} comparison of the differential cross-section,
leptonic channels.]
{Comparison of the differential EW cross-sections, [pb], leptonic channels.
First  row -- {\tt ZFITTER}, second row -- {\tt eeffLib}.}
\label{table4}
\vspace*{5mm}
\centering
\begin{tabular} {||c|l|l|l|l|l|l||}
\hline
\hline
${\sqrt s}$    &  100 GeV  & 200 GeV  & 300 GeV  & 400 GeV  & 700 GeV  & 1000 GeV \\
\hline
$\cos\vartheta$&           &          &          &          &          &          \\
\hline
\hline
\multicolumn{7}{||c||}{$\nu\bar{\nu}$ channel, $m_\nu=0$}                         \\
\hline
$-0.9$         &  49.100086& 0.579630 & 0.198147 & 0.101625 & 0.030297 & 0.014381 \\
               &  49.100085& 0.579630 & 0.198147 & 0.101625 & 0.030297 & 0.014381 \\
\hline
$0$            &  31.834491& 0.358941 & 0.122362 & 0.062455 & 0.018431 & 0.008648 \\
               &  31.834490& 0.358941 & 0.122362 & 0.062455 & 0.018431 & 0.008648 \\
\hline
$0.9$          &  66.267213& 0.739640 & 0.234000 & 0.115572 & 0.032285 & 0.014601 \\
               &  66.267213& 0.739640 & 0.234000 & 0.115572 & 0.032285 & 0.014601 \\
\hline
\hline
\multicolumn{7}{||c||}{$\fem\fep$ channel, $m_e=0$}                               \\
\hline
$-0.9$         &   7.991109& 0.533194 & 0.286463 & 0.169610 & 0.058032 & 0.028672 \\
               &   7.991110& 0.533194 & 0.286463 & 0.169609 & 0.058032 & 0.028672 \\
\hline
$0$            &  19.773604& 1.121915 & 0.474004 & 0.262276 & 0.084408 & 0.040933 \\
               &  19.773604& 1.121915 & 0.474004 & 0.262276 & 0.084408 & 0.040933 \\
\hline
$0.9$          &  63.501753& 3.504349 & 1.466665 & 0.811430 & 0.263378 & 0.128882 \\
               &  63.501750& 3.504348 & 1.466665 & 0.811430 & 0.263378 & 0.128882 \\
\hline
\hline
\multicolumn{7}{||c||}{$\mu^+\mu^-$ channel, $m_\mu =0.106$ GeV}                  \\
\hline
$-0.9$         &   7.991040& 0.533195 & 0.286463 & 0.169610 & 0.058032 & 0.028672 \\ 
               &   7.991028& 0.533195 & 0.286463 & 0.169610 & 0.058032 & 0.028672 \\
\hline
$0$            &  19.773498& 1.121915 & 0.474004 & 0.262276 & 0.084408 & 0.040933 \\
               &  19.773506& 1.121915 & 0.474004 & 0.262276 & 0.084408 & 0.040933 \\
\hline
$0.9$          &  63.501436& 3.504346 & 1.466665 & 0.811430 & 0.263378 & 0.128882 \\
               &  63.501422& 3.504345 & 1.466664 & 0.811430 & 0.263378 & 0.128882 \\
\hline
\hline
\multicolumn{7}{||c||}{$\tau^+\tau^-$ channel, $m_\tau =1.77705$ GeV}             \\
\hline
${\sqrt s}$    &  100 GeV  & 200 GeV  & 300 GeV  & 400 GeV  & 700 GeV  & 1000 GeV \\
\hline
$-0.9$         &   7.971611& 0.533509 & 0.286519 & 0.169627 & 0.058034 & 0.028673 \\
               &   7.968192& 0.533295 & 0.286477 & 0.169613 & 0.058032 & 0.028672 \\
\hline
$0$            &  19.743430& 1.121827 & 0.473992 & 0.262273 & 0.084407 & 0.040933 \\
               &  19.745968& 1.121978 & 0.474021 & 0.262282 & 0.084408 & 0.040933 \\
\hline
$0.9$          &  63.412131& 3.503720 & 1.466558 & 0.811398 & 0.263375 & 0.128882 \\
               &  63.408973& 3.503524 & 1.466520 & 0.811385 & 0.263373 & 0.128881 \\
\hline
\hline
\end{tabular}
\end{table}
\clearpage


\begin{table}[t]
\caption[{\tt eeffLib}--{\tt ZFITTER} comparison of the differential cross-section,
quarkonic channels.]
{Comparison of the differential EW cross-sections, [pb], quarkonic channels.
First  row -- {\tt ZFITTER}, second row -- {\tt eeffLib}.}
\label{table4p}
\vspace*{5mm}
\centering
\begin{tabular} {||c|l|l|l|l|l|l||}
\hline
\hline
${\sqrt s}$    &  100 GeV  & 200 GeV  & 300 GeV  & 400 GeV  & 700 GeV  & 1000 GeV \\
\hline
$\cos\vartheta$&           &          &          &          &          &          \\
\hline
\hline
\multicolumn{7}{||c||}{$u\bar{u}$ channel, $m_u=0.1$ GeV}                         \\
\hline
$-0.9$         &  45.404742& 0.386966 & 0.225923 & 0.138065 & 0.048621 & 0.024156 \\
               &  45.404602& 0.386966 & 0.225923 & 0.138065 & 0.048621 & 0.024156 \\
\hline
$0$            &  60.382423& 1.882835 & 0.771939 & 0.421410 & 0.133475 & 0.064245 \\
               &  60.382566& 1.882837 & 0.771939 & 0.421410 & 0.133475 & 0.064245 \\
\hline
               &  173.467517&6.450000 & 2.510881 & 1.346620 & 0.417295 & 0.198842 \\
$0.9$          &  173.467551&6.450000 & 2.510881 & 1.346620 & 0.417295 & 0.198842 \\
\hline
\hline
\multicolumn{7}{||c||}{$d\bar{d}$ channel, $m_d=0.1$ GeV}                          \\
\hline
$-0.9$         &  86.554414& 0.430807 & 0.136720 & 0.069644 & 0.020899 & 0.009978 \\
               &  86.554110& 0.430807 & 0.136720 & 0.069644 & 0.020899 & 0.009978 \\
\hline
$0$            &  72.820806& 1.180211 & 0.419410 & 0.219070 & 0.066312 & 0.031351 \\
               &  72.820908& 1.180212 & 0.419411 & 0.219070 & 0.066312 & 0.031351 \\
\hline            
$0.9$          & 176.717336& 3.770861 & 1.469194 & 0.796291 & 0.255118 & 0.124714 \\
               & 176.717376& 3.770861 & 1.469194 & 0.796291 & 0.255118 & 0.124714 \\
\hline
\hline
\multicolumn{7}{||c||}{$d\bar{d}$ channel, $m_b=0.1$ GeV}                         \\
\hline
$-0.9$         &  84.970331& 0.416155 & 0.136467 & 0.071144 & 0.020786 & 0.009868 \\
               &  84.970030& 0.416155 & 0.136467 & 0.071144 & 0.020786 & 0.009868 \\
\hline
$0$            &  71.762397& 1.185284 & 0.426009 & 0.219777 & 0.067063 & 0.031698 \\
               &  71.762496& 1.185285 & 0.426009 & 0.219777 & 0.067063 & 0.031698 \\
\hline
$0.9$          & 174.716111& 3.819650 & 1.442852 & 0.764945 & 0.249501 & 0.123086 \\
               & 174.716145& 3.819651 & 1.442852 & 0.764945 & 0.249501 & 0.123086 \\
\hline
\hline
\multicolumn{7}{||c||}{$d\bar{d}$ channel, $m_b=4.7$ GeV}                         \\
\hline
$-0.9$         &  84.350009& 0.416665 & 0.136590 & 0.071186 & 0.020792 & 0.009870 \\
               &  84.084519& 0.415686 & 0.136435 & 0.071137 & 0.020786 & 0.009868 \\
\hline
$0$            &  71.232935& 1.183490 & 0.425748 & 0.219707 & 0.067058 & 0.031697 \\
               &  71.248102& 1.184186 & 0.425882 & 0.219745 & 0.067061 & 0.031697 \\
\hline
$0.9$          & 173.420486& 3.812786 & 1.441766 & 0.764642 & 0.249478 & 0.123084 \\
               & 173.041374& 3.811860 & 1.441592 & 0.764577 & 0.249464 & 0.123077 \\
\hline
\end{tabular}
\end{table}
\clearpage


\begin{table}[h]
\caption[{\tt eeffLib}--{\tt ZFITTER} comparison of the total EW cross-sections.]
{Comparison of the total EW cross-sections, [pb].
First row --  {\tt ZFITTER}, second row -- {\tt eeffLib}.} 
\label{table5}
\vspace*{5mm}
\centering
\begin{tabular}{||c|c||c|c||c|c||}
\hline
\hline
\multicolumn{2}{||c||}{$100$ GeV}&\multicolumn{2}{|c||}{200 GeV}  
                                 &\multicolumn{2}{|c||}{300 GeV}                  \\
\hline
$\sigma_{\rm{tot}}$&$\sigma_{\sss{\rm{FB}}}$&$\sigma_{\rm{tot}}$&$\sigma_{\sss{\rm{FB}}}$
                                  &$\sigma_{\rm{tot}}$&$\sigma_{\sss{\rm{FB}}}$   \\
\hline
\multicolumn{6}{||c||}{$\nu\bar{\nu}$ channel, $m_\nu=0$}                         \\
\hline
  84.81710   &  9.509864   &  0.963362   &  0.089665  &   0.320985   &  0.021592  \\
  84.81710   &  9.509865   &  0.963362   &  0.089665  &   0.320985   &  0.021592  \\
\hline
\multicolumn{6}{||c||}{$\fem\fep$ channel, $m_e=0$}                               \\
\hline
  52.61662   &  30.78899   &  2.980668   &  1.654673  &   1.276008   &  0.648414  \\
  52.61662   &  30.78899   &  2.980667   &  1.654673  &   1.276008   &  0.648414  \\
\hline
\multicolumn{6}{||c||}{$\mu^+\mu^-$ channel, $m_\mu =0.106$ GeV}                  \\
\hline
  52.61634   &  30.78885   &  2.980667   &   1.654671  &  1.276008   &  0.648414  \\
  52.61634   &  30.78885   &  2.980667   &   1.654671  &  1.276008   &  0.648414  \\
\hline
\multicolumn{6}{||c||}{$\tau^+\tau^-$ channel, $m_\tau =1.77705$ GeV}             \\
\hline
  52.53632   &  30.75010   &  2.980435   &   1.654149  &  1.275972   &  0.648324  \\
  52.53661   &  30.75024   &  2.980443   &   1.654156  &  1.275974   &  0.648326  \\  
\hline
\multicolumn{6}{||c||}{$u\bar{u}$ channel, $m_u =0.1$ GeV}                        \\
\hline
  160.8980   &  70.98406   &  5.021808   &   3.360848  &  2.031754   &  1.269556  \\
  160.8981   &  70.98416   &  5.021810   &   3.360848  &  2.031754   &  1.269556  \\
\hline
\multicolumn{6}{||c||}{$d\bar{d}$ channel, $m_d =0.1$ GeV}                        \\
\hline
  193.7658   &  50.03208   &  3.120724   &   1.867871  &  1.149479   &  0.725581  \\
  193.7658   &  50.03227   &  3.120725   &   1.867871  &  1.149479   &  0.725581  \\
\hline
\multicolumn{6}{||c||}{$b\bar{b}$ channel, $m_b =0.1$ GeV}                        \\
\hline
  191.0416   &  49.76543   &   3.134547  &   1.892628  &  1.149243   &  0.720476  \\
  191.0416   &  49.76562   &   3.134547  &   1.892629  &  1.149243   &  0.720476  \\
\hline
\multicolumn{6}{||c||}{$b\bar{b}$ channel, $m_b =4.7$ GeV}                        \\
\hline
  189.6321   &  49.39098   &   3.129824  &   1.888530  &  1.148541   &  0.719805  \\
  189.3855   &  49.32791   &   3.129858  &   1.888542  &  1.148565   &  0.719802  \\
\hline
\end{tabular}
\end{table}
We conclude this subsection with a comment about the technical precision of our calculations.
We do not use the {\tt looptools} package~\cite{Looptools}.
For all PV functions but one, namely the $D_0$ function, we use our own coding, where we can
control precision internally and, typically, we can guarantee 11 digits precision.
For the $D_0$ function we use, instead, {\tt REAL*16 TOPAZ0} coding~\cite{Montagna:1999kp} and 
the only way that is accessible to us to control the precision is to compare our results with 
those computed with the {\tt looptools} package.
This was done for typical $D_0$ functions entering the $ZZ$ box contributions.
We obtained an agreement within 14 or 15 digits between these two versions for all 
$\sqrt{s}=400$--$10000$ GeV and $\cos\vartheta=0.99,\,0,\,-0.99$.
\clearpage 

\subsection{Comparison with a code generated by {\tt s2n.f}}
Here we present a numerical comparison of the complete scalar form factors of \eqn{soular_ff}, 
extracted from two independently created codes: the `manually written' {\tt eeffLib}
and a code `automatically generated' by the {\tt s2n.f} software.
We use a special input parameter set here: all lepton masses $\alpha$ and a conversion factor
from GeV$^{-1}$ to pb are taken from the 2000 Particle Data Tables while for quark and
photon and gauge boson masses we use:
\begin{align}
&m_{u,d,c,s,t,b} = 0.062,\;0.083,\;1.50,\;0.215,\;173.8,\;4.70\;\mbox{GeV},
\nl
&\lambda = 1\;\mbox{GeV},\;\mzl=91.1867\;\mbox{GeV},\;\mwl=80.4514958\;\mbox{GeV}.
\label{SIPS}
\end{align}

As seen from \tbn{t_com6}, the numbers agree within 11--13 digits, i.e. {\tt REAL*8} computational
precision is saturated.
The form factors $F_{\sss{LD,QD}}$ are multiplied by $10^4$ to make more digits visible.

\begin{table}[h]
\tiny
\caption[EWFF for the process $\fep\fem\to\ft\bar{t}$. 
Comparison of {\tt eeffLib}--{\tt s2n.f}.]
{EWFF for the process $\fep\fem\to\ft\bar{t}$. 
First row -- {\tt eeffLib}, second row -- {\tt s2n.f}. 
\label{t_com6}}
\vspace*{3mm}
\centering
\begin{tabular}{||c|c|c|c||}
\hline
\hline
\multicolumn{2}{||c|}{$\sqrt{s}$}& 400 GeV & 700 GeV \\
\hline
$\cos\vartheta$&FF&  & \\
\hline
\hline
--0.9&$F_{\sss{LL}}$
 &$68.36399900074-i1.24743850729   $&$79.63957322115-i20.53758995637  $\\
&&$68.36399900068-i1.24743850728   $&$79.63957322113-i20.53758995637  $\\
    &$F_{\sss{QL}}$
 &$75.12465846647+i34.81991916400  $&$76.19283172015+i28.44336684106  $\\
&&$75.12465846641+i34.81991916400  $&$76.19283172013+i28.44336684106  $\\
    &$F_{\sss{LQ}}$
 &$81.01546270426+i19.81343626967  $&$82.67283873006+i13.79952080171  $\\
&&$81.01546270420+i19.81343626968  $&$82.67283873004+i13.79952080171  $\\
    &$F_{\sss{QQ}}$
 &$225.63977621858+i154.37838168488$&$207.09189805263+i133.45188150116$\\
&&$225.63977621832+i154.37838168491$&$207.09189805254+i133.45188150117$\\
    &$F_{\sss{LD}}$
 &$ -0.57522852857+i0.34010611241~~$&$ -0.33030593699+i0.14897150833~~$\\
&&$ -0.57522852857+i0.34010611241~~$&$ -0.33030593699+i0.14897150833~~$\\
    &$F_{\sss{QD}}$
 &$ 0.16677424366-i0.34326069364   $&$  0.29925308488-i0.14107543098  $\\
&&$ 0.16677424366-i0.34326069364   $&$  0.29925308488-i0.14107543098  $\\
\hline
 0.0&$F_{\sss{LL}}$
 &$48.42950001713+i8.26103890366   $&$ 28.23570422021+i2.43705570966  $\\
&&$48.42950001707+i8.26103890367   $&$ 28.23570422019+i2.43705570966  $\\
&$F_{\sss{QL}}$
 &$68.02678564355+i37.08805801477  $&$ 58.00469565609+i33.82433896562 $\\
&&$68.02678564349+i37.08805801477  $&$ 58.00469565607+i33.82433896562 $\\
&$F_{\sss{LQ}}$
 &$73.37133716227+i22.69397728402  $&$ 62.40775508619+i20.75544388763 $\\
&&$73.37133716220+i22.69397728403  $&$ 62.40775508616+i20.75544388764 $\\
&$F_{\sss{QQ}}$
 &$196.60425612149+i162.74818773960$&$132.63279537966+i152.68259938740$\\
&&$196.60425612123+i162.74818773963$&$132.63279537957+i152.68259938741$\\
&$F_{\sss{LD}}$
 &$ -0.56319765502+i0.33645326768~~$&$ -0.29067043403+i0.13992893252~~$\\
&&$ -0.56319765502+i0.33645326768~~$&$ -0.29067043403+i0.13992893252~~$\\
&$F_{\sss{QD}}$
 &$ 0.15893936555-i0.37254018572   $&$  0.26429138671-i0.15437851127  $\\
&&$ 0.15893936555-i0.37254018572   $&$  0.26429138671-i0.15437851127  $\\
\hline
 0.9&$F_{\sss{LL}}$
 &$35.17736865724+i14.84038724783\,$&$\;0.21531292996+i13.66645015866 $\\
&&$35.17736865718+i14.84038724784\,$&$\;0.21531292994+i13.66645015866 $\\
    &$F_{\sss{QL}}$
 &$61.03099608330+i39.09196533610  $&$ 40.77942026097+i37.94118444135 $\\
&&$61.03099608324+i39.09196533611  $&$ 40.77942026095+i37.94118444136 $\\
    &$F_{\sss{LQ}}$
 &$66.08215572935+i25.04151178684  $&$ 44.50915974057+i25.51875704261 $\\
&&$66.08215572929+i25.04151178685  $&$ 44.50915974055+i25.51875704261 $\\
    &$F_{\sss{QQ}}$
 &$167.63393504156+i170.36384103672$&$~59.87568281297+i168.13599380718$\\
&&$167.63393504130+i170.36384103675$&$~59.87568281288+i168.13599380719$\\
    &$F_{\sss{LD}}$
 &$ -0.56772633347+i0.34299744419~~$&$ -0.32035310873+i0.14419510235~~$\\
&&$ -0.56772633347+i0.34299744419~~$&$ -0.32035310873+i0.14419510235~~$\\
    &$F_{\sss{QD}}$
 &$ 0.18031346246-i0.40091423652   $&$  0.34968026058-i0.16945266925  $\\
&&$ 0.18031346246-i0.40091423652   $&$  0.34968026058-i0.16945266925  $\\
\hline
\hline
\end{tabular}
\vspace*{-7mm}
\end{table}
\clearpage

The next \tbn{t2_p2} shows a comparison of {\tt eeffLib}--{\tt s2n.f} for the complete 
one-loop differential cross-sections ${d\sigma^{(1)}}/{d\cos\vartheta}$, for the standard input parameter set
\eqn{SIPS}. As seen, numbers agree within 12 or 13 digits.

\begin{table}[h]
\vspace*{-5mm}
\caption[$\frac{\ds d\sigma^{(1)}}{\ds d\cos\vartheta}$ for the process 
$\fep\fem\to\ft\bar{t}$. Comparison of {\tt eeffLib}--{\tt s2n.f}.]
        {$\frac{\ds d\sigma^{(1)}}{\ds d\cos\vartheta}$ for the process 
$\fep\fem\to\ft\bar{t}$. {\tt eeffLib}--{\tt s2n.f} comparison.}
\label{t2_p2}
\vspace*{3mm}
\centering
\begin{tabular}{||c|c|c|c||}
\hline
\hline
 $\sqrt{s}$   &   400 GeV   &   700 GeV   &   1000 GeV        \\
\hline
$\cos\vartheta$ &  &  &        \\
\hline
 --0.9   &   0.22357662754774  &    0.06610825350063  &  0.02926006442715   \\
$\qquad$ &   0.22357662754769  &    0.06610825350063  &  0.02926006442715
\\                         
\hline                                                      
   0.0   &   0.34494634728716  &    0.14342802645636  &  0.06752160108814   \\
$\qquad$ &   0.34494634728707  &    0.14342802645634  &  0.06752160108813
\\
\hline                                                              
   0.9   &   0.54806778978208  &    0.33837133344667  &  0.16973989931024   \\
$\qquad$ &   0.54806778978194  &    0.33837133344664  &  0.16973989931023
\\
\hline 
\hline 
\end{tabular} 
\vspace*{-5mm}
\end{table}

\subsection{About a comparison with the other codes}
As is well known, the one-loop differential cross-section of $\fep\fem\to\ft\bar{t}$
may be generated with the aid of the FeynArts system~\cite{Hahn:2000jm}.
Previous attempts to compare with FeynArts are described in~\cite{eett_subm}.
In December 2001, we were provided with the numbers computed with the FeynArts 
system~\cite{FynArts:2000} for $d\sigma/d\cos\vartheta$, 
with and without QED contributions,
at $\sqrt{\sman}=700$ GeV and three values of $\cos\vartheta=0.9,\;0,\;-0.9$.
With the current versions of {\tt eeffLib} and {\tt s2n.f} we have 11 digits agreement 
for both the tree-level and one-loop-corrected cross-sections.

\begin{table}[h]
\vspace*{-5mm}
\caption[$\frac{\ds d\sigma^{(1)}}{\ds d\cos\vartheta}$ for the process 
$\fep\fem\to\ft\bar{t}$ with soft photons, $E^{\rm max}_\gamma = \sqrt{s}/10$.]
{ $\frac{\ds d\sigma^{(1)}}{\ds d\cos\vartheta}$ for the process 
$\fep\fem\to\ft\bar{t}$ with soft photons, $E^{\rm max}_\gamma = \sqrt{s}/10$.}
\label{t8}
\vspace*{3mm}
\centering
\begin{tabular}{||c|c|c|c||}
\hline
\hline
 $\sqrt{s}$   &  400 GeV   &    700 GeV   &   1000 GeV        
\\
\hline
$\cos\vartheta$ &  &  &        \\
\hline
--0.9 &  0.17613018248935  &    0.05199100267864 &  0.02310170508071 \\
\hline
--0.5 &  0.21014509428358  &   0.06560630503586  &  0.02882301902010 \\
\hline
 0.0  &  0.27268108572063  &   0.11496514450150  &  0.05495088904853 \\
\hline
 0.5  &  0.35592722356682  &   0.19615154401629  &  0.09941700898317 \\
\hline
 0.9  &  0.43637377538440  &   0.27915043976042  &  0.14426233253975 \\
\hline 
\hline 
\end{tabular} 
\end{table}

Recently, the Bielefeld--Zeuthen team~\cite{Zeuthen:2001}
performed an alternative calculation using the DIANA system~\cite{DIANA}.
Working in close contact with this team, we managed to perform several high-precision
comparisons, reaching for separate contributions an agreement up to 10 digits.

The results of a comparison between FeynArts and the Bielefeld--Zeuthen team are 
presented in detail in \cite{K-BZcomparison:2002}.

 As another example we present in \tbn{t8} the same cross-section
$\big[\frac{ d\sigma}{ d\cos\vartheta}\big]_{\sss {\rm SM}}$ as given in the tables 
of~\cite{K-BZcomparison:2002}.
 For the complete cross-section, including soft photons, we agree with the 
Bielefeld--Zeuthen calculations within 8 digits. (See also~\cite{BZ_Japan}.)

\begin{figure}[!h]
\vspace*{-1.5cm}
\setlength{\unitlength}{0.1mm}
\begin{picture}(1800,1800)
\put(-220,0){\makebox(0,0)[lb]{\epsfig{file=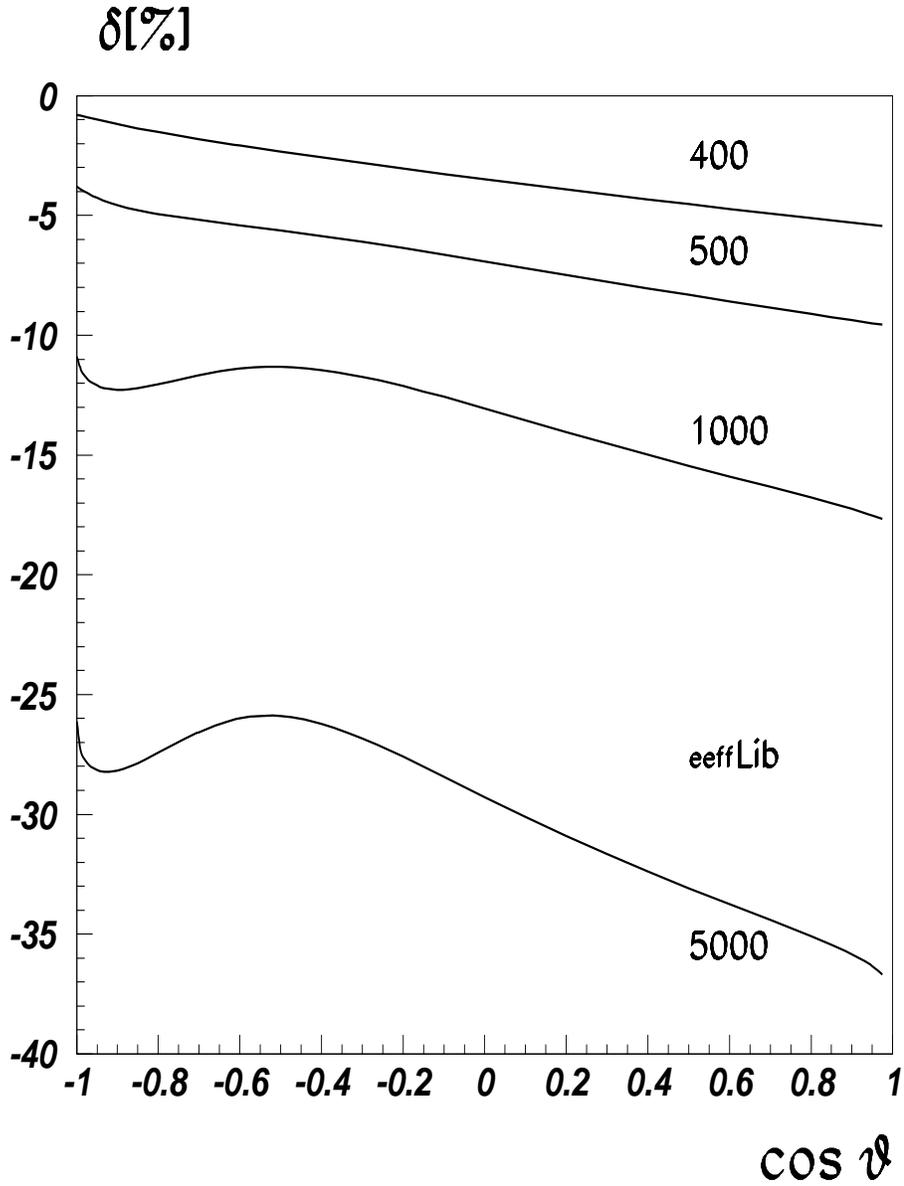,width=140mm,height=170mm}}}
\end{picture}
\caption[Relative EWRC to the $e^+e^-\to\ft\bar{t}$ differential cross-section.] 
{Relative EWRC~~$\delta(\sqrt{s},\cos\vartheta)=
\ds\frac{d\sigma^{(1)}(\sman,\tman)/d\tman}{d\sigma^{(0)}(\sman)/d\tman}-1$
to the $e^+e^-\to\ft\bar{t}$ differential cross-section.
Numbers near the curves show $\sqrt{\sman}$ in GeV.\label{fig:delta-double}}
\end{figure}

\begin{figure}[!h]
\vspace*{-1.5cm}
\setlength{\unitlength}{0.1mm}
\begin{picture}(1800,1800)
\put(-220,0){\makebox(0,0)[lb]{\epsfig{file=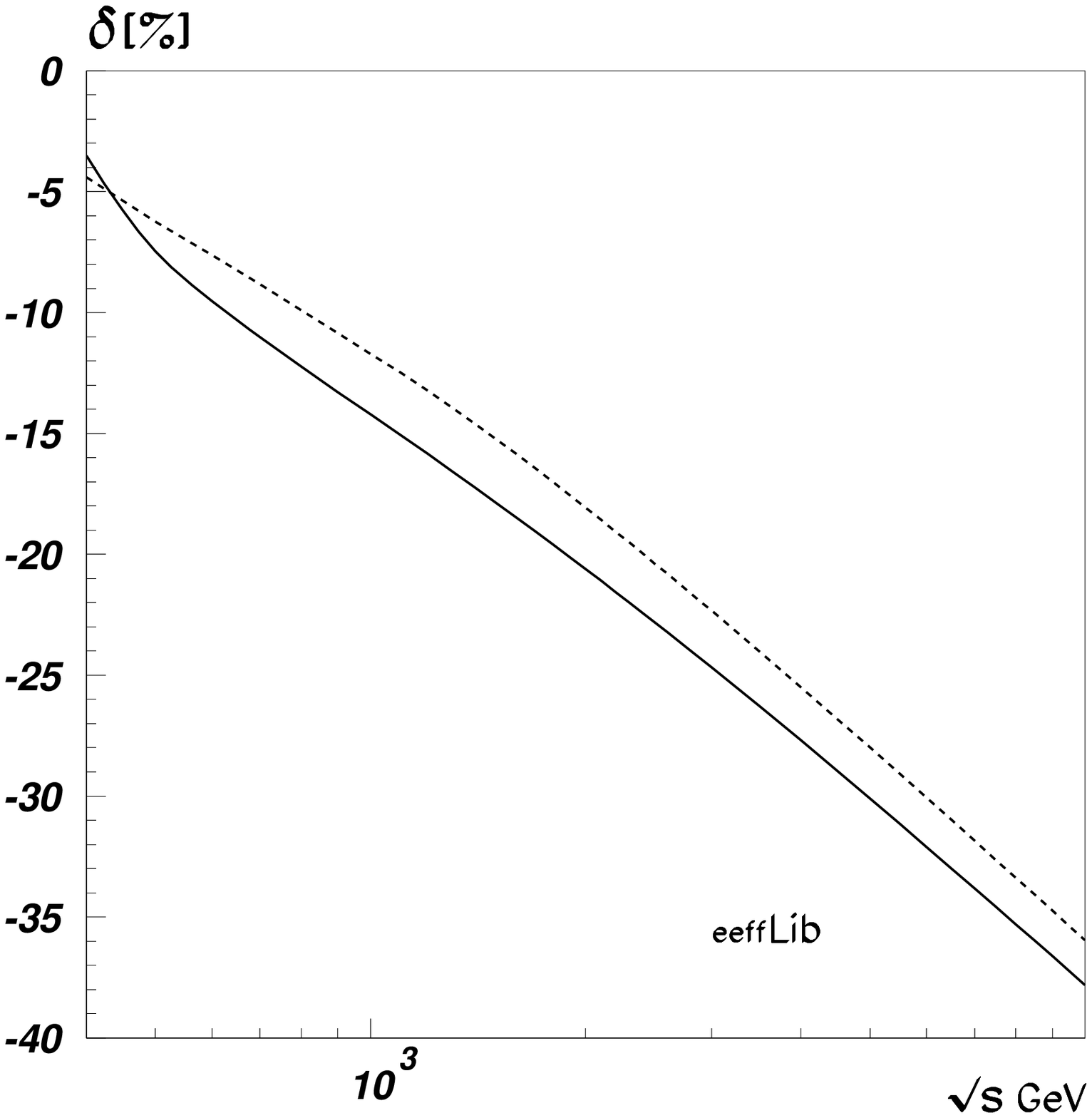,width=140mm,height=170mm}}}
\end{picture}
\caption[
 Relative EWRC to $e^+e^-\to\ft\bar{t}$ total cross-section.]
{Relative EWRC to $e^+e^-\to\ft\bar{t}$ for $\mhl=100$ GeV (solid line) and $\mhl=1000$ GeV
(dashed line).\label{fig:delta-integrated}}
\end{figure}

Two more graphical examples of the differential and the total EWRC for the process $e^+e^-\to\ft\bar{t}$
are shown in the figures \ref{fig:delta-double} and \ref{fig:delta-integrated}, correspondingly.

\addcontentsline{toc}{section}{Acknowledgments}
\section*{Acknowledgements}
 We are indebted to G. Passarino for valuable discussions.
 We would like to thank W.~Hollik and C.~Schappacher for a discussion of issues of the 
comparison with FeynArts.
 We acknowledge a common work on numerical comparison with J.~Flei\-scher, A.~Leike, T.~Riemann, 
and A.~Werthenbach, which helped us to debug our `manually written' code {\tt eeffLib}.
 We also wish to thank G.~Altarelli for extending to us the hospitality of the CERN TH Division 
at various stages of this work.

\clearpage
\addcontentsline{toc}{section}{References}
{

}

\end{document}